\documentclass[oldversion]{aa}

\usepackage{amsmath}
\usepackage{natbib}
\usepackage{epsfig}
\usepackage{txfonts}

\newcommand{\id}{{\,\rm d}}

\newcommand{\beq}{\begin{equation}}   %

\newcommand{\eeq}{\end{equation}}   %

\newcommand{\beqa}{\begin{eqnarray}}   %

\newcommand{\eeqa}{\end{eqnarray}}   %

\newcommand{\beal}{\begin{align}}

\newcommand{\enal}{\end{align}}

\newcommand{\bspl}{\begin{split}}

\newcommand{\espl}{\end{split}}

\newcommand{\bsub}{\begin{subequations}}

\newcommand{\esub}{\end{subequations}}

\newcommand{\bmulti}{\begin{multline}}   %

\newcommand{\beqm}{\begin{mathletters}}   %

\newcommand{\eeqm}{\end{mathletters}}   %

\newcommand{\Tg}{T_{\gamma}}

\newcommand{\Abl}[2]{\frac{{\rm d} #1}{{\rm d} #2}}

\newcommand{\pot}[2]{#1 \times 10^{#2}}

\newcommand{\xD}{{{x_{\rm D}}}}

\newcommand{\tauS}{{\tau_{\rm S}}}

\newcommand{\nbb}{{n_{\rm pl}}}

\usepackage{color}

\newcommand{\change}[1]{{#1}}

\newcommand{\changeI}[1]{{#1}}

\newcommand{\changeII}[1]{{#1}}

\begin{document}

\titlerunning{Ly $\alpha$ escape during cosmological hydrogen recombination}
\title{Ly $\alpha$ escape during cosmological hydrogen recombination:\\ the 3d-1s and 3s-1s two-photon processes}

\author{J. Chluba\inst{1,2} \and R.A. Sunyaev\inst{2,3}}
\authorrunning{Chluba \and Sunyaev}

\institute{Canadian Institute for Theoretical Astrophysics, 60 St. George Street,
Toronto, ON M5S 3H8, Canada 
\and 
Max-Planck-Institut f\"ur Astrophysik, Karl-Schwarzschild-Str. 1,
85741 Garching bei M\"unchen, Germany
\and
Space Research Institute, Russian Academy of Sciences, Profsoyuznaya 84/32,
117997 Moscow, Russia
}

\offprints{J. Chluba, 
\\ \email{jchluba@cita.utoronto.ca}
}

\date{Received / Accepted}

\abstract
{
We give a formulation of the radiative transfer equation for Lyman $\alpha$ photons which allows us to include the two-photon corrections for the 3s-1s and 3d-1s \changeII{decay} channels during cosmological hydrogen recombination.
We use this equation to compute the corrections to the Sobolev escape probability for Lyman $\alpha$ photons during hydrogen recombination, which then allow us to calculate the changes in the free electron fraction and CMB temperature and polarization power spectra.
We show that the effective escape probability changes by $\Delta P/P\sim +11\%$ at $z\sim 1400$ in comparison with the one obtained using the Sobolev approximation. 
This \changeII{speeds up of hydrogen recombination by} $\Delta N_{\rm e}/N_{\rm e}\sim-1.6\%$ at $z\sim 1190$, implying $|\Delta C_l/C_l | \sim 1\%-3\%$ at $l\gtrsim 1500$ with \changeII{shifts in the positions of the maxima and minima} in the CMB power spectra. These corrections will be important for the analysis of future CMB data.
The total correction is the result of the superposition of three independent processes, related to (i) {\it time-dependent} aspects of the problem,  (ii) corrections due to quantum mechanical deviations in the {\it shape} of the \changeII{emission and absorption profiles in the vicinity of the Lyman $\alpha$ line} from the normal Lorentzian, and (iii) a {\it thermodynamic} correction factor, \changeII{which occurs to be very important}. 
All these corrections are neglected in the Sobolev-approximation, but they are important in the context of future CMB observations.
\changeII{All three can be naturally obtained in the two-photon formulation of the Lyman $\alpha$ absorption process. However, the corrections (i) and (iii) can also be deduced in the normal '$1+1$' photon language, without necessarily going to the two-photon picture.}
\changeII{Therefore only} (ii) is really related to the quantum mechanical aspects of the two-photon process. 
\changeII{We show here that} (i) and (iii) lead to the largest separate contributions to the result, however they partially cancel each other close to $z\sim 1100$. At $z\sim 1100$ the \changeII{modification} due to the shape of the line profile contributes about $\Delta N_{\rm e}/N_{\rm e}\sim-0.4\%$, while the sum of the other \changeII{two} contributions gives $\Delta N_{\rm e}/N_{\rm e}\sim-0.9\%$.
}

\maketitle

\section{Introduction}
\label{sec:Intro}
After the seminal works of \citet{Zeldovich68} and \citet{Peebles68} on cosmological recombination,
and the improvements in the theoretical modeling of this epoch introduced later \citep[e.g.][]{Jones1985, Seager2000}, leading to the widely used standard recombination code {\sc Recfast} \citep{SeagerRecfast1999}, over the past few years the detailed physics of cosmological recombination has again been reconsidered by several independent groups \citep[e.g.][]{Dubrovich2005, Chluba2006, Kholu2006, Jose2006, Switzer2007I, Wong2007}. 
It is clear that understanding the cosmological ionization history at the level of $\sim 0.1\%$ \citep[e.g. see][for a more detailed overview of the different previously neglected physical processes 
that are important at this level of accuracy]{Sunyaev2008,Fendt2008} will be very important for accurate theoretical predictions of the Cosmic Microwave Background
(CMB) temperature and polarization angular fluctuations \citep[e.g. see][]{Hu1995, Seljak2003} in the context of the {\sc Planck}
Surveyor\footnote{www.rssd.esa.int/Planck}, which will be launched later this year. 

Also for a precise calibration of the {\it acoustic horizon} at recombination and the possibility to constrain dark energy using baryonic acoustic oscillation \citep[e.g.][]{EisensteinRev2005}, it is crucial to understand the physics of cosmological recombination at a high level of accuracy.
Ignoring percent-level corrections to the ionization history at last scattering ($z\sim 1100$) may therefore also result in significant biases to the cosmological parameters deduced using large catalogs of galaxies \citep[e.g.][]{Eisenstein2005, Gert2006}, as for example recently demonstrated for more speculative additions to the cosmological recombination scenario \citep{deBernardis2009} related to the possibility of {\it delayed recombination} \citep{Peebles2000}.

Among all the additional physical mechanisms during cosmological recombination that have been addressed so far, the problems connected with the {\it radiative transfer} of \ion{H}{i} Lyman $\alpha$ photons, including {\it partial frequency redistribution} and {\it atomic recoil} due to multiple resonance scattering, {\it electron scattering}, and corrections due to {\it two-photon processes} during \ion{H}{i} recombination ($z\sim 800-1600$), have still not been solved at full depth. 
Here we will focus on the inclusion of  two-photon corrections to the 3s-1s and 3d-1s emission and absorption process.

The potential importance of two-photon transitions from highly excited levels in hydrogen and helium was first pointed out by \citet{Dubrovich2005}.
They predicted a $\sim5\%$ decrease in the free electron fraction at
$z\sim1200$.  However, in their computations of the effective two-photon decay
rates for the $n$s and $n$d-levels they only included the first non-resonant
term (i.e. due to the dipole matrix element connecting $n{\rm s}/n{\rm
d}\rightarrow n{\rm p}$) into the infinite sum over intermediate states.
Also in their approach they neglected any possible transfer or reabsorption of photons in the vicinity of the Lyman $\alpha$ resonance, but simply assumed that {\it all} the photons accounted for by the inferred effective two-photon decay rate can directly escape.
Using rate coefficients for the vacuum two-photon decays of the 3s and 3d-levels in
hydrogen, as computed by \citet{Cresser1986}, \citet{Wong2007} concluded that
\citet{Dubrovich2005} overestimated the impact of two-photon transitions on
the ionization history by about one order of magnitude.
However, the calculation of \citet{Cresser1986} was incomplete, since in their attempt to separate the '$1+1$' photon contributions to the two-photon formula\footnote{This expression was first derived by \citet{Mayer1931}} from the 'pure' two-photon decay terms, without clear justification they neglected the first non-resonant term \citep{Chluba2008a}.
%
Physically it seems very difficult to {\it separate} the
`pure' two-photon decay rate from the '$1+1$' resonant contributions \citep[see
discussions in][]{Chluba2008a, Hirata2008, Karshenboim2008}, e.g. because of
non-classical interference effects.
In a complete analysis these contributions should be taken into account.
In addition, \citet{Wong2007} also neglected radiative transfer aspects of the problem.

Slightly later this problem was reinvestigated in more detail \citep{Chluba2008a}, showing that due to two-photon decays during hydrogen recombination a decrease of more than $\sim 0.3-0.5\%$ in the free electron fraction at $z\sim1150$ can still be expected.
This estimate was obtained by taking into account departures of the full $n$s-1s and $n$d-1s
two-photon line profiles from the Lorentzian shape in the very distant,
{\it optically thin} part of the red wing of the Lyman $\alpha$ line.
In these regions it can be assumed that {\it all} released photons can directly escape, and hence lead to a successful settling of the electron in the ground state. No radiative transfer formulation is needed to estimate this fraction of transitions, however as mentioned in their work the corrections coming from regions with significant radiative transfer can still be important.
According to their computations, the two-photon decays from s-states seem to
slow hydrogen recombination down, while those from d-states speed it up. In
addition it was shown that the slight net acceleration of hydrogen
recombination seems to be dominated by the 3s and 3d contribution
\citep{Chluba2008a}.

\changeII{Another} investigation of the two-photon aspects of the recombination problem was recently performed by \citet{Hirata2008}. He gave a formulation of the photon transfer problem simultaneously including {\it all} two-photon corrections during hydrogen recombination related to $n$s-1s, $n$d-1s, c-1s transitions and {\it Raman scattering} processes, also taking into account stimulated processes in the ambient CMB blackbody radiation field.
In order to solve this complicated problem two approaches were used. In the first the two-photon continuum was discretized and turned into an effective multilevel-atom with virtual states related to the energy of the photons. 
In the second approach the corrections were analytically modeled as effective modifications of the Lyman $\alpha$ and Lyman $\beta$ decay rates.
In addition, in both approaches a {\it distinction} between regions with '$1+1$' photon contributions and those with pure two-photon contributions was introduced to avoid the {\it double-counting} problem (see Sec. III.C of \citet{Hirata2008}) for the decay rates.
As pointed out this distinction is not unique, but the results were shown to be independent of the chosen parameters \citep{Hirata2008}, in total yielding $\Delta N_{\rm e}/N_{\rm e}\sim +1.3\%$ at $z\sim 900$ and  $\Delta N_{\rm e}/N_{\rm e}\sim -1.3\%$ at $z\sim 1300$.
%

Given the delicate complexity of the two-photon transfer problem it is very important to {\it independently} cross-validate the results obtained by different groups. 
In this paper we offer another approach to this problem in which we take into account the two-photon nature of the 3s-1s and 3d-1s decay channels, without introducing any criterion distinguishing between 'pure' two-photon decays and '$1+1$' resonant contributions.
%
%
We give a formulation of modified rate equations for the different hydrogen levels and the photon transfer equation, which we then use to compute the effective \ion{H}{i} Lyman $\alpha$ photon escape probability including these corrections.

Although it is clear that in particular the atomic recoil effect speeds up hydrogen recombination up at the percent-level \citep{Grachev2008, Chluba2008b} and also partial frequency redistribution will lead to some additional modifications\footnote{\changeII{As already mentioned in \citet{Chluba2008b}, our current version of the code already contains the corrections due to line diffusion on thermal atoms, atomic recoil and electron scattering. In good agreement with  \citet{Grachev2008} we found that atomic recoil is most important, but partial frequency redistribution only leads to an additional small modification. In \citet{Chlubaprep} we will present the results of these computations, also taking the 3d-1s and 3s-1d two-photon corrections into account.}}, here like in \citet{Hirata2008} we will neglect the frequency redistribution of photons due to resonance scattering and work in the {\it no line scattering} approximation. 
As explained in several previous works \citep{Chluba2008b, Switzer2007I, Jose2008, Hirata2008} for conditions in our Universe (practically no collisions) this is a much better description than the assumption of {\it complete redistribution}, which is used in the derivation of the Sobolev escape probability.
We also take into account stimulated 3s-1s and 3d-1s two-photon emission, finding this process to be sub-dominant. However, until now we do not include the effect connected with Raman scattering in this paper.

Instead of solving the obtained coupled system of equations simultaneously, we assume that the corrections will be small, so that each of them can be considered as a {\it perturbation} of the normal '$1+1$' photon result. 
Therefore we can use pre-computed solutions\footnote{We used the output of the latest version of our multilevel code \citep{Jose2006, Chluba2007}.} for the populations of the different hydrogen levels as a function of time to obtain the time-dependent photon emission rate for the different decay channels.  
%
%
This approach allows us to solve the \ion{H}{i} Lyman $\alpha$ radiative transfer equation {\it semi-analytically} also including the 3s-1s and 3d-1s two-photon corrections. 
Using the \changeII{obtained} solution for the spectral distortion at different redshifts one can then compute the {\it effective Lyman $\alpha$ escape probability} as a function of time.
\changeII{This value} can be directly compared to the normal Sobolev escape probability which then also allows to deduce the expected modification in the cosmological ionization history and CMB temperature and polarization power spectra.

Here we show that the effective escape probability changes by $\Delta P/P\sim +11\%$ at $z\sim 1400$ in comparison with the value derived in the Sobolev approximation (see Fig.~\ref{fig:DP_P.final}). 
As we explain in detail, this total correction is the result of the superposition of three independent processes, related to (i) {\it time-dependent} aspects of the problem,  (ii) corrections due to deviations in the {\it shape} of the \changeII{emission and absorption profiles in the vicinity of the Lyman $\alpha$ line} from the normal Lorentzian, and (iii) a {\it thermodynamic correction factor}. 
All these corrections are neglected in the cosmological recombination problem, but for the analysis of future CMB data they should be taken into account.

\changeII{In the '$1+1$' photon picture the} {\it purely time-dependent} correction was already discussed earlier \citep{Chluba2008b}, showing that changes in the state of the medium (e.g. number densities and Hubble expansion rate) cannot be neglected in the computation of the Lyman $\alpha$ escape probability. 
This is because only a very small fraction ($\sim 10^{-4}-10^{-3}$) of all interactions with the Lyman $\alpha$ resonance lead to a {\it complete redistribution} of photons over the whole line profile. As a consequence only the region inside the Doppler core reaches full equilibrium with the photon occupation number at the line center and can be considered using {\it quasi-stationary} conditions. 
However, outside the Doppler core time-dependent aspects of the problem have to be taken into account \citep{Chluba2008b}.

The second correction is related to {\it quantum mechanical} modifications in the {\it shape} of the line profiles describing the $n$s-1s and $n$d-1s two-photon decay channels.
\changeII{As we explain here, this is the only correction that can only be obtained when using the two-photon picture.}
As already discussed earlier \citep[e.g.][]{Chluba2008a}, this leads to deviations of the corresponding profiles from the normal Lorentzian. 
One consequence of this is that, depending on the considered process, {\it more} (for $n$d-1s transitions) or {\it fewer} (for $n$s-1s transitions) photons will directly reach the very distant red wing ($\xD\lesssim -1000$ Doppler width), where they can immediately escape. 
This correction was already estimated earlier \citep{Chluba2008a}, but here it will now be possible to refine these computations, also extending to regions closer to the line center, where radiative transfer effects are important.
Similarly, modifications in the blue wing emission can be taken into account using the approach presented here. Most importantly, because of the correct inclusion of energy conservation, the two-photon profiles will not extend to arbitrarily high frequencies. This will avoid the low redshift {\it self-feedback} that was recently seen in a time-dependent formulation of the Lyman $\alpha$ escape problem \citep{Chluba2008b}, and which here can be modeled more consistently.

The last and also most important correction discussed in this paper is related to a  {\it  frequency-dependent asymmetry} between the line emission and absorption process, that is normally neglected in the derivation of the Sobolev escape probability.
As pointed out earlier \citep{Chluba2008b}, within the normal '$1+1$' photon formulation for the line emission and absorption process especially in the damping wings of the Lyman $\alpha$ line a blackbody spectrum is {\it not exactly} conserved in full thermodynamic equilibrium.
This leads to the requirement of an additional factor, $f_\nu$, inside the absorption coefficient, which in the '$1+1$' photon picture can be deduced using the detailed balance principle (see Appendix~\ref{app:DB_f}).
However, within the two-photon formulation this correction {\it naturally} appears in connection with the two-photon absorption process, where one photon is taken from close to the Lyman $\alpha$ resonance and the other is drawn from the ambient CMB blackbody photon field at frequency following from energy conservation\footnote{For the 1s-3d two-photon absorption this will be $\nu'=\nu_{31}-\nu$, where $\nu_{31}$ is the corresponding 1s-3d transition frequency and $\nu$ denotes the frequency of the photon that is absorbed in the vicinity of the Lyman $\alpha$ resonance.} (see Sect.~\ref{sec:thermodyn} and in particular Sect.~\ref{sec:tau_f_corr}).

We will henceforth refer to $f_\nu$ as the {\it thermodynamic correction factor}.
It results in a {\it suppression} of the line absorption probability in the red, and an {\it enhancement} in the blue wing of the Lyman $\alpha$ resonance. 
This asymmetry becomes {\it exponentially} strong at large distances from the resonance.
In most astrophysical applications one is not interested in the photon distribution very far away from the Lyman $\alpha$ line center, so that this correction usually can be neglected. 
However, for the cosmological recombination problem even details at distances of $\sim 10^3-10^4$ Doppler width do matter \citep{Chluba2008b}, so that such an inconsistency in the formulation of the transfer problem has to be resolved.
As we will show here the associated correction is very important, leading to a significant speed-up of \ion{H}{i} recombination.

We also demonstrate that including all three modifications to the escape probability, the number density of free electrons is expected to change by $\Delta N_{\rm e}/N_{\rm e}\sim -1.3\%$ (see Fig.~\ref{fig:DN_N.final}). close to the maximum of the Thomson visibility function \citep{Sunyaev1970} at $z\sim 1100$, which matters most in connection with the CMB power spectra.
The 3s-1s and 3d-1s two-photon corrections (related to the shape of the profiles and the thermodynamic factor alone) yield $\Delta N_{\rm e}/N_{\rm e}\sim -2.4\%$ at $z\sim 1110$. A large part ($\sim 1.1\%$ at $z=1100$) of this correction is canceled by the contributions from the time-dependent aspect of the problem (see Fig.~\ref{fig:DN_N.final} for details). 
Our results seem to be rather similar to those of \citet{Hirata2008} for the contributions from high level two-photon decays alone \footnote{Note that this is only part of the total correction \changeII{which} was presented in \citet{Hirata2008}.}. 

We also compute the final changes in the CMB temperature and polarization power spectra when simultaneously including all processes under discussion here (see Fig.~\ref{fig:DCl}).
The corrections in the E-mode power spectrum are particularly impressive, reaching a peak to peak amplitude of $\sim 2\% - 3\%$ at $l\gtrsim 1500$, and significant shifts in the positions of the maxima in the CMB power spectra. Taking these corrections into account will be important for the future analysis of CMB data.
The paper is structured as follows: in Sect.~\ref{sec:main_eq} we give the equation for the modified Lyman $\alpha$ transfer problem. There we {\it infer} the equations by generalizing the normal '$1+1$' photon transfer equation in order to account for the mentioned processes. 
In the Appendix~\ref{sec:channels} we give a more rigorous derivation using the two-photon formulae, also generalizing the rate equations for the different hydrogen levels.
We then give the solution of the transfer equation in Sect.~\ref{sec:sol_tra} and show how to use it to compute the effective Lyman $\alpha$ escape probability (Sect.~\ref{sec:Nbar_abs}).
We explain the main physical differences and expectations for the corrections in comparison with the '$1+1$' photon formulation in Sect.~\ref{sec:sources_of_corr}.
Then we include 'step by step' the different correction terms and explain the changes in the results for the spectral distortion around the Lyman $\alpha$ line (Sect.~\ref{sec:F}) and the effective escape probability (Sect.~\ref{sec:Pesc}). 
In Sect.~\ref{sec:Ne} we then give the results for the ionization history and the CMB temperature and polarization power spectra.
We conclude in Sect.~\ref{sec:conc}.

\section{Two-Photon corrections to the Lyman $\alpha$ emission and absorption process}
\label{sec:main_eq}
\change{
The aim of this Section is to write down the line-emission and
absorption terms describing the evolution of the photon field in the
vicinity of the Lyman $\alpha$ resonance including the 3s-1s and 3d-1s two-photon
corrections.
Here we will try to motivate the form of this equation in terms of the additional physical aspects of the problem that should be incorporated. We refer the interested reader to Appendix~\ref{sec:channels} in which we provide the actual derivation of this equation using a two-photon formulation. There the central ingredient is that the photon distribution around the Balmer $\alpha$ line is given by the CMB blackbody. This fact makes it possible to rewrite the two-photon transfer equation as effective equation for one photon, as presented here.

In this Section we also give the solution of the modified transfer equation (Sect.~\ref{sec:sol_tra}) and explain how one can use it to compute the effective escape probability for the Lyman $\alpha$ photons (Sect.~\ref{sec:Nbar_abs}).
}

\subsection{Modified equation describing the emission and death of Lyman $\alpha$ photons}
Within the semi-classical formulation of the Lyman $\alpha$ transfer equation
every relevant physical process is envisioned as a single step process involving one
photon of the photon field. This leads to the introduction of \change{photon} {\it death} and
{\it scattering} probabilities that {\it only} depend on redshift \citep[e.g. see][]{Chluba2008b}.
Also in the single photon picture the line profiles for the different
Lyman $\alpha$ emission and absorption channels under the assumption of {\it complete redistribution} are all identical.
For example, it will not make difference if the
electron reaches the 2p-state and then goes to the 3s, 3d or continuum.
In all three cases the absorption profile will be given by the usual Voigt-profile.
As explained earlier \citep{Chluba2008b}, in the normal '$1+1$' photon language the Lyman $\alpha$ line-emission and absorption terms can be cast into the form
\beal
\label{eq:real_em_abs_11}
\frac{1}{c}\left.\Abl{N_{\nu}}{t}\right|^{\rm 1\gamma}_{\rm Ly-\alpha}
&=\frac{\phi_{\rm V}(\nu)}{4\pi\,\Delta\nu_{\rm D}}\times\left[p_{\rm em}^{\rm 1\gamma}\,
R^{+}_{\rm 2p}
-
p^{\rm 1\gamma}_{\rm d}\,h\nu_{\rm 21}\,B_{12}\, N_{1\rm s}\,N_{\nu}\right]. 
\end{align}
Here $\phi_{\rm V}(\nu)$ is the usual Voigt-profile (see Appendix~\ref{app:profiles} for definition), with normalization
$\int_0^\infty \frac{\phi(\nu)}{4\pi\,\Delta\nu_{\rm
    D}}\id\nu\id\Omega\equiv 1$, where $\Delta\nu_{\rm D}$ is the
Doppler-width of the Lyman $\alpha$ line.
Furthermore, $p_{\rm em}^{\rm 1\gamma}$ is the Lyman $\alpha$ emission probability in
the '$1+1$' photon picture\footnote{For formal consistency we included the factors $1+\nbb(\nu_{21})$ due to stimulated emission in the ambient CMB blackbody field in the definition of $p_{\rm em}^{\rm 1\gamma}$ although during recombination $\nbb(\nu_{21})\ll 1$.}, as given by Eq.~\eqref{eq:p_2p1s}, and $p_{\rm d}^{\rm 1\gamma}=1-p_{\rm em}^{\rm 1\gamma}$ the corresponding death probability.
$R^{+}_{\rm 2p}$ describes the rate at which fresh electrons are added to the 2p-state, and is defined by Eq.~\eqref{eq:Rp2p}. 

\subsubsection{Introducing the thermodynamic correction factor}
\label{sec:thermodyn}
\change{
As mentioned in the introduction, in this form Eq.~\eqref{eq:real_em_abs_11} does {\it not exactly} conserve a blackbody spectrum in the case of 
full thermodynamic equilibrium. 
Knowing the '$1+1$' photon line emission term and using the detailed balance principle one can obtain the {\it thermodynamic correction factor}\footnote{We added a short derivation for $f_\nu$ in Appendix~\ref{app:DB_f}.}
\beal
\label{eq:f_fac}
f_\nu(z)=\frac{\nu_{21}^2}{\nu^2}\,e^{h[\nu-\nu_{21}]/k\Tg(z)}
\end{align}
which is necessary to avoid this problem. 
\changeII{This factor was introduced in \citet{Chluba2008b} already.}
Inserting it into Eq.~\eqref{eq:real_em_abs_11} we then have
\beal
\label{eq:real_em_abs_11_tilde}
\frac{1}{c}\left.\Abl{N_{\nu}}{t}\right|^{\rm 1\gamma}_{\rm Ly-\alpha}
&=\frac{\phi_{\rm V}(\nu)}{4\pi\,\Delta\nu_{\rm D}}\left[p_{\rm em}^{\rm 1\gamma}\,
R^{+}_{\rm 2p}
-
p^{\rm 1\gamma}_{\rm d}\,h\nu_{\rm 21}\,B_{12}\, N_{1\rm s}\,f_\nu\,N_{\nu}\right]. 
\end{align}
%
In the standard '$1+1$' photon formulation $f_\nu$ has no direct physical interpretation. It is simply a consequence of thermodynamic requirements on the form of the equations.
However, as shown in Appendix~\ref{sec:channels} the same factor $f_\nu$ naturally appears in a two-photon formulation of the problem. 
It is actually related to the shape of the photon distribution in the vicinity of the second photon that is involved in the Lyman $\alpha$ absorption process (Sect.~\ref{sec:tau_f_corr}). 
This is due to the fact that the photon which enables the 2p-electron to reach
the 3s, 3d, or continuum is drawn from the ambient radiation field, which in the cosmological recombination problem is given by the CMB blackbody.
}

\subsubsection{Including the corrections due to the profiles of the different decay channels}
\label{sec:profile_corrections}
\change{
As a next step we want to take the differences between the line profiles of the different absorption and emission channels into account. 
One can see that in Eq.~\eqref{eq:real_em_abs_11_tilde} there is {\it no distinction} made between the different routes the electron took before or after entering the ${\rm 1s}\leftrightarrow {\rm 2p}$ transition.
However, as mentioned in the introduction, the line-emission profiles
depend on how the fresh electron reached the 2p-state via channels other than the Lyman $\alpha$ transition. 

In order to distinguish between the different possibilities (e.g. ${\rm 1s}\leftrightarrow {\rm 2p}\leftrightarrow {\rm 3s/3d/c}$), one should allow for profiles, $\phi_{i}(\nu)$, that depend on the channel $i$. 
Also the partial rate at which electrons enter the 2p-state will depend on $i$, leading to the replacement $R^{+}_{\rm 2p}\rightarrow R^{i,+}_{\rm 2p}$ with $R^{+}_{\rm 2p}=\sum_i R^{i,+}_{\rm 2p}$, where the sum runs over all possible '$1+1$' photon channels via which the number of
Lyman $\alpha$ photons can be affected.
Furthermore, the probability with which electrons are absorbed will become channel-dependent, so that $p^{\rm 1\gamma}_{\rm d}\rightarrow
p^{i}_{\rm d}$ with $p^{\rm 1\gamma}_{\rm d}=\sum_i p^{i}_{\rm d}$.

Here it is important that $R^{i,+}_{\rm 2p}$ and  $p^{i}_{\rm d}$ both will only depend on time but not on frequency. This is because microscopically it is assumed that the absorption process leads to a complete redistribution over the profile $\phi_{i}(\nu)$.
With this comment it is also clear that the factor $f_\nu$ should be independent of the channel, since otherwise detailed balance for each process cannot be achieved.

With this in mind it is clear that the more general form of Eq.~\eqref{eq:real_em_abs_11_tilde}  should
read
%
\beal
\label{eq:real_em_abs_11_gen}
\frac{1}{c}\left.\Abl{N_{\nu}}{t}\right|_{\rm Ly}
&\!\!\!
=\!\sum_{i}\frac{\phi_{i}(\nu)}{4\pi\,\Delta\nu_{\rm D}}\left[p_{\rm em}^{1\gamma}\,
R^{i,+}_{\rm 2p}
-
p^{i}_{\rm d}\,h\nu_{\rm 21} B_{12}\, N_{1\rm s}\,f_\nu\,N_{\nu}\right]. 
\end{align}
In Appendix~\ref{sec:channels} we argue that both $R^{i,+}_{\rm 2p}$ and $p^{i}_{\rm d}$ can be given using the normal '$1+1$' photon values for the different rates. 
We also specify how to compute the profiles, $\phi_{i}(\nu)$, including stimulated two-photon emission \changeII{(Sect.~\ref{app:2gamma_profiles})}.
However, in what follows it is only important that non of these depends on the solution of the problem for the photon field. This is because we assume that the readjustments in the populations of the different level or number density of free electrons is small and hence can be neglected to lowest order.
Numerically one can include the correction to the correction iteratively, but we leave this for a future paper.
}

It is important to mention that because for two-photon transitions $n{\rm s/d}\rightarrow 1{\rm s}$ from $n>3$ also photons connected with the other Lyman series are emitted, Eq.~\eqref{eq:real_em_abs_11_gen} in principle can be used to describe the simultaneous evolution of all Lyman series photons. Similarly, one can account for the two-photon corrections due to transitions from the continuum ${\rm c}\rightarrow 1{\rm s}$, simultaneously  including the Lyman continuum and all other continuua.
However, in this case one can no longer clearly distinguish between the different Lyman series. Also the equation will simultaneously describe the process of Ly-$n$ feedback \citep{Chluba2007b}, in addition accounting for its exact time-dependence.
To avoid these complications, below we will first only take into account the two-photon corrections for the 3s-1s and 3d-1s channel, but leave the others unchanged. In this case it is possible to directly compare the results with the Lyman $\alpha$ problem. \changeII{In Sect.~\ref{sec:conc}} we briefly discuss the expected effect of this approximation, but leave a detailed analysis for another paper.

\subsection{Solution of the transfer equation}
\label{sec:sol_tra}
For a given ionization history, the solution of Eq.~\eqref{eq:real_em_abs_11_gen} in the
expanding Unverse can be readily found, using the proceedure described
in \citet{Chluba2008b}.
If we introduce the effective absorption optical depth as
\bsub
\label{sec:tau_abs_all}
\beal
\label{sec:tau_abs_all_a}
\tau_{\rm abs}(\nu, z', z)
&=\int_{z}^{z'}
p^{1\gamma}_{\rm d}\,
\frac{c\,\sigma_{\rm r}\,N_{\rm 1s}}{H (1+\tilde{z})}\,\phi_{\rm abs}(x[1+\tilde{z}], \tilde{z})\id \tilde{z}
\\[1mm]
\label{sec:tau_abs_all_b}
\phi_{\rm abs}(\nu, z)
&=
 f_\nu(z) 
 \sum_{i} \frac{p^{i}_{\rm d}}{p^{1\gamma}_{\rm d}}\, \phi_{i}(\nu, z)=f_\nu(z)\, \phi^\ast_{\rm abs}(\nu, z)
\end{align}
\esub
with $p^{1\gamma}_{\rm d}=1-p^{1\gamma}_{\rm em}$, $\sigma_{\rm r}=\frac{h\nu_{21}}{4\pi}\,\frac{B_{12}}{\Delta\nu_{\rm D}}$ and the dimensionless frequency $x=\nu/(1+z)$, and define the effective emission profile
\beal
\label{sec:phi_em}
\phi_{\rm em}(\nu, z)
&=
 \sum_{i} \frac{R^{i,+}_{\rm 2p}}{R^{+}_{\rm 2p}}\, \phi_{i}(\nu, z)
\end{align}
then Eq.~\eqref{eq:real_em_abs_11_gen} takes the simple form
\bsub
\label{eq:real_em_abs_simp}
\beal
\label{eq:real_em_abs_simp_a}
\frac{1}{c}\left.\Abl{N_{\nu}}{t}\right|_{\rm Ly}
&\!=\!
p^{1\gamma}_{\rm d}\,\sigma_{\rm r}\,N_{\rm 1s}\,\phi_{\rm abs}(\nu, z)
\left\{N^{\rm em}_\nu-N_{\nu}\right\}
\\
\label{eq:real_em_abs_simp_b}
N^{\rm em}_\nu
&\!=\!\frac{2\nu_{21}^2}{c^2}\frac{g_{\rm 1s}}{g_{\rm 2p}}
\frac{R^{+}_{2 \rm  p}}{R^{-}_{2 \rm p}N_{\rm 1s}}\times\frac{\phi_{\rm em}(\nu, z)}{\phi_{\rm abs}(\nu, z)}
\equiv N_{\rm em}\,\frac{\phi_{\rm em}(\nu, z)}{\phi_{\rm abs}(\nu, z)},
\end{align}
\esub
where $N_{\rm em}$ is only redshift dependent.

The solution of this equation in the expanding Universe can be directly given \citep[see][]{Chluba2008b}
\bsub
\label{app:kin_abs_em_Sol_phys_asym}
\beal
\label{app:kin_abs_em_Sol_phys_asym_a}
\Delta N^{\rm asym}_{\nu}(z)
&=[N_{\rm em}(z)-N^{\rm pl}_{\nu_{21}}]\times F(\nu, z).
\end{align}
Here the function $F_\nu$ represents the frequency dependent part of the solution for the spectral distortion, which is defined by
\beal
\label{app:kin_abs_em_Sol_phys_asym_F}
F(\nu, z)&=\!\!\int_{z_{\rm s}}^z \!\!\Theta^{\rm a}(z, z')
\,
\partial_{z'}e^{-\tau_{\rm abs}(\nu, z', z)}\id z'
\\
\label{app:kin_abs_em_Sol_phys_asym_b}
\Theta^{\rm a}(z, z')
&=\frac{
\tilde{N}_{\rm em}(z')\times\frac{\phi_{\rm em}(\nu', z')}{\phi_{\rm abs}(\nu',z')}-\tilde{N}^{\rm pl}_x}{\tilde{N}_{\rm em}(z)-\tilde{N}^{\rm pl}_{x_{21}}}
\\
&\equiv
\frac{1}{f_{\nu'}}
\frac{
\tilde{N}_{\rm em}(z')\times\frac{\phi_{\rm em}(\nu',z')}{\phi^\ast_{\rm abs}(\nu',z')}-\tilde{N}^{\rm pl}_{x'_{21}} }{\tilde{N}_{\rm em}(z)-\tilde{N}^{\rm pl}_{x_{21}}},
\end{align}
\esub
where $\Delta N_{\nu}=N_{\nu}-N^{\rm pl}_\nu$, $\nu'=x[1+z']$ and at $z>z_{\rm s}$ the CMB spectrum is assumed to be given by a pure blackbody spectrum $N^{\rm pl}_\nu$.
Furthermore, $\tilde{N}_{\rm em}(z)=N_{\rm em}(z)/[1+z]^2$, $x_{21}=\nu_{21}/[1+z]$, $x'_{21}=\nu_{21}/[1+z']$, and $\tilde{N}^{\rm pl}_{x}=\frac{2}{c^2}\frac{x^2}{e^{hx/kT_0}-1}$, with $T_0=2.725\,$K.
Note that $\tilde{N}^{\rm pl}_{x}$ does not explicitly depend on redshift.
Also we have used that $f_{\nu'}(z') \tilde{N}^{\rm pl}_x\equiv \tilde{N}^{\rm pl}_{x'_{21}}$.

\subsection{Number of absorbed photons and the effective Lyman $\alpha$ escape probability}
\label{sec:Nbar_abs}
With the solution~\eqref{app:kin_abs_em_Sol_phys_asym} one can directly compute the number of absorbed photons. \change{For this we define the mean of $N_\nu=I_\nu/h\nu$ over the absorption profile}
\beal
\label{eq:N_bar_abs_def}
\bar{N}_{\rm abs}(z)
&=\int \frac{\phi_{\rm abs}(\nu,z)}{4\pi\,\Delta\nu_{\rm D}}N_{\nu}\id\nu\id\Omega
=\int \frac{\phi^\ast_{\rm abs}(\nu, z)}{4\pi\,\Delta\nu_{\rm D}} \,f_\nu \,N_{\nu}\id\nu\id\Omega
\nonumber\\[1mm]
&=
N^{\rm pl}_{\nu_{21}}\,\bar{\varphi}^\ast_{\rm abs}
+\int \varphi^\ast_{\rm abs}(\nu, z)\,f_\nu(z) \,\Delta N_{\nu}\id\nu.
\end{align}
where we have set $\varphi^\ast_{\rm abs}(\nu, z)=\phi^\ast_{\rm abs}(\nu, z)/\Delta\nu_{\rm D}$ and introduced the norm of $\varphi^\ast_{\rm abs}$ as 
$\bar{\varphi}^\ast_{\rm abs}=\int \varphi^\ast_{\rm abs}(\nu, z)\id\nu
$.
If we now insert the solution~\eqref{app:kin_abs_em_Sol_phys_asym} into this expression we can write
\bsub
\beal
\label{eq:DN_bar_abs_def}
\Delta \bar{N}^{\rm asym}_{\rm abs}(z)
&=[N_{\rm em}(z)-N^{\rm pl}_{\nu_{21}}][1-P]
\\[1mm]
\label{eq:def_P}
P&=
1-
\!\int \!\varphi^\ast_{\rm abs}(\nu)\, f_\nu(z)\, F_\nu \id\nu.
\end{align}
\esub
Here $P$ will later be interpreted as the main part of the effective escape probability \changeII{(see Sects.~\ref{sec:DF_DP} and \ref{sec:Pesc})}. 

Similar to $\bar{N}_{\rm abs}(z)$ one can also define
\beal
\label{eq:N_bar_em_def}
\bar{N}_{\rm em}(z)
&=\int \frac{\phi_{\rm abs}(\nu, z)}{4\pi\,\Delta\nu_{\rm D}}N^{\rm em}_{\nu}\id\nu\id\Omega
\stackrel{!}{\equiv} \int \frac{\phi_{\rm em}(\nu, z)}{4\pi\,\Delta\nu_{\rm D}} N_{\rm em}(z)\id\nu\id\Omega
\nonumber
\\[1mm]
&=
N_{\rm em}\,\bar{\varphi}_{\rm em},
\end{align}
so that with the transfer equation~\eqref{eq:real_em_abs_simp} it follows
\bsub
\beal
\label{eq:dNgdt}
\left.\Abl{N_{\gamma}}{t}\right|_{\rm Ly}
&=\frac{1}{c} \, \int\left.\Abl{N_{\nu}}{t}\right|_{\rm Ly}\id\nu\id\Omega
\nonumber\\
&=
p^{1\gamma}_{\rm d}\,h\nu_{21}B_{12}\,N_{\rm 1s}
\left\{
N_{\rm em}\,\bar{\varphi}_{\rm em}
-N^{\rm pl}_{\nu_{21}}\,\bar{\varphi}^\ast_{\rm abs}
\right.
\nonumber
\\
&\qquad\qquad\qquad\qquad\quad
\left.
-[N_{\rm em}-N^{\rm pl}_{\nu_{21}}][1-P]
\right\}
\nonumber\\[1mm]
&=
p^{1\gamma}_{\rm d}\,h\nu_{21}B_{12}\,N_{\rm 1s}
\left\{
P_{\rm eff}\,[N_{\rm em}-N^{\rm pl}_{\nu_{21}}]
\right\},
\end{align}
with 
\beal
\label{eq:DP_eff}
P_{\rm eff}&=P+\Delta P_{\rm ind}
\\
\label{eq:DP_ind}
\Delta P_{\rm ind}&=
\frac{N_{\rm em}\,\Delta\bar{\varphi}_{\rm em}-N^{\rm pl}_{\nu_{21}}\,\Delta\bar{\varphi}^\ast_{\rm abs}}
{N_{\rm em}-N^{\rm pl}_{\nu_{21}}}
\end{align}
\esub
where $\Delta\bar{\varphi}_{\rm em}=\bar{\varphi}_{\rm em}-1$ and $\Delta\bar{\varphi}^\ast_{\rm abs}=\bar{\varphi}^\ast_{\rm abs}-1$. 
As we explain below, with these definitions the effective escape probability, $P_{\rm eff}$, can now be directly compared with the value in the normal '$1+1$' photon formulation and the Sobolev escape probability.

\subsubsection{Range of integration over the profiles}
\label{sec:integration}
 In the above derivation we have not specified the range of integration. Since the 3s and 3d two-photon profile include both the Balmer $\alpha$ and Lyman $\alpha$ photons, by carrying out the integrals over the frequency interval $(0,\infty)$, one would count $2\gamma$ per transition. 
In order to avoid this problem, we can simply restrict the range of integration to $\nu\geq \nu_{31}/2$, but leave all the other definitions unaltered. Since $\nu_{31}/2$ is far away from the Lyman $\alpha$ resonance this does not lead to any significant problem regarding the normalization of the normal Voigt-function\footnote{$\nu_{31}/2$ corresponds to $\xD\approx -\pot{1.7}{4}\left[\frac{1+z}{1100}\right]^{-1/2}$ Doppler width, so that the absolute error in the normalization of $\phi_{\rm V}$ is $\sim \pot{1.6}{-8}$.}. 
 In addition, for the quasi-stationary approximation the contribution to the value of the escape probability from this region are completely negligible.
 Therefore this restriction does not lead to any bias in the result, but does simplify the numerical integration significantly.

\subsubsection{Relating the corrections in the spectral distortion to the corrections in the effective escape probability}
\label{sec:DF_DP}
We now want to understand how differences in $F_\nu$ \changeII{and $\Abl{N_{\gamma}}{t}$} relate to corrections in the effective escape probability.
For this we first want to emphasize that in the normal '$1+1$' photon picture, under the assumption of quasi-stationarity and in the no line scattering approximation, following the \changeII{derivation of the previous Section} one would find \citep{Chluba2008b} 
\beal
\label{eq:Ng_P_d}
\left.\Abl{N^{\rm d}_{\gamma}}{t}\right|_{\rm Ly}=
p^{1\gamma}_{\rm d}\,h\nu_{21}B_{12}\,N_{\rm 1s}\,
P_{\rm d}\,[N_{\rm em}-N^{\rm pl}_{\nu_{21}}]
\end{align}
with $P_{\rm d}=\frac{1-e^{-\tau_{\rm d}}}{\tau_{\rm d}}$ and $\tau_{\rm d}=p^{1\gamma}_{\rm d}\tau_{\rm S}$.

It is clear that $\Abl{N^{\rm d}_{\gamma}}{t}\,\Delta t$ represents the effective change in the total number density of photons involved in the Lyman $\alpha$ evolution over a short time interval $\Delta t$, and hence \changeII{is directly} related to the change in the total number of electrons that settle in the ground state via the Lyman $\alpha$ channel. By comparing $P_{\rm d}$ with $P_{\rm eff}$, as defined by Eq.~\eqref{eq:DP_eff}, one can therefore deduce the required effective correction to the Sobolev escape probability, that is normally used in the formulation of the recombination problem.
Following the arguments of \citet{Chluba2008b} this correction should be given by
\beal
\label{eq:DP_S}
\Delta P_{\rm S}=\frac{p^{1\gamma}_{\rm d}\,P_{\rm eff}}{1-p^{1\gamma}_{\rm em}\,P_{\rm eff}}-P_{\rm S},
\end{align}
 where $P_{\rm S}=\frac{1-e^{-\tau_{\rm S}}}{\tau_{\rm S}}$ is the standard Sobolev escape probability, with the usual Sobolev optical depth, $\tau_{\rm S}$.

%

%
\section{Main sources of corrections to the Lyman $\alpha$ spectral distortion}
\label{sec:sources_of_corr}
Using the solution~\eqref{app:kin_abs_em_Sol_phys_asym} one can already identify the main sources for the corrections to the photon distribution in comparison with the quasi-stationary approximation.
These can be split up into those acting as a time and frequency dependent {\it emissivity}, which is characterized by $\Theta^{\rm a}$, and those just affecting the {\it absorption optical depth}, $\tau_{\rm abs}$.
Below we explain how the two-photon aspect of the problem enters here, and which effects are expected. In Sect.~\ref{sec:F} and \ref{sec:Pesc} we discuss the corrections to the Lyman $\alpha$ spectral distortion and the effective escape probability in comparison with the standard '$1+1$' photon formulation in more detail.

\begin{figure}
\centering 
\includegraphics[width=0.98\columnwidth]
{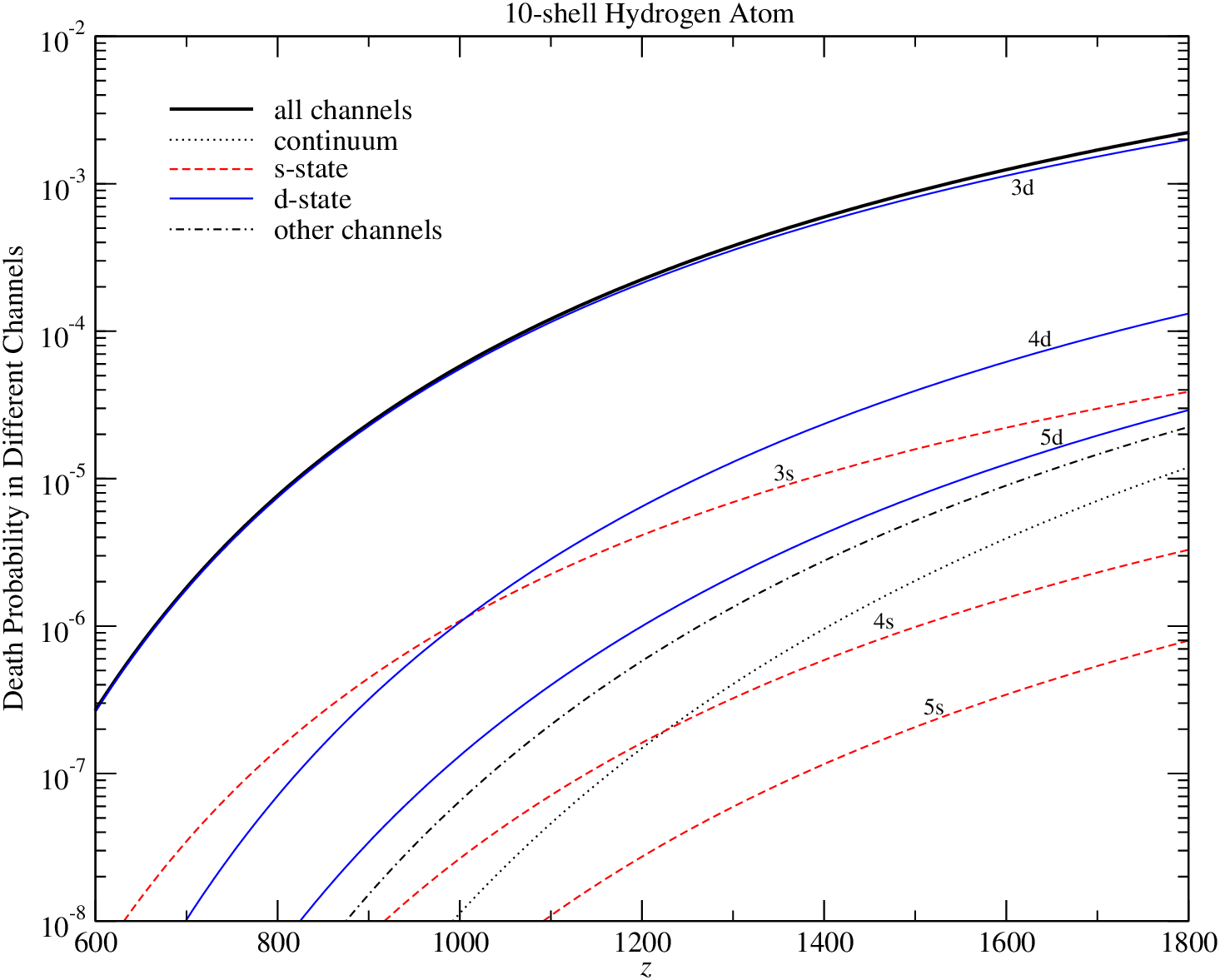}
\caption
{The death probabilities for different Lyman $\alpha$ absorption channels. We used a 10-shell hydrogen atom. The thick solid line shows the total death probability, $p_{\rm d}$ \changeI{(for definition see Appendix~\ref{sec:channels})}. }
\label{fig:p_d}
\end{figure}

\begin{figure*}
\centering 
\includegraphics[width=0.95\columnwidth]
{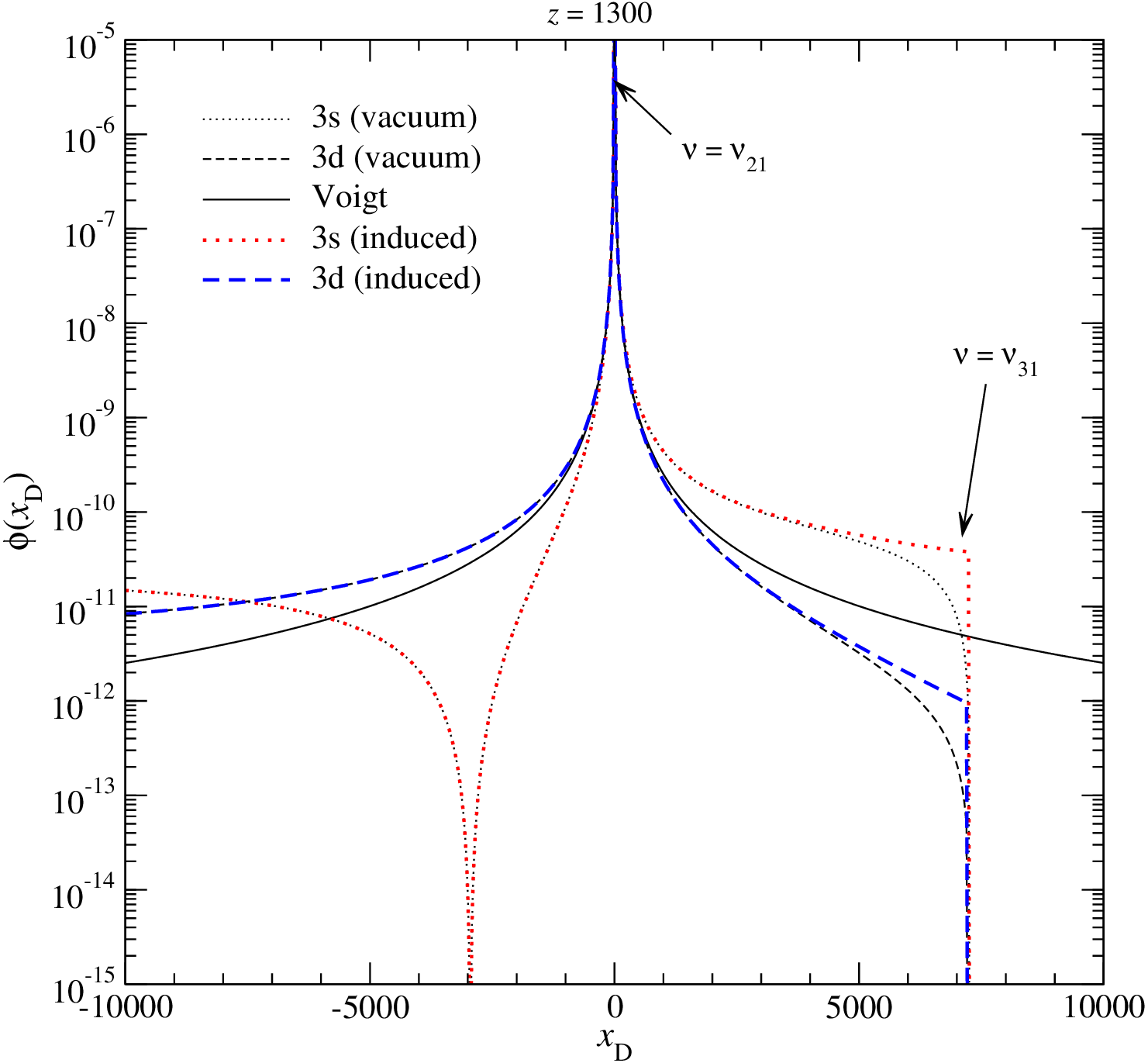}
\hspace{4mm}
\includegraphics[width=0.95\columnwidth]
{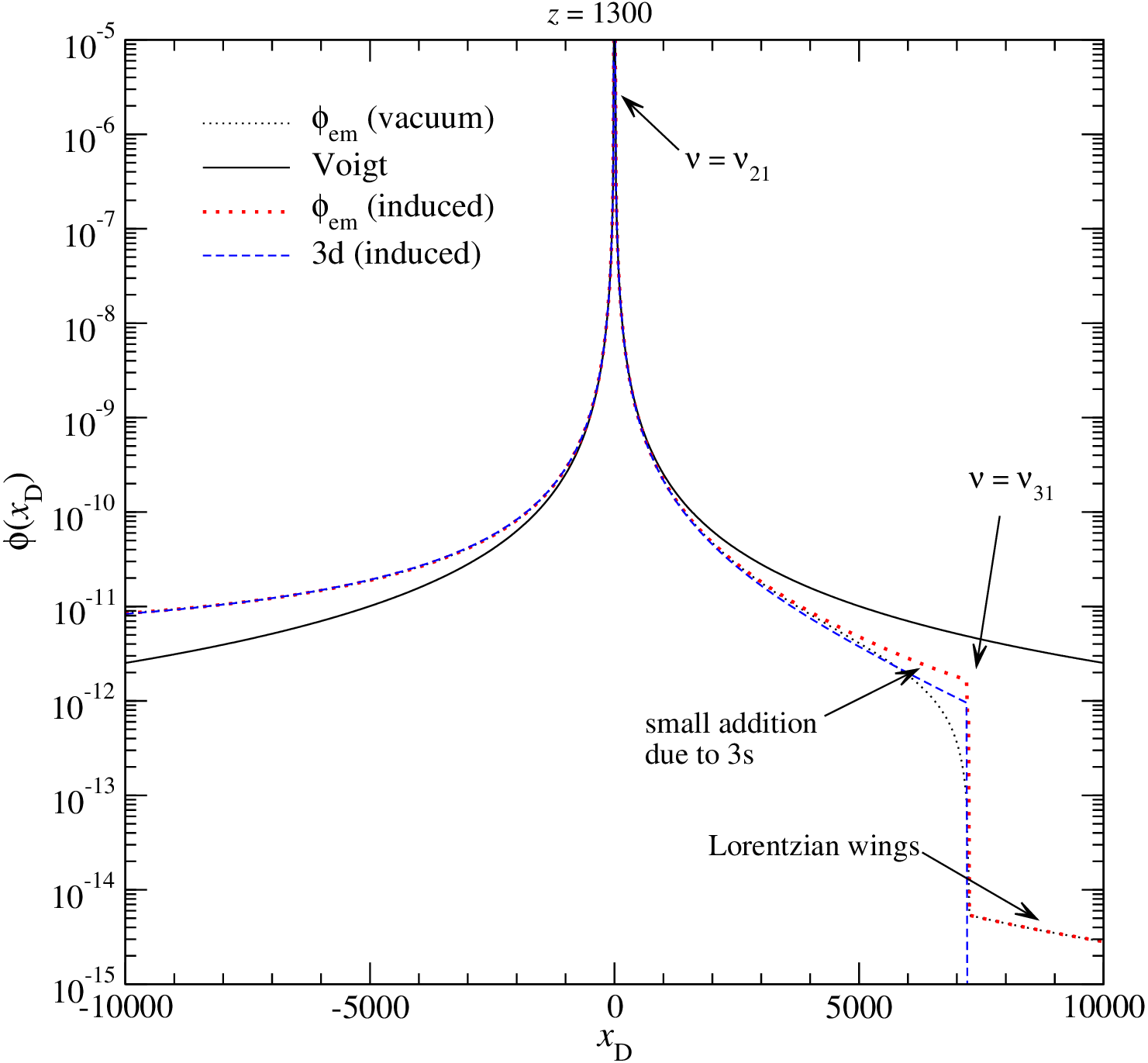}
\caption
{Different line profiles in the vicinity of the Lyman $\alpha$ resonance at redshift $z=1300$. The left panel shows the 3s and 3d emission profiles in comparison with the normal Voigt profile. In the right panel we show the effective emission profile \change{for a 3 shell hydrogen atom}, as defined by Eq.~\eqref{sec:phi_em}, in comparison with the 3d emission profile and the Voigt profile. The curves labeled 'induced' include the effect of stimulated two-photon emission \change{due to enhancement connected with the second photon released at low frequencies. This process is only important close to $\nu\sim\nu_{31}$ and eventually leads to a sub-dominant correction of $\Delta N_{\rm e}/N_{\rm e}\lesssim 0.1\%$ in the ionization history. On the other hand, the overall asymmetry  in the effective emission profile \changeI{(red wing stronger that blue wing)} has important implications for the hydrogen recombination problem \citep[see][]{Chluba2008a}.}}
\label{fig:phi_abs}
\end{figure*}
\subsection{Relative importance of the different Lyman $\alpha$ absorption channels}
\label{sec:channels_i}
Before looking at the solution of the transfer equation in more detail it is important to understand, which channels \change{on average} contribute most to the absorption of Lyman $\alpha$ photons.
In Fig.~\ref{fig:p_d} we present the partial death probabilities for different channels, as defined in the Appendix~\ref{sec:channels}. At all considered redshifts more than $\sim 90\%$ of the absorbed Lyman $\alpha$ photons disappear from the photon distribution in 1s-3d two-photons transition. 
In contrast to this, only about $2\%$ of all transitions end in the 3s-state. This is because the ratio of the 2p-3s and 2p-3d transition rates is about $g_{\rm 3s}A_{\rm 3s2p}/g_{\rm 3d}A_{\rm 3s2p}\sim 1/50$.
One can also see that in general the 1s-$n$d channels are more important than the 1s-$n$s channels, and that the contributions of 1s-3s and 1s-4d two-photon channels are comparable, where at high redshifts the 1s-4d channels contributes slightly more ($\sim 2\%$ versus $\sim 7\%$). 
However, less than $\sim 0.5\%$ of photons are directly absorbed to the continuum. 

\change{Assuming that the final modification in the ionization history is $\Delta N_{\rm e}/N_{\rm e}\sim 1\%$ when {\it only} including the two-photon aspects for the 3d-1s channel, then the above numbers suggest that: 
(i) the additional correction is expected to be similar to $\Delta N_{\rm e}/N_{\rm e}\sim 0.1\%$ when also taking the two-photon character of the 1s-3s, 1s-4d, and 1s-5d channels into account; 
(ii) neglecting the two-photon character for the transition to the continuum should lead to an uncertainty of $\Delta N_{\rm e}/N_{\rm e}\lesssim 0.1\%$. 
These simple conclusions seem to be in good agreement with the computations of \citet{Hirata2008}.
This also justifies the fact that here as a first step we only consider the two-photon corrections to the 3s-1s and 3d-1s channel. However, we plan to take the other two-photon corrections into account in a future paper.
}

\subsection{Effective Lyman $\alpha$ emission and absorption profile}
\label{sec:profile_i}
As we have seen in the previous section, the main channel for Lyman $\alpha$ absorption is due to the 1s-3d two-photon transition. This implies that the effective absorption profile, $\phi_{\rm abs}^\ast$, will be very close to the one following from the 3d-1s channel alone. 
In Fig.~\ref{fig:phi_abs} we give the spectral dependence of different line profiles in the vicinity of the Lyman $\alpha$ resonance at redshift $z=1300$.
For comparison we also show the Voigt profile, $\phi_{\rm V}$ (see Appendix~\ref{app:profiles}). One can clearly see the asymmetry of the two-photon profiles around the Lyman $\alpha$ line center and the deviations from the Lorentzian shape in the distant damping wings.

In the right panel we also show the effective emission profile, $\phi_{\rm em}$, for the 3 shell atom, as defined by Eq~\eqref{sec:phi_em}. In the computations we only included the 3s and 3d two-photon profiles, but assumed that in the continuum channel ($\rm 1s\leftrightarrow 2p \leftrightarrow c$) photons are emitted according to the normal Voigt profile.
As one can see the effective emission profile indeed is very close to the 3d-1s two-photon profile, including stimulated emission. Only at $\nu\geq\nu_{31}$ one can see the small Lorentzian contribution from the continuum channel.
Close the $\nu_{31}$ one can also see the small admixture of the 3s-1s two-photon profile. As can be deduced from the left panel in Fig.~\ref{fig:phi_abs}, at $\nu\sim \nu_{31}$ the stimulated 3s-1s two-photon profile is about $\sim 40$ times larger than the 3d-1s two-photon profile. With appropriate renormalization one can also obtain this factor using the approximation~\eqref{eq:phi_3s2d_0}.
Although $R^{\rm 3s, +}_{\rm 2p}\sim R^{\rm 3d, +}_{\rm 2p}/50$, due to this factor at $\nu\sim \nu_{31}$ the 3s channel adds about $\frac{4}{5}\,\phi_{\rm 3d}$, or $\sim 44\%$ to the effective emission profile.

\begin{figure}
\centering 
\includegraphics[width=0.9\columnwidth]
{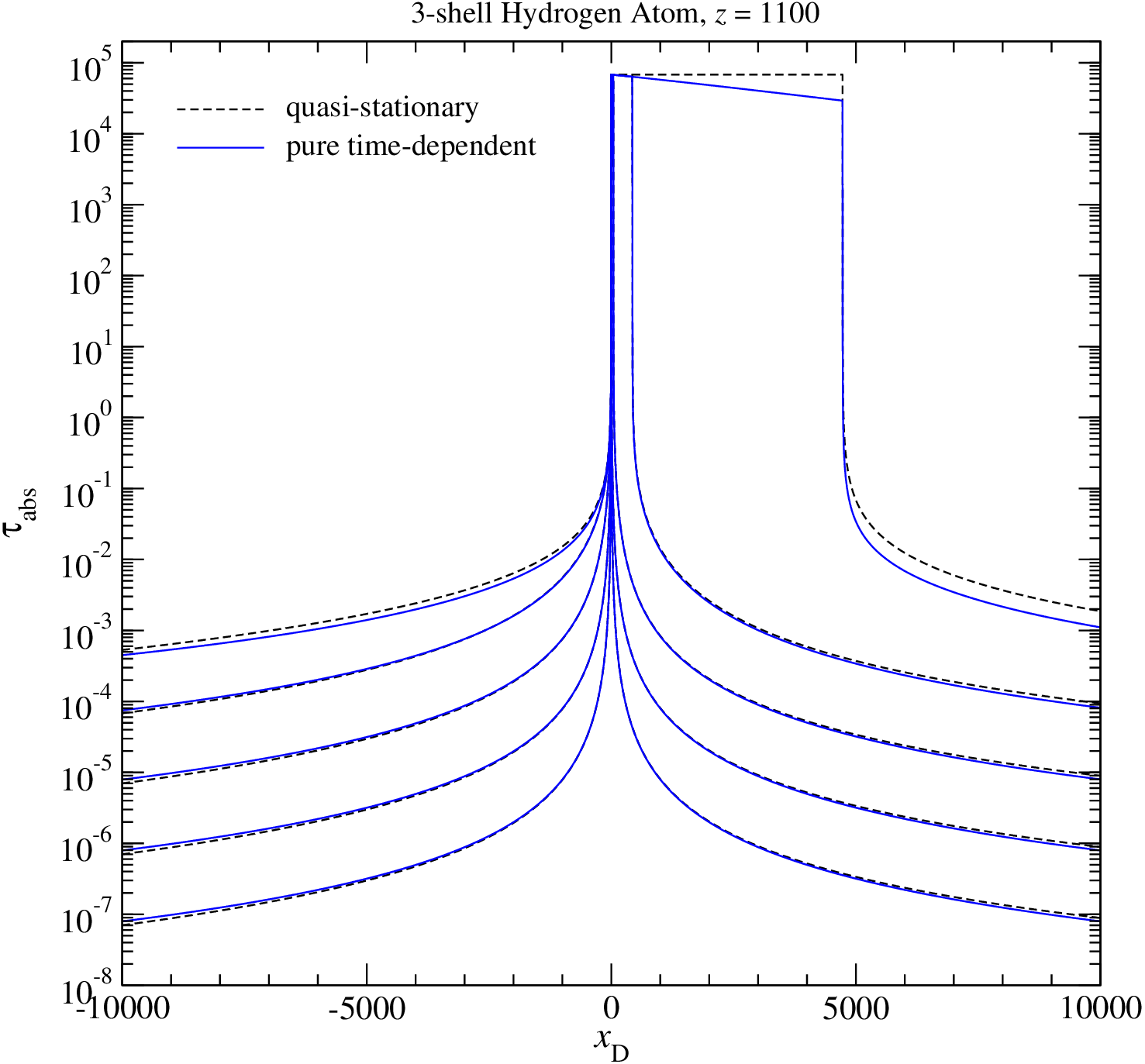}
\\[1mm]
\includegraphics[width=0.9\columnwidth]
{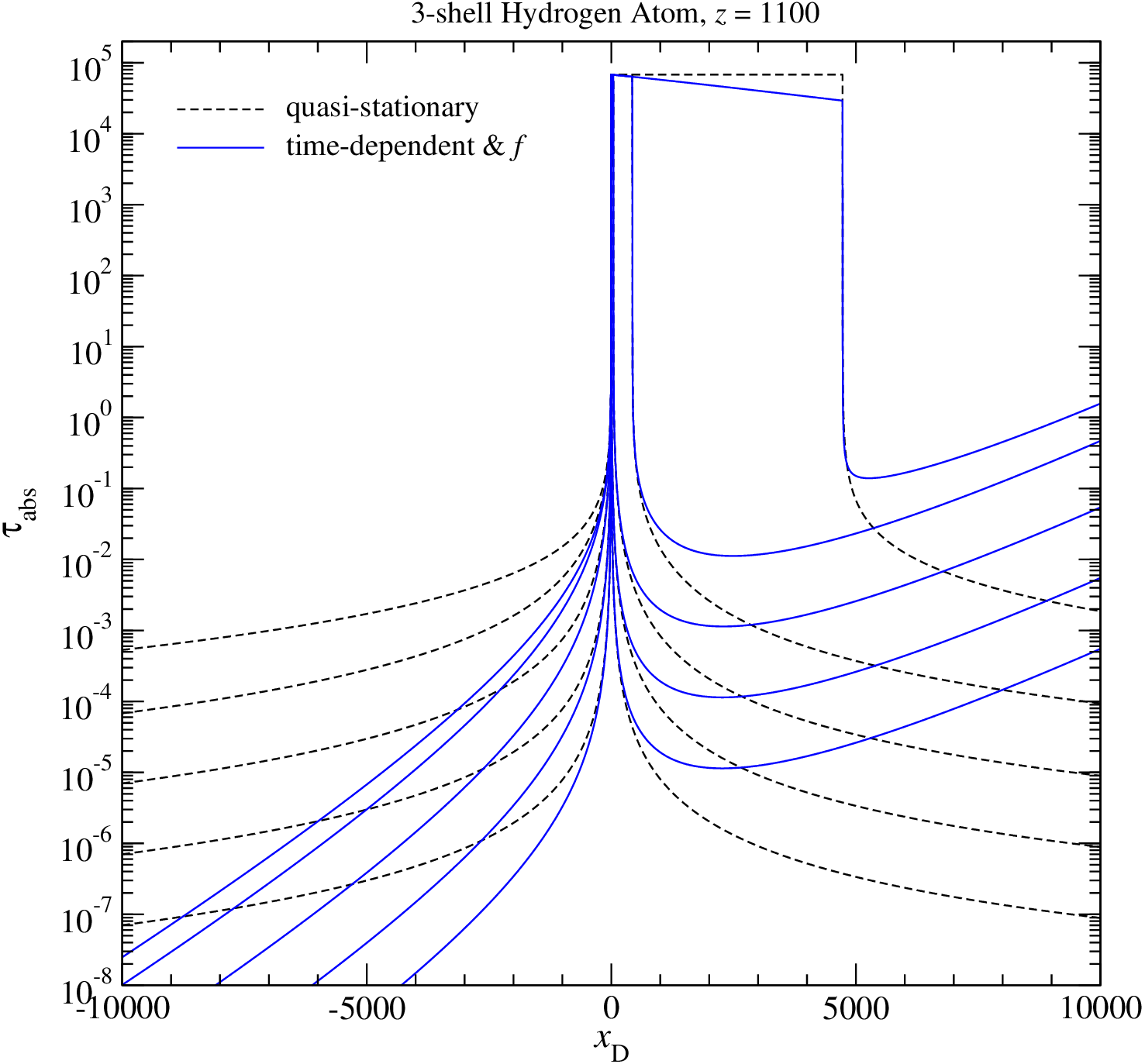}
\\[1mm]
\includegraphics[width=0.9\columnwidth]
{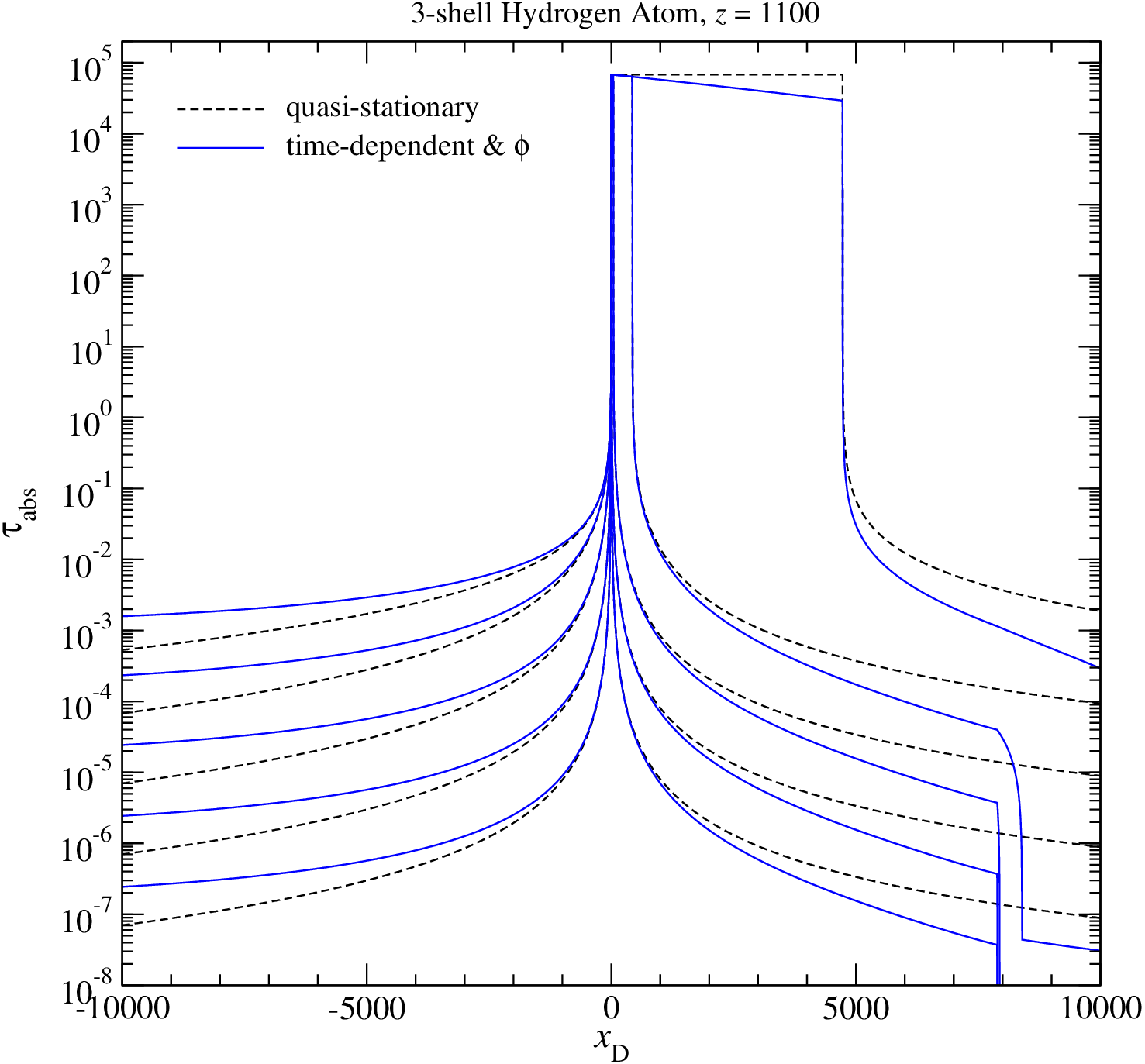}
\caption
{Modifications in the absorption optical depth $\tau_{\rm abs}(\xD, z, z-\Delta z)$ for $z=1100$. 
Here $\xD=[\nu-\nu_{21}]/\Delta\nu_{\rm D}$, where $\Delta\nu_{\rm D}$ is the Doppler width of the Lyman $\alpha$ resonance.
In each plot we show a sequence (lower to upper set of curves) of $\Delta z/z = 10^{-5}, 10^{-4}, 10^{-3}, 10^{-2}$ and $0.1$. 
For detailed explanation see Sect.~\ref{sec:sources_of_corr}.}
\label{fig:tau_abs}
\end{figure}

\subsection{Time and frequency dependence of the absorption optical depth}
\label{sec:tau_all}
In the definition of $F(\nu, z)$, Eq.~\eqref{app:kin_abs_em_Sol_phys_asym_F}, the function $\Theta^{\rm a}$ accounts for the frequency and time dependence of the emission process. For $\Theta^{\rm a}=1$ the shape of the solution for the spectral distortion depends only on the absorption optical depth, $\tau_{\rm abs}$, as defined by Eq.~\eqref{sec:tau_abs_all_a}.
In this case one can directly write
%
\beal
\label{eq:F0}
F_0(\nu, z)=1-e^{-\tau_{\rm abs}(\nu, z_{\rm s}, z)}.
\end{align}
Separating this part of the solution turns out the be very useful for numerical purposes. However, as we will see in  \changeII{Sect.~\ref{sec:Theta_a_therm}} $F_0$ does {\it not} describe the main behavior of the spectral distortion when including the thermodynamic correction factor $f_\nu$.

\subsubsection{Pure time-dependent correction to $\tau_{\rm abs}$}
\label{sec:tau_time}
If we neglect the two-photon corrections to the 3s and 3d profiles ($\phi_i=\phi_{\rm V}$) and set $f_\nu\equiv 1$ then we can look at the pure time-dependent correction to $\tau_{\rm abs}$.
As explained earlier \citep{Chluba2008b}, the dependencies of $p_{\rm d}$, $N_{\rm 1s}$, and $H$ on redshift lead to deviations of the solution for the spectral distortion from the quasi-stationary case.
Here the most important aspects are that, depending on the emission redshift, the total absorption optical depth until the time of observation (here $z$), is effectively lower (for $z_{\rm em}\gtrsim 1400$), or greater (for $z_{\rm em}\lesssim 1400$) than in the quasi-stationary case.
In addition the deviation from the quasi-stationary case depends on the initial frequency of the considered photon, since close to the line center photons travel a much shorter distance before getting absorbed than in the very distant wings, implying that time-dependent corrections are only important for photons that are emitted outside the Doppler core \citep[for more details see][]{Chluba2008b}. 

\change{
In Fig.~\ref{fig:tau_abs} we illustrate these effects on $\tau_{\rm abs}$ for emission redshift $z=1100$. We show the optical depth as a function of the initial frequency for different $\Delta z$.
In the upper panel we show the results for the case under discussion here \changeI{(solid line)}. For comparison we show the values of the optical depth using the normal quasi-stationary optical depth (dashed lines) for which one has
\beal
\label{eq:tau_qs_normal}
\tau^{\rm qs}_{\rm d}(\nu, z, z') 
&\approx \tau_{\rm d}(z)
\int_{\nu'}^{\nu}
\phi_{\rm V}(\tilde{\nu})\frac{\id \tilde{\nu}}{\Delta\nu_{\rm D}},
\end{align}
with $ \tau_{\rm d}(z)=p^{1\gamma}_{\rm d}\,\tauS$, where $\tauS$ is the normal Sobolev optical depth, $\nu'=\nu\frac{1+z'}{1+z}$ and $z'=z-\Delta z$.

For very small $\Delta z/z$ one expects no significant difference between the full numerical result for $\tau_{\rm abs}$ and this approximation. However, looking at the cases $\Delta z/z=10^{-5}, 10^{-4}$ and $10^{-3}$ one can see that even then there is a small difference in the distant red and blue wings of the line. This is not due to time-dependent corrections but due to the fact that, as usual, in Eq.~\eqref{eq:tau_qs_normal} we neglected the factor $\nu_{21}/\nu$ which appears in the definition of $\tau_{\rm  abs}$, leading to $\tau_{\rm  abs}/\tau^{\rm qs}_{\rm d}\sim \nu_{21}/\nu \gtrsim 1$ on the red, and $\tau_{\rm  abs}/\tau^{\rm qs}_{\rm d}\lesssim 1$ on the blue side of the resonance.

For the cases $\Delta z/z=0.01$ and $0.1$ we start to see the corrections due to the time-dependence.
\changeI{Here most interestingly for $\Delta z/z=0.1$ in both wings $\tau_{\rm abs}\lesssim \tau^{\rm qs}_{\rm d}$, is because the photons were released at $z\lesssim 1400$, so that $\tau_{\rm d}(z)$ decreases while the photons travel  \citep{Chluba2008b}.
This means that the integral over different redshifts $\tau_{\rm d}(\nu, z, z')\approx \int_{\nu'}^{\nu}
 \tau_{\rm d}(\tilde{z}) \phi_{\rm V}(\tilde{\nu})\frac{\id \tilde{\nu}}{\Delta\nu_{\rm D}}$ cannot reach the value for $\tau^{\rm qs}_{\rm d}(z)$.
Note that comparing with the value $\tau^{\rm qs}_{\rm d}(z')\lesssim \tau^{\rm qs}_{\rm d}(z)$ at the absorption redshift $z'=z-\Delta z<z$  one finds $\tau_{\rm d}(\nu, z, z')\gtrsim \tau^{\rm qs}_{\rm d}(z')$ following a similar argument. Usually this is the comparison which is made when talking about the escape probability at redshift $z$, so that the role of $z$ and $z'$ is simply interchanged.
}

The difference due to the time-dependence is not yet very visible for $\Delta z/z=0.01$ (the changes should be $|\Delta \tau/\tau|\sim |\Delta z/z|$), but one can see it in the region $0\lesssim \xD\lesssim x^{\rm c}_{\rm D}\sim 500$. There it is clear that the emitted photons will reach the Doppler core over a period that is shorter than the chosen $\Delta z/z$. For the case $\Delta z/z=0.1$ this region is $0\lesssim \xD\lesssim x^{\rm c}_{\rm D}\sim 4800$. 
Depending on how far the photon initially was emitted from the Doppler core the time it will travel before reaching $\xD\sim 0$ will grow with increasing $\xD$. This implies that at the redshift $z_{\rm c}<z\lesssim 1400$ of Doppler core crossing $\tau_{\rm d}(z_{\rm c})\lesssim \tau_{\rm d}(z)$, leading to the slope seen in the regions $0\lesssim \xD\lesssim x^{\rm c}_{\rm D}$.

Note that in the final result the time-dependent correction \changeII{to $\tau_{\rm abs}$} is not \changeI{so} important, only leading to modifications in the escape probability by $|\Delta P/P|\sim 1\% -3 \%$.
\changeII{The time-dependence of $\Theta^{\rm t}$ is much more relevant (see Sect.~\ref{sec:Pesc} for more details).}
}

\subsubsection{Effect of Thermodynamic correction factor on $\tau_{\rm abs}$}
\label{sec:tau_f_corr}
If we now include the {\it thermodynamic} correction factor $f_{\nu}$, as given by  Eq.~\eqref{eq:f_fac},  in the computation of $\tau_{\rm abs}$, then it is clear that for photons appearing at a given time on the {\it red side} of the Lyman $\alpha$ resonance, the total absorption optical depth over a fixed redshift interval will be {\it lower} than in the standard approach, independent of the emission redshift. 
Since $h\nu_{21}/k\Tg\sim 40 \left[\frac{1+z}{1100}\right]^{-1}$ one has $h[\nu-\nu_{21}]/k\Tg\sim \frac{\xD}{10^3}\,\left[\frac{1+z}{1100}\right]^{-1/2}$. 
Due to the exponential dependence of $f_\nu$ on the distance to the line center this implies that at $\xD\lesssim -10^3\,\left[\frac{1+z}{1100}\right]^{1/2}$ photons even directly escape, without any further reabsorption. 
This is in stark contrast to the standard approximation ($f_\nu =1$) for which even at distances $\sim -10^4$ some small fraction of photons \change{(comparable to $10^{-3}$ at $z\sim 1100$)} still disappears.
\change{
We illustrate this behavior in the central panel of Fig.~\ref{fig:tau_abs}, where at large distances on the red side of the resonance the value of $\tau_{\rm abs}$ is many orders of magnitude smaller than in the quasi-stationary approximation.
As we will see below (e.g. Sect.~\ref{sec:normal}), the thermodynamic factor leads to the largest correction discussed in this paper, \changeI{and in fact it is  this {\it red wing suppression of the absorption cross section} that contributes most}.
} 

\changeI{As mentioned in Sect.~\ref{sec:thermodyn}}, physically this behavior reflects the fact that the photon which enables the 2p-electron to reach the 3s and 3d is drawn from the ambient CMB radiation field. For photons on the red side of the Lyman $\alpha$ resonance ($\nu<\nu_{21}$) a photon with $\nu'>\nu_{32}$ is necessary \changeI{for a 1s electron} to reach the third shell. Since during \ion{H}{i} recombination the Balmer $\alpha$ line already is in the Wien tail of the CMB, this means the that relative to the Balmer $\alpha$ line center the amount of photons at $\nu'>\nu_{32}$ is {\it exponentially} smaller, depending on how large the detuning is.
Denoting the frequency of the second photon (absorbed close to the Balmer $\alpha$ resonance) with $\nu'=\nu_{31}-\nu$, by taking the ratio of the photon occupation numbers $n'/\nbb(\nu_{32})\approx\nbb(\nu_{31}-\nu)/\nbb(\nu_{32})\approx e^{h[\nu-\nu_{21}]/k\Tg}$ we again can confirm the exponential behavior of $f_\nu$.
Note that the same factor will appear even when thinking about two-photon transitions towards higher levels with $n>3$ or the continuum. It is a result of thermodynamic requirements, which should be independent of the considered process, as long as the second photon is drawn from the CMB blackbody.

On the other hand, for photons that are released on the {\it blue side} of the Lyman $\alpha$ line the the total absorption optical depth is {\it larger} than in the standard approximation \change{(see Fig.~\ref{fig:tau_abs} central panel for illustration)}. Due to the exponential dependence of $f_\nu$ on frequency for $\phi_{\rm abs}^\ast=\phi_{\rm V}$ this even leads to an {\it arbitrarily} large absorption optical depth in the very distant blue wing. 
Again this behavior can be understood when thinking about the second photon as drawn from the CMB blackbody. However, now there are exponentially more photons available than at the Balmer $\alpha$ line center.

This very strong increase in the absorption optical depth implies that photons are basically reabsorbed {\it quasi-instantaneously}, so that the usual quasi-stationary approximation for the computation of $\tau_{\rm abs}$ should be possible, like inside the Doppler core. In this case one therefore has
\beal
\label{eq:tau_qs}
\tau^{\rm qs}_{\rm abs}(\nu, z_{\rm s}, z) &\approx \tau_{\rm d}(z)\int_\nu^{\nu_{\rm s}}\phi_{\rm abs}(\nu')\,\frac{\nu_{21}\,\id \nu'}{\Delta\nu_{\rm D}\,\nu'} 
\nonumber\\[1mm]
&\approx \tau_{\rm d}(z)\,f_\nu(z)
\int_\nu^{\nu_{\rm s}}
\phi^\ast_{\rm abs}(\nu')\frac{\id \nu'}{\Delta\nu_{\rm D}},
\end{align}
with $ \tau_{\rm d}(z)=p^{1\gamma}_{\rm d}\,\tauS$, where $\tauS$ is the normal Sobolev optical depth, and $\nu_{\rm s}=\nu\frac{1+z_{\rm s}}{1+z}$.
This approximation for $\tau_{\rm abs}$ will also be very accurate close to the line center, but is very crude in the red wing.
Note that for $f_{\nu}=1$ and $\phi^\ast_{\rm abs}=\phi_{\rm V}$, $\tau^{\rm qs}_{\rm abs}$ recovers the approximation for the normal absorption optical depth in the quasi-stationary approximation, Eq.~\eqref{eq:tau_qs_normal}.

For $z_{\rm s}\rightarrow \infty$, $\nu\gg \nu_{21}$ and assuming that $\phi^\ast_{\rm abs}=\phi_{\rm V}$ one has
\beal
\label{eq:tau_blue}
\tau^{\rm blue}_{\rm abs}(\nu, z_{\rm s}, z) 
&\approx \tau_{\rm d}(z)\,f_{\nu}(z) \frac{a}{\pi \, \xD}.
\end{align}
With this equation it is possible to estimate the position on the blue side of the Lyman $\alpha$ resonance at which $\tau_{\rm abs}\sim 1$. Above that point $F_0\rightarrow 1$, however this does not represent the main behavior of $F_\nu$ for the given assumptions, since also the factor $1/f_{\nu'}$ in $\Theta^{\rm a}$ becomes important, so that $F_\nu$ instead actually scales like $1/f_{\nu}$ at large $\xD$ (see Sect.~\ref{sec:Theta_a_therm}).

\subsubsection{Effect of line absorption profile on $\tau_{\rm abs}$}
It is clear that also the {\it shape} of the absorption profile has an effect on the frequency dependence of the the absorption optical depth.
As we explained in Sect.~\ref{sec:profile_i} the effective absorption profile, $\phi^\ast_{\rm abs}$ is very close the two-photon emission profile of the 3d-level (see Fig.~\ref{fig:phi_abs}).
For simplicity assuming that $\phi^\ast_{\rm abs}\equiv \phi_{\rm 3d\rightarrow1s}$, it is clear that at $\nu\geq\nu_{31}$ no photons can be absorbed in the Lyman $\alpha$ transition, since there $\phi_{\rm 3d\rightarrow1s}=0$. This is in stark contrast to the case of a normal Voigt profile, for which in principle some photons can be absorbed at arbitrarily large frequencies.
Considering photons that reach the frequency interval $\nu_{21}\leq\nu\leq\nu_{31}$, the fact that there $\phi_{\rm 3d\rightarrow1s}\lesssim \phi_{\rm V}$ (see Fig.~\ref{fig:phi_abs}) implies that the contribution to the total absorption optical depth coming from this region is {\it smaller} than in the standard '$1+1$' photon formulation.
Similarly, at $\nu\leq\nu_{21}$ the contribution to the total absorption optical depth becomes {\it larger} than in the standard case, because there $\phi_{\rm 3d\rightarrow1s}\gtrsim \phi_{\rm V}$.
%
 
 \change{
In Fig.~\ref{fig:tau_abs}, lower panel, we illustrate these effects on $\tau_{\rm abs}$ for the 3 shell hydrogen atom. However, here we used the full absorption profile , $\phi^\ast_{\rm abs}$, which at  $\nu\gtrsim\nu_{31}$ has a small contribution from the Voigt profile that is used to model the continuum channel (${\rm 1s}\leftrightarrow{\rm 2p}\leftrightarrow{\rm c}$). Therefore the optical depths does not vanish at $\nu\gg \nu_{31}$.
The additional differences in the values of the optical depth seen in Fig.~\ref{fig:tau_abs} confirm the above statements.
Comparing with the case for the thermodynamic factor (central panel) it is clear that the correction to $\tau_{\rm abs}$ due to the shape of the absorption profile is not as important.
}
 
\change{One should also mention that setting $\Theta^{\rm a}=1$ and $f_\nu=1$ we obtain the solution $F_0^\phi$ as given by Eq.~\eqref{eq:F0}. With the comments made above, one therefore expects a strong drop in the value of $F_0^\phi$ for $\nu\rightarrow \nu_{31}$, since there $\tau_{\rm abs}\rightarrow 0$. Numerically we indeed find this behavior (see Sect.\ref{sec:F}). 
 }

\subsection{Time and frequency dependence of the effective emissivity}
If we look at the definition of $\Theta^{\rm a}$, Eq.~\eqref{app:kin_abs_em_Sol_phys_asym_b}, and rewrite it like
%
\bsub
\label{app:Theta_a}
\beal
\label{app:Theta_a_I}
\Theta^{\rm a}(z, z')
&=\frac{1}{f_{\nu'}}
\left[
\Theta^{\rm t}
+
\Theta^{\rm \phi}\right]
\\
\label{app:Theta_a_II}
\Theta^{\rm t}(z, z')&=
\frac{
\tilde{N}_{\rm em}(z')-\tilde{N}^{\rm pl}_{x'_{21}} }{\tilde{N}_{\rm em}(z)-\tilde{N}^{\rm pl}_{x_{21}}}
\\
\label{app:Theta_a_III}
\Theta^{\rm \phi}(z, z')&=\frac{
\tilde{N}_{\rm em}(z')}
{\tilde{N}_{\rm em}(z)-\tilde{N}^{\rm pl}_{x_{21}}}
\times\left[\frac{\phi_{\rm em}(\nu',z')}{\phi^\ast_{\rm abs}(\nu',z')} -1\right],
\end{align}
\esub
we can clearly see that there are also three sources for the corrections to the effective emissivity. The first is  due to the {\it purely time-dependent} correction ($\Theta^{\rm a}=\Theta^{\rm t}$), the second due to the {\it thermodynamic correction factor} ($\Theta^{\rm a}=1/f_{\nu'}$), and the last due to the {\it quantum mechanical asymmetry\footnote{More clearly here one should refer to the mixture of quantum mechanical processes important for the emission and absorption profile.}} between the emission and absorption profile ($\Theta^{\rm a}=\Theta^{\phi}$). Below we now shortly discuss the expected consequences of each of these.

\subsubsection{Pure time-dependent correction to $\Theta^{\rm a}$}
For $\Theta^{\rm a}=\Theta^{\rm t}$ we are looking at the pure time-dependent correction to the emission coefficient. This correction was already discussed in detail earlier \citep{Chluba2008b}. 
For quasi-stationary conditions one would have $\Theta^{\rm t}=1$.
However, in the cosmological recombination problem $\Theta^{\rm t}\neq1$ most of the time.
This fact leads to significant changes in the shape of the spectral distortion at different redshifts, where at frequencies $\xD\lesssim 0$ only $\Theta^{\rm t}\neq 1$ is able to affect the distortion \citep{Chluba2008b}.

\subsubsection{Effect of Thermodynamic correction factor in $\Theta^{\rm a}$} 
\label{sec:Theta_a_therm}
If we only include the correction due to the thermodynamic factor $f_\nu$ then we have $\Theta^{\rm a}=1/f_{\nu'}$. 
Since for $\nu'=\nu\,\frac{1+z'}{1+z}\neq\nu_{21}$ one has $f_{\nu'}\neq 1$, so that due to $f_{\nu}$ one expects similar effect on the shape of the distortion like from $\Theta^{\rm t}$.
However, since $f_{\nu'}\gg1$ at large detuning blue-ward of the line center, it turns out that this correction can be very large. 
\change{As mentioned in Sect.~\ref{sec:tau_all}, from Eq.~\eqref{eq:F0}, one naively 
expects $F_0\rightarrow 1$, but when including the factor $1/f_{\nu'}$ in $\Theta^{\rm a}$ instead we find $F_\nu\sim 1/f_{\nu}$ at large $\xD$.
}

\change{To show this,} it is illustrative to look at the solution for $F(\nu, z)$ in this case, assuming that the quasi-stationary approximation ($\Delta z/z\ll 1$ and $\Delta \nu/\nu\ll 1$ between the emission and absorption redshift of the photons) is possible. Introducing the new variable $\chi(\nu)=\int_0^\nu \phi_{\rm abs}^\ast(\nu')\id \nu'$, and using $\tau_{\rm abs}^{\rm qs}\approx \tau_{\rm d}(z)\,f_{\nu}(z)\,[\chi'-\chi]$, where $\chi'=\chi(\nu')$, one has
\bsub
\label{app:F_QS_f}
\beal
\label{app:F_QS_f_a}
F^{\rm qs}_{\rm f}(\nu, z)&=
\!\!\int_z^{z_{\rm s}} \frac{c\,\sigma_{\rm r}(z')\,N_{\rm 1s}(z')}{H(z')(1+z')}\,\phi^\ast_{\rm abs}(\nu', z')
\,
e^{-\tau_{\rm abs}(\nu, z', z)}\,\id z'
\\
&\approx \tau_{\rm d}(z)\int^{\chi_{\rm s}}_0\id\chi'\,e^{-\tau_{\rm d}\,f_{\nu}\,[\chi'-\chi]}
=\frac{1-e^{-\tau_{\rm d}\,f_{\nu}[\chi_{\rm s}-\chi]}}{f_\nu(z)}
\end{align}
\esub
Since $\tau_{\rm d}\,f_{\nu}[\chi_{\rm s}-\chi]\rightarrow \infty$ for $\xD\gg1$,  there one has $F(\nu, z)\rightarrow 1/f_\nu(z)$. In addition one expects $F(\nu, z)\sim 1/f_\nu(z)$ for $\xD\lesssim 1$.
As we will show below, for the correction due to the thermodynamic factor the scaling $F(\nu, z)\sim 1/f_\nu(z)$ indeed is correct at $\xD\gg 1$ and $\xD\sim 1$. However, at $\xD\rightarrow-\infty$ one finds $F(\nu, z)\sim 1$ instead.

\begin{figure*}
\centering 
\includegraphics[width=0.9\columnwidth]{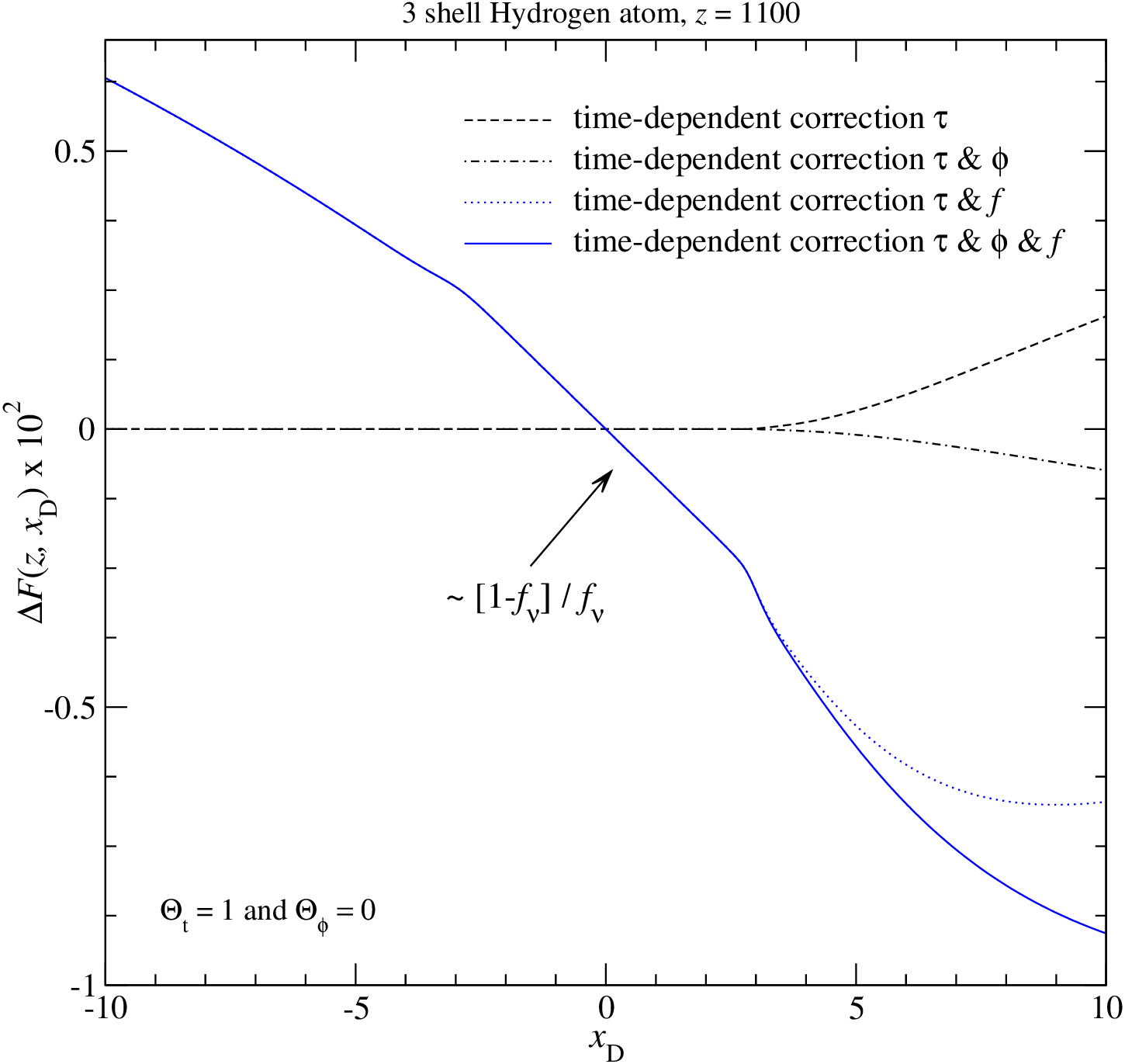}
\hspace{4mm}
\includegraphics[width=0.9\columnwidth]{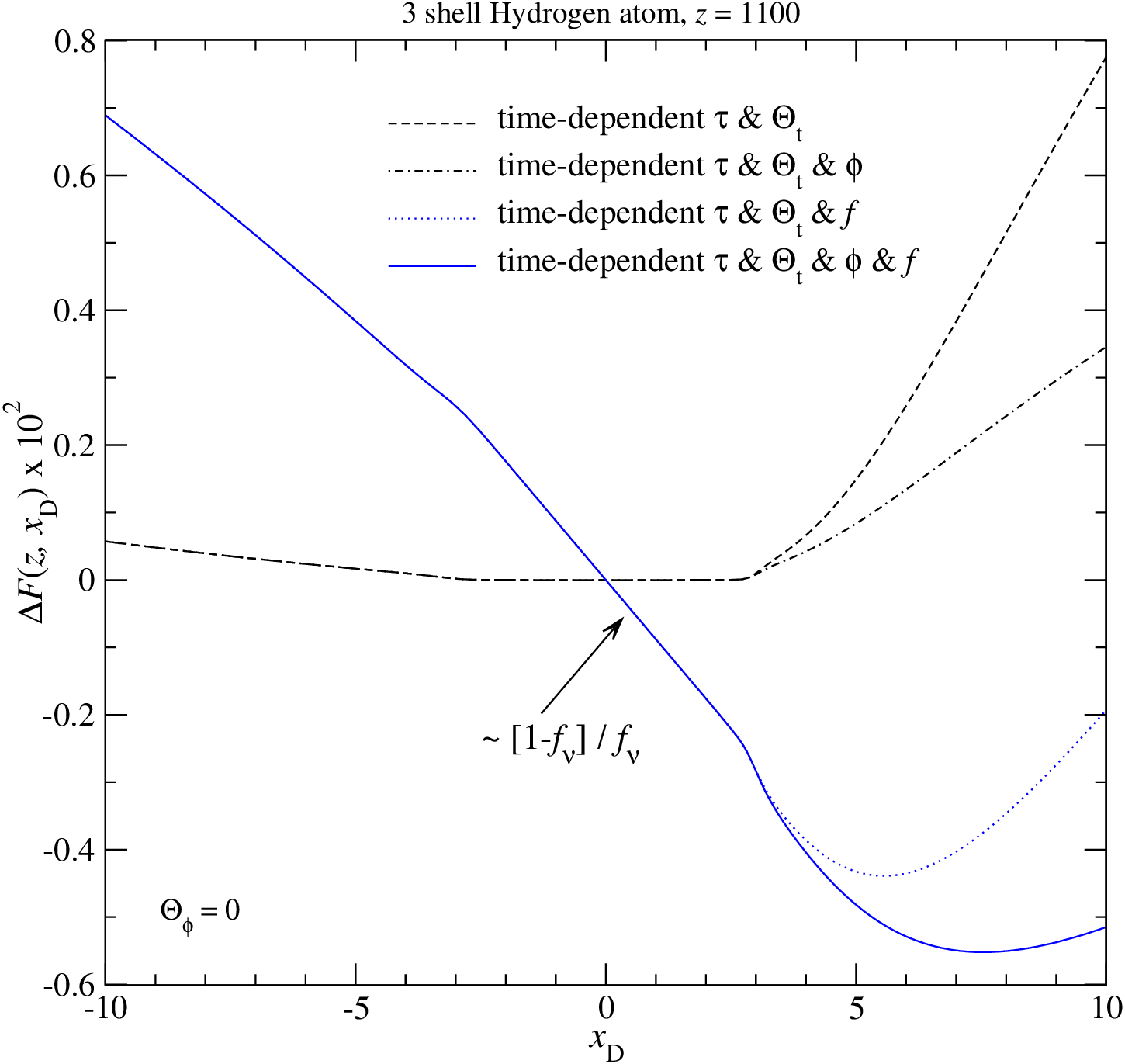}
\caption
{Difference in the Lyman $\alpha$ spectral distortion with respect to the quasi-stationary solution in the no redistribution approximation, $F^{\rm qs}(\nu, z)$ as given by Eq.~\eqref{eq:FQS},  at $z=1100$ close to the line center. For all computations shown in the left panel we set $\Theta^{\rm t}=1$ and $\Theta^{\phi} =0$, while in the right we only set $\Theta^{\phi} =0$. The cases labeled with $f$ are computed using $\Theta^{\rm a}=1/f_{\nu'}$ (left panel) and $\Theta^{\rm a}=\Theta^{\rm t}/f_{\nu'}$ (right panel), while for the others we set $f_{\nu}=1$. In addition the quoted correction factors were included in the computation of $\tau_{\rm abs}(\nu, z_{\rm s}, z')$. We assumed a 3 shell hydrogen atom.}
\label{fig:DFf.core.1100.z}
\label{fig:DFt.core.1100.z}
\end{figure*}
%
\subsubsection{Correction due to the quantum-mechanical asymmetry between emission and absorption profile}
\label{sec:mixture_asym}
Since in general the admixture the different transition channels to the emission and absorption profile is {\it not} identical one does expect $\phi^\ast_{\rm abs}\neq \phi_{\rm em}$. 
%
%
\changeII{For this we} can also look at the difference between $\phi_{\rm abs}^\ast$ and $\phi_{\rm em}$, which will be given by
\beal
\label{sec:Dphi_em}
\Delta\phi(\nu)=\phi_{\rm abs}^\ast-\phi_{\rm em}
&=
 \sum_{i} \left[\frac{R^{i,-}_{\rm 2p}}{R^{-}_{\rm 2p}}-\frac{R^{i,+}_{\rm 2p}}{R^{+}_{\rm 2p}}\right]\, 
 \phi_{i}(\nu).
\end{align}
Since in full thermodynamic equilibrium $(R^{i,+}_{\rm 2p})^{\rm eq}\equiv (R^{i,-}_{\rm 2p}\,N_{\rm 2p})^{\rm eq}$, it is clear that $(R^{i,-}_{\rm 2p}/R^{-}_{\rm 2p})^{\rm eq}\equiv (R^{i,+}_{\rm 2p}/R^{+}_{\rm 2p})^{\rm eq}$, so that $\Delta\phi\equiv 0$.
On the other hand it is known \citep[e.g. see][]{Chluba2007} that in the cosmological recombination problem $R^{i,-}_{\rm 2p}$ and $R^{i,+}_{\rm 2p}$ should always be very close to their equilibrium values, so that one expects $\Delta\phi/\phi\ll 1$.
Only at low redshifts ($z\lesssim 800$) this condition may not be fulfilled.
However, as we will see below in the context of CMB power spectrum computations this aspect of the problem never becomes significant (see Sect.~\ref{sec:F}).

\section{Changes in the Lyman $\alpha$ spectral distortion}
\label{sec:F}
In this Section we show the detailed dependence of the resulting Lyman $\alpha$ spectral distortion on the different corrections that are taken into account. 
As explained above there are three types of corrections that are considered here: (i) the time-dependent correction, (ii) the thermodynamic correction factor $f_{\nu}$, and (iii) the dependence on the detailed shape of the effective line emission and absorption profile.
We start our discussion by first only including these corrections in the computation of $\tau_{\rm abs}$ but setting $\Theta^{\rm t}=1$ and $\Theta^{\phi} =0$ (Sect.~\ref{sec:F_tau10}). 
%
%
In Sect.~\ref{sec:F_p0} we then also allow for $\Theta^{\rm t}\neq 1$, but still set $\Theta^{\phi} =0$.
Finally, we also include the correction due to $\Theta^{\phi} \neq 0$ (Sect.~\ref{sec:F_all}), but this aspect of the problem turns out to be not very important.
It should be possible to deduce all the other combinations from these cases.

\subsection{Corrections related to $\tau_{\rm abs}$ for $\Theta^{\rm t}=1$ and $\Theta^{\phi} =0$}
\label{sec:F_tau10}
As a first case we study the effect of different corrections to the absorption optical depth. For this we set  $\Theta^{\rm t}=1$ and $\Theta^{\phi} =0$, meaning that in the emission coefficient $\Theta^{\rm a}$ we ignore the pure time-dependent correction and the one related to the shape of the profile. 
However, depending on the considered case we do allow for these corrections in the computation of $\tau_{\rm abs}$.
In addition, we also discuss the effect of the thermodynamic correction factor, $f_\nu$. but here we include it in both $\tau_{\rm abs}$ and $\Theta^{\rm a}$ at the same time.
As explained \changeI{Sect.~\ref{sec:Theta_a_therm}}, if one would only include $f_{\nu}$ for $\tau_{\rm abs}$ or $\Theta^{\rm a}$ separately, the corresponding spectral distortion is physically not very meaningful. Therefore we omit this case here.

\begin{figure*}
\centering 
\includegraphics[width=0.9\columnwidth]{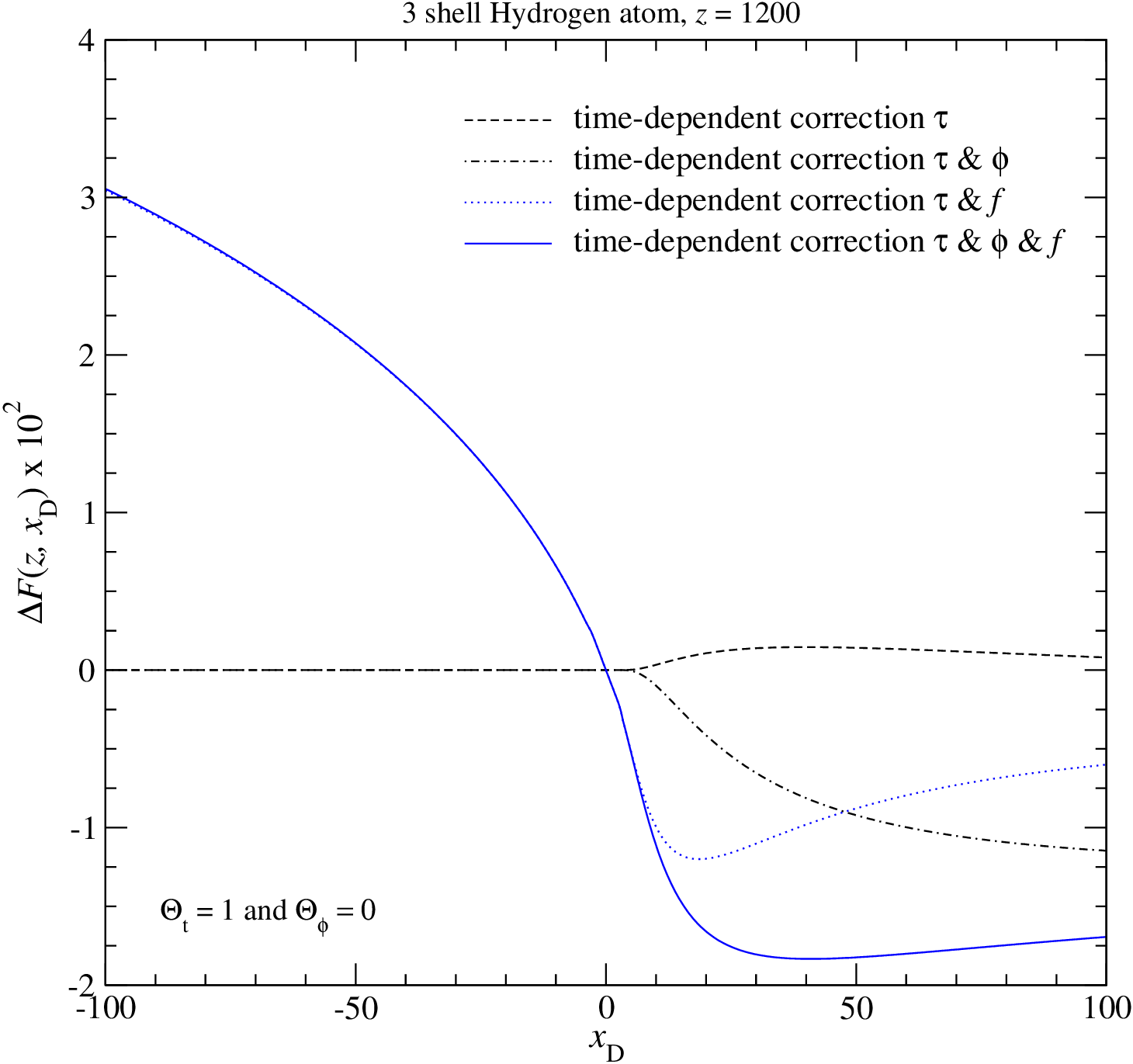}
\hspace{2mm}
\includegraphics[width=0.9\columnwidth]{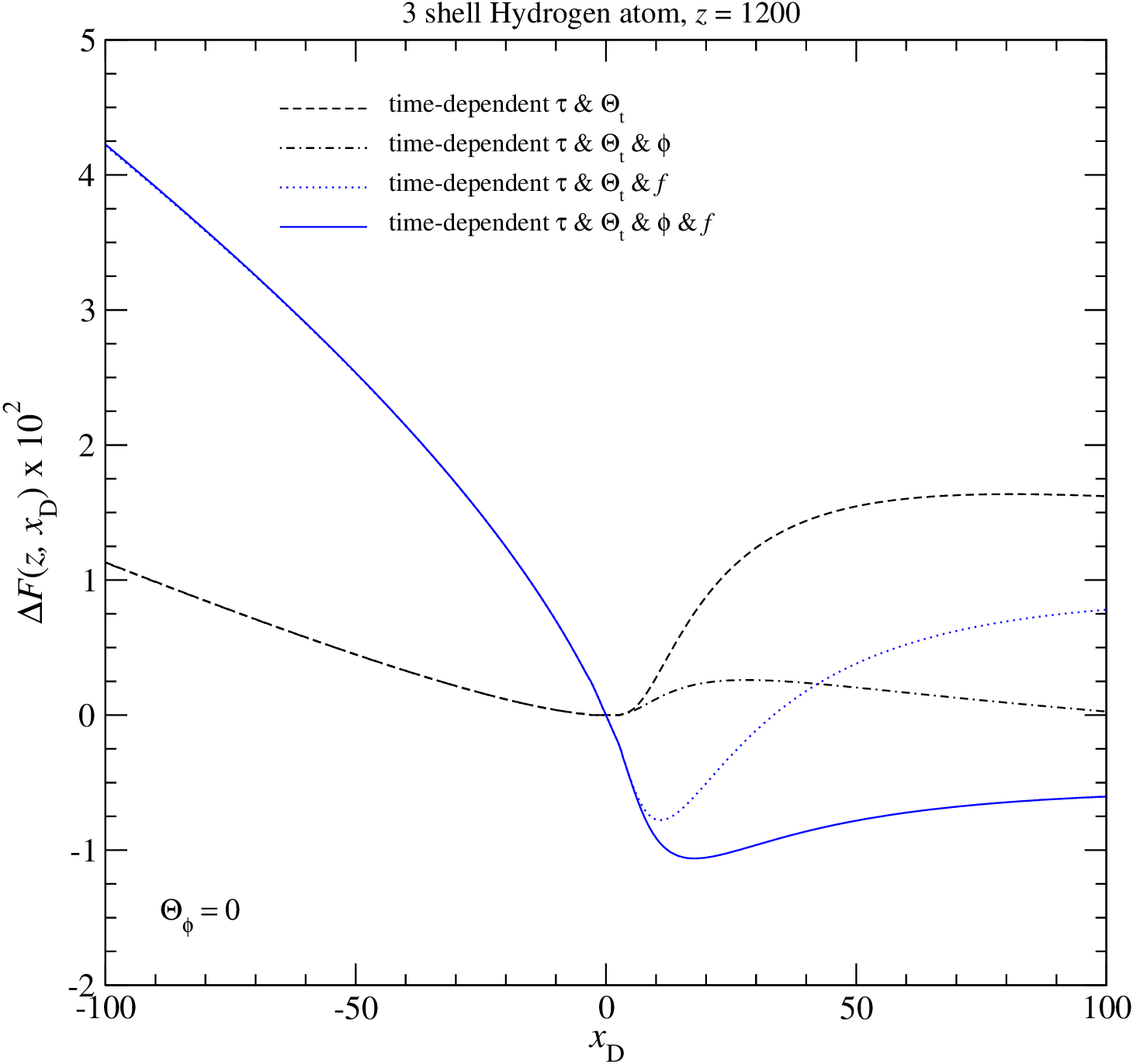}
\\
\includegraphics[width=0.9\columnwidth]{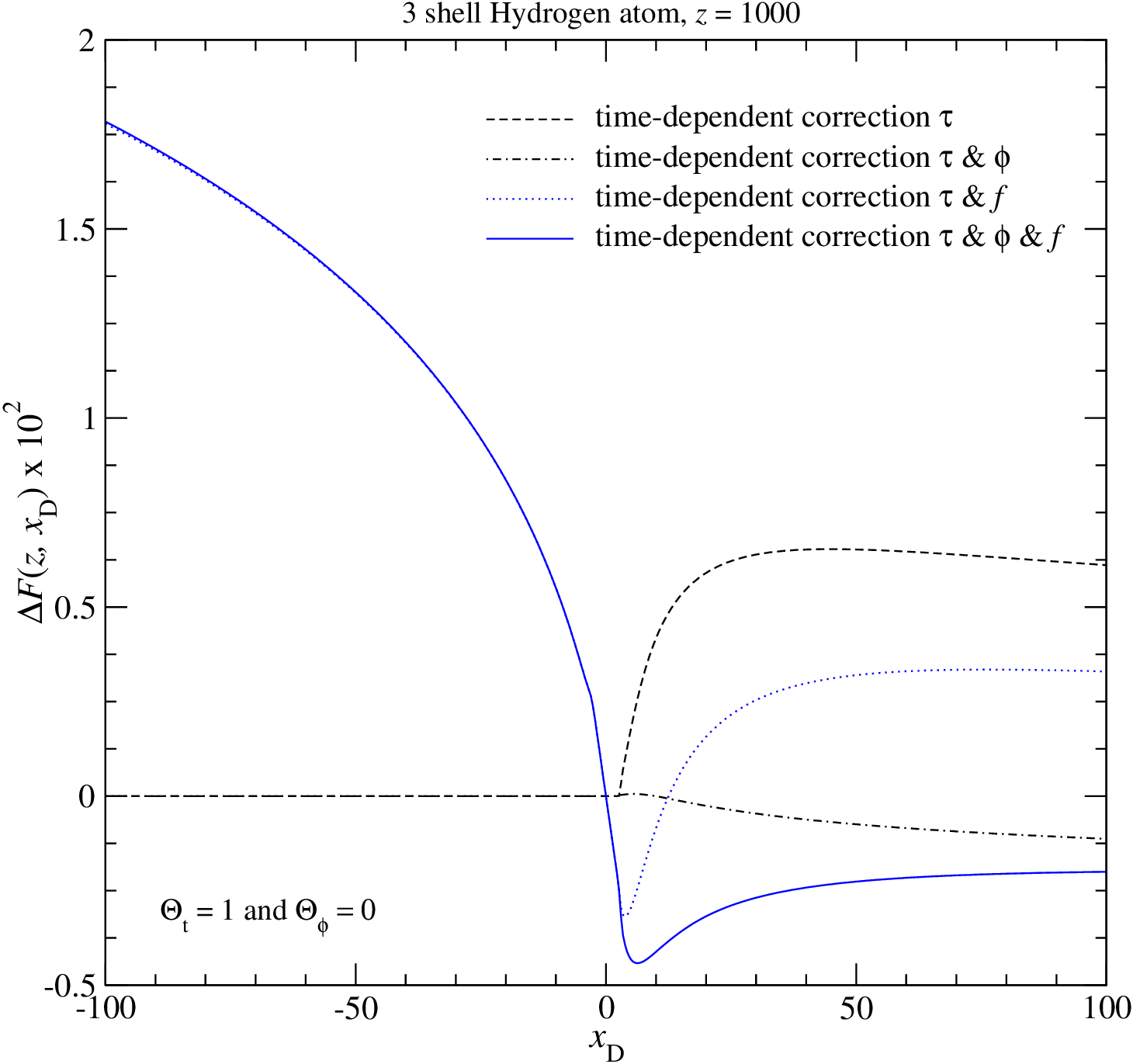}
\hspace{2mm}
\includegraphics[width=0.9\columnwidth]{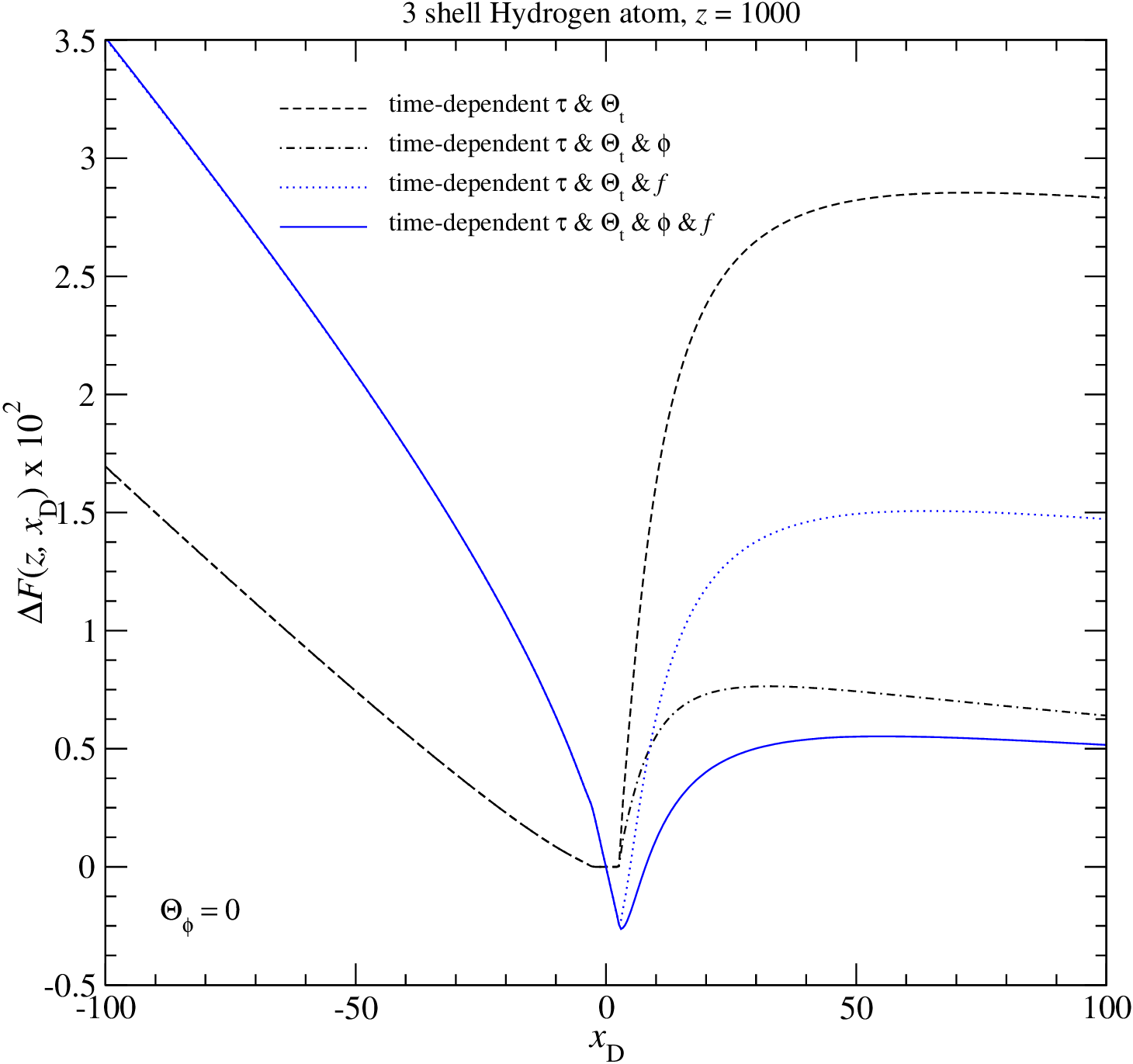}
\\
\includegraphics[width=0.9\columnwidth]{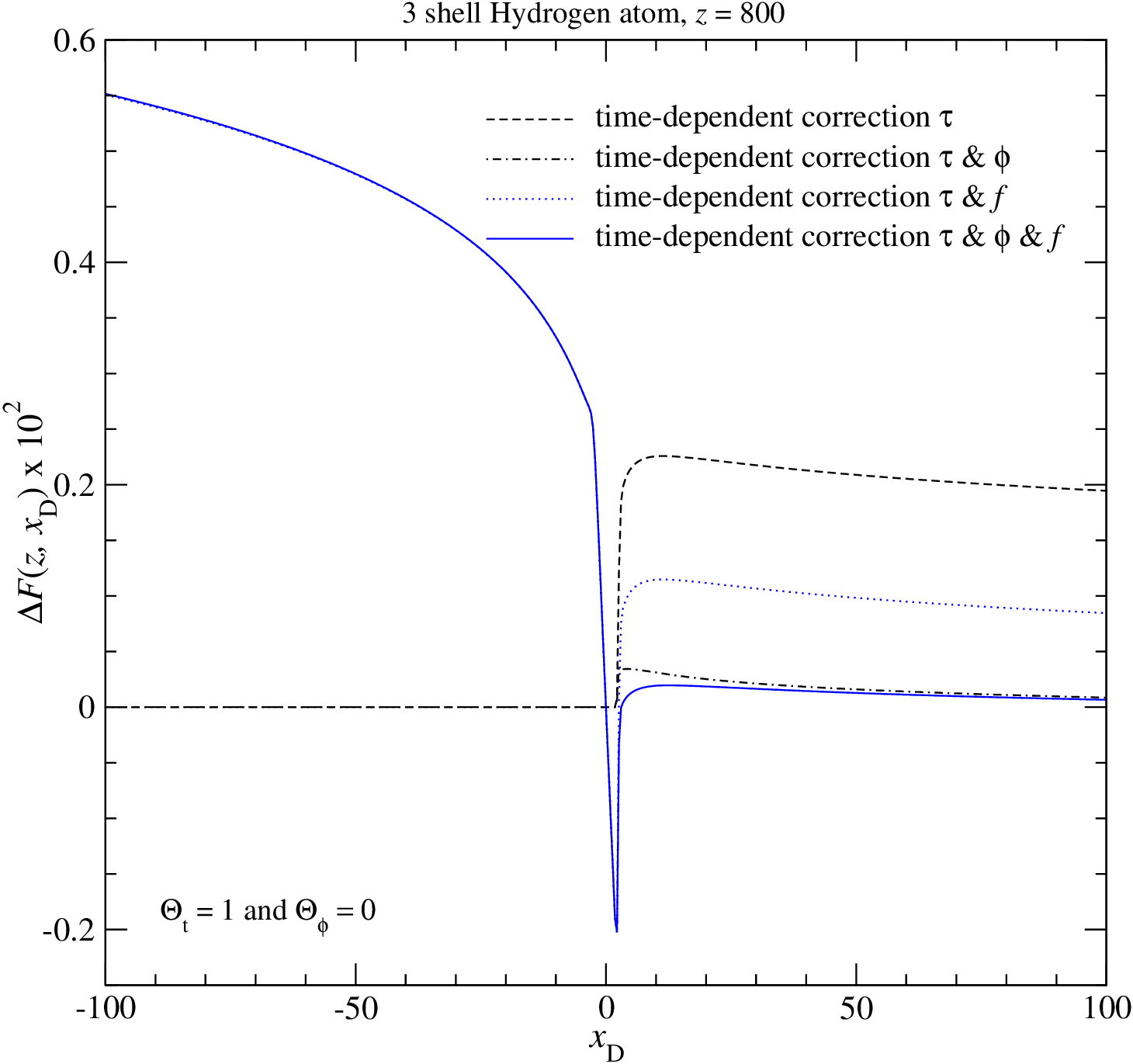}
\hspace{2mm}
\includegraphics[width=0.9\columnwidth]{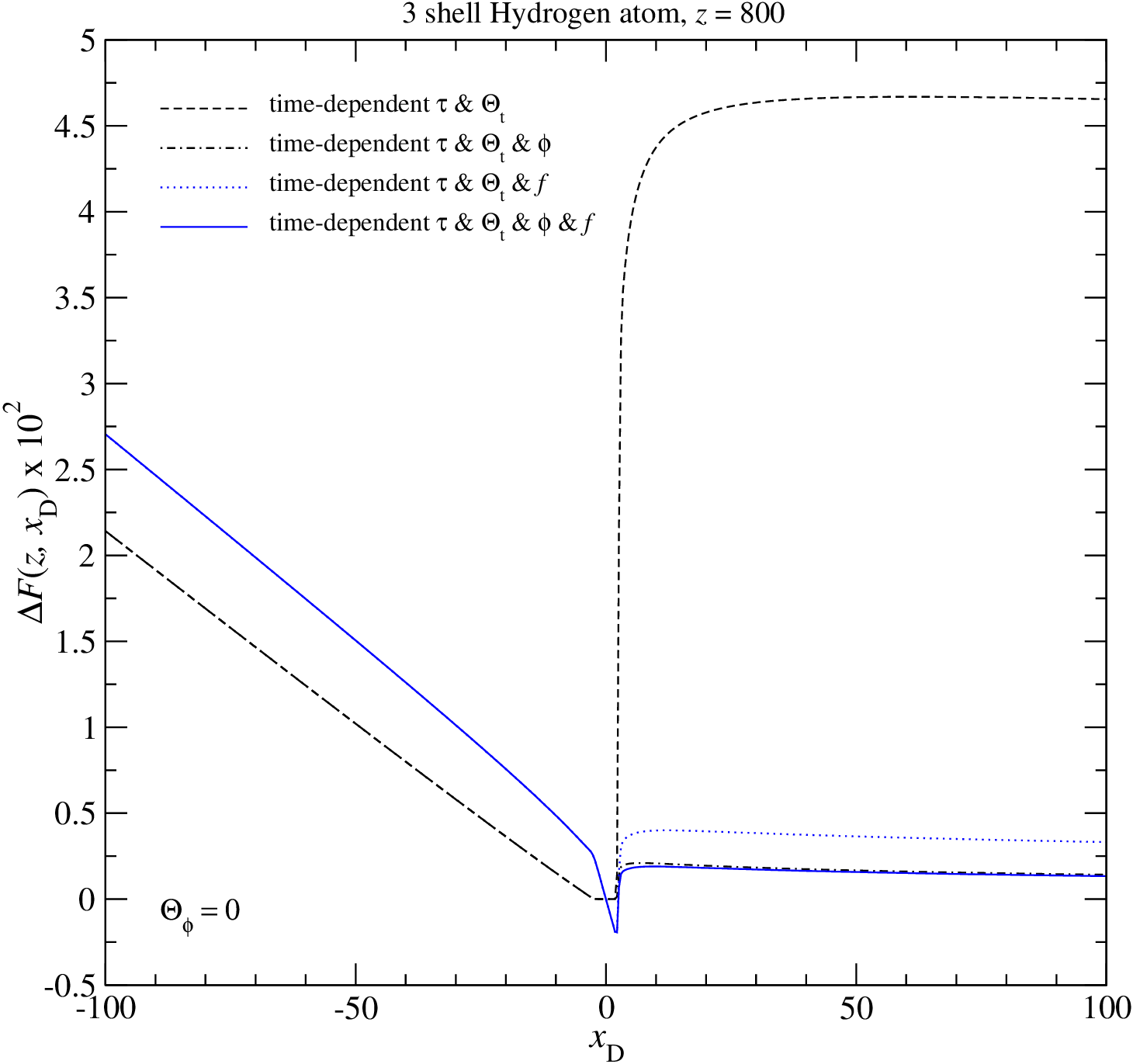}
\caption
{Difference in the Lyman $\alpha$ spectral distortion with respect to the quasi-stationary solution in the no redistribution approximation, $F^{\rm qs}(\nu, z)$, as given by Eq.~\eqref{eq:FQS}, at several redshift close to the line center. For the all computations shown in the left column we set $\Theta^{\rm t}=1$ and $\Theta^{\phi} =0$, while in the right we only set $\Theta^{\phi} =0$. The cases labeled with $f$ are computed using $\Theta^{\rm a}=1/f_{\nu'}$ (left column) and $\Theta^{\rm a}=\Theta^{\rm t}/f_{\nu'}$ (right column), while for the others we set $f_{\nu}=1$. In addition the quoted correction factors were included for $\tau_{\rm abs}(\nu, z_{\rm s}, z')$. We assumed a 3 shell hydrogen atom.}
\label{fig:DFf.z}
\label{fig:DFt.z}
\end{figure*}

\begin{figure*}
\centering 
\includegraphics[width=0.9\columnwidth]{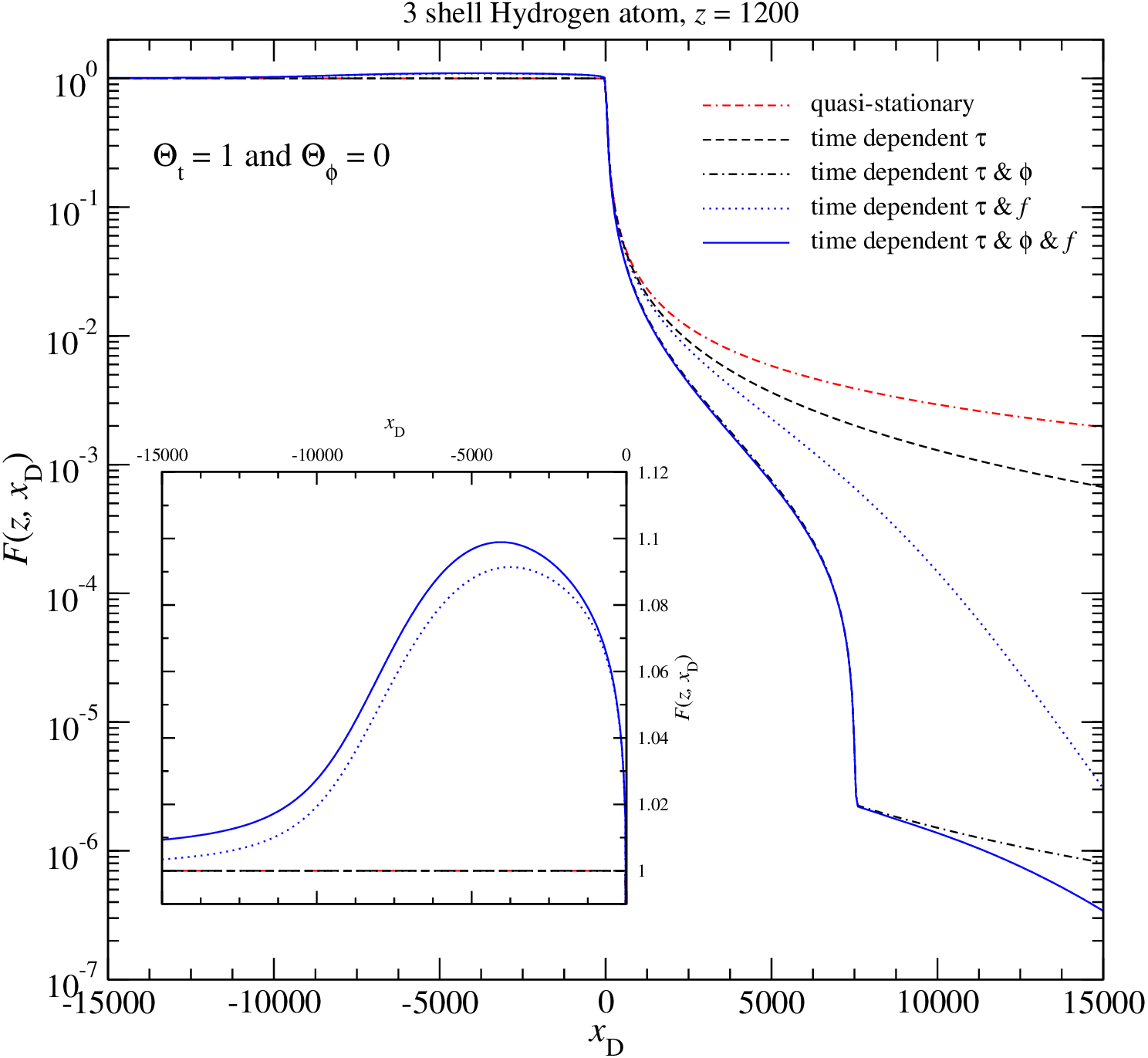}
\hspace{2mm}
\includegraphics[width=0.9\columnwidth]{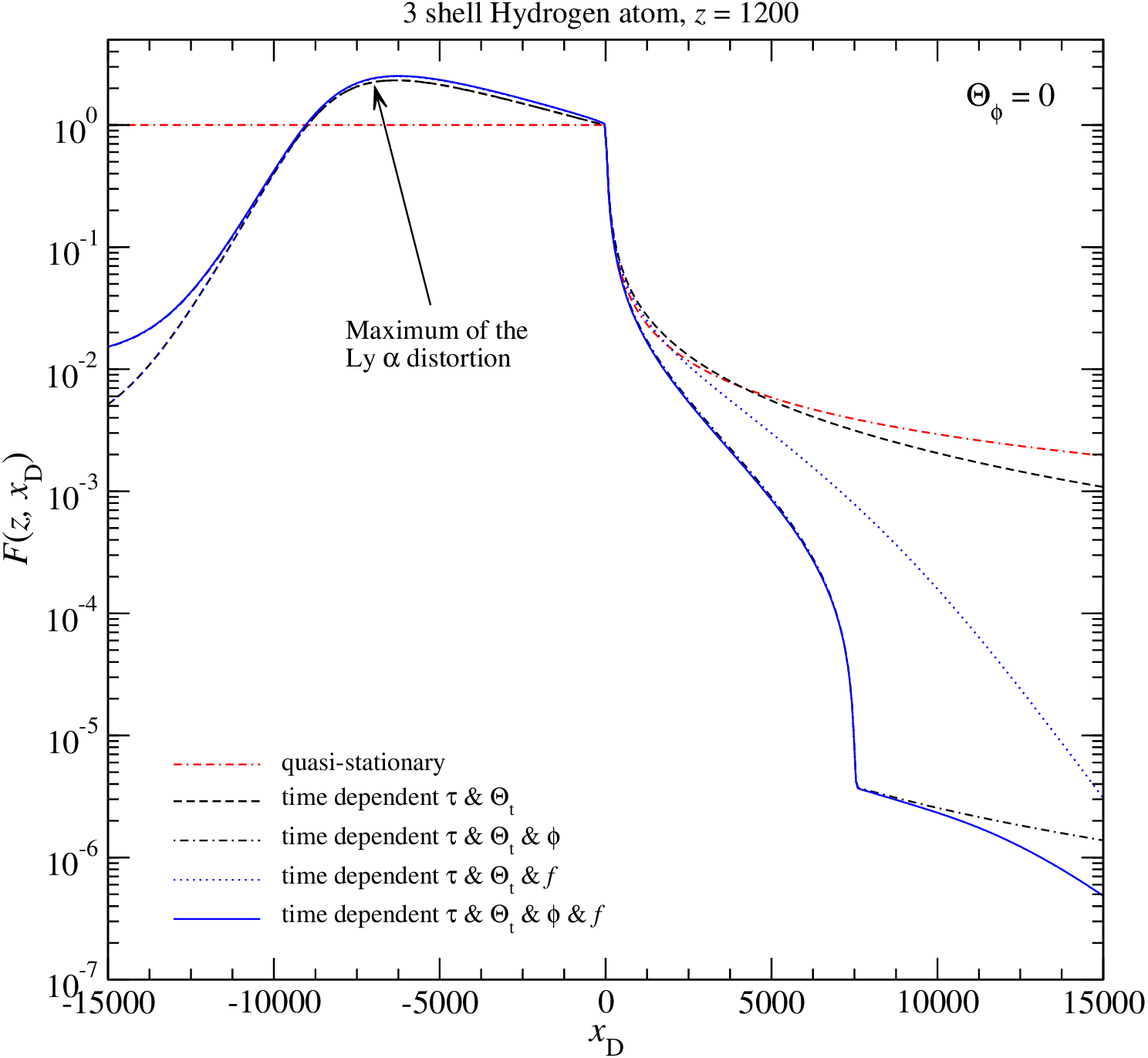}
\\
\includegraphics[width=0.9\columnwidth]{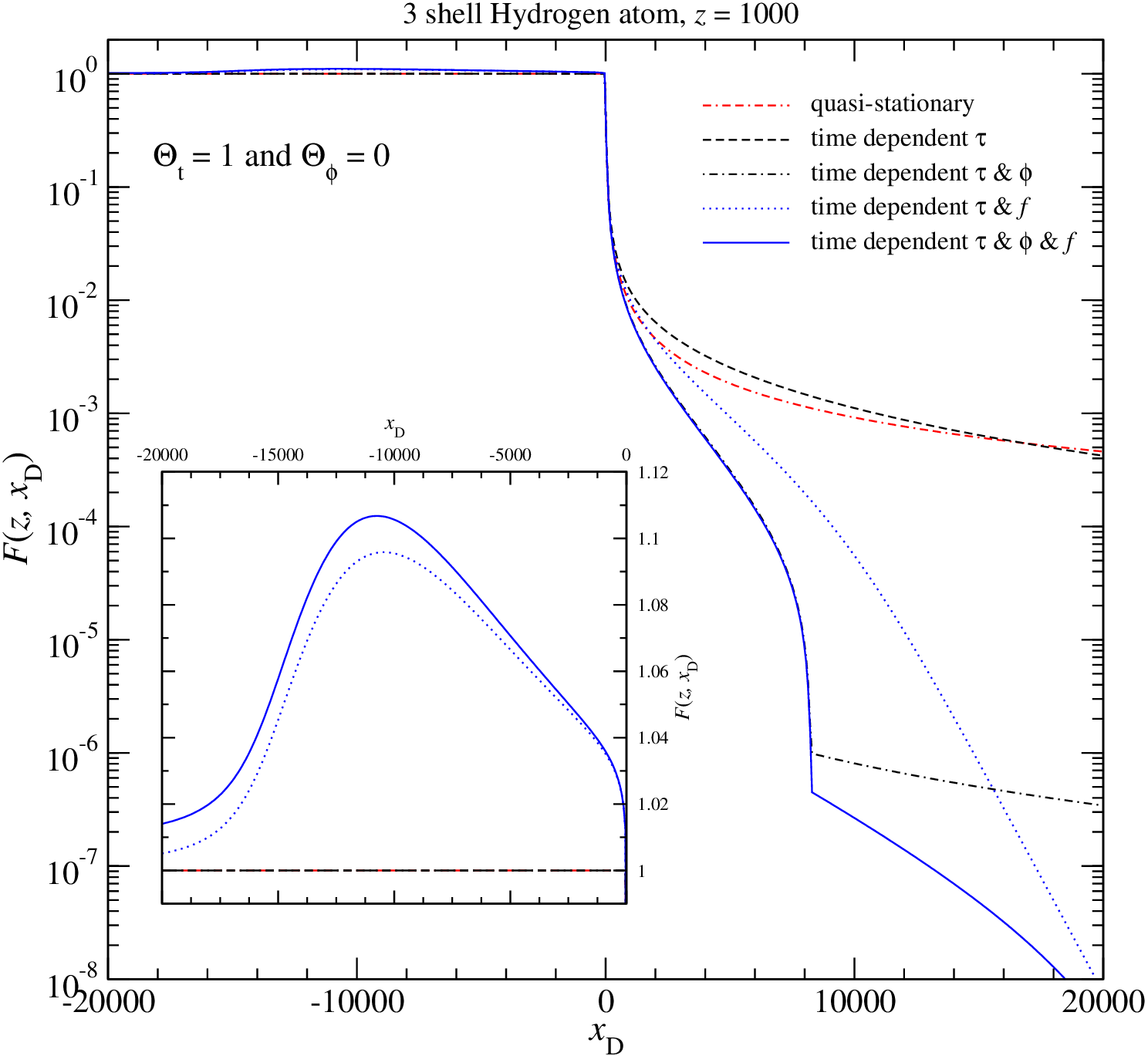}
\hspace{2mm}
\includegraphics[width=0.9\columnwidth]{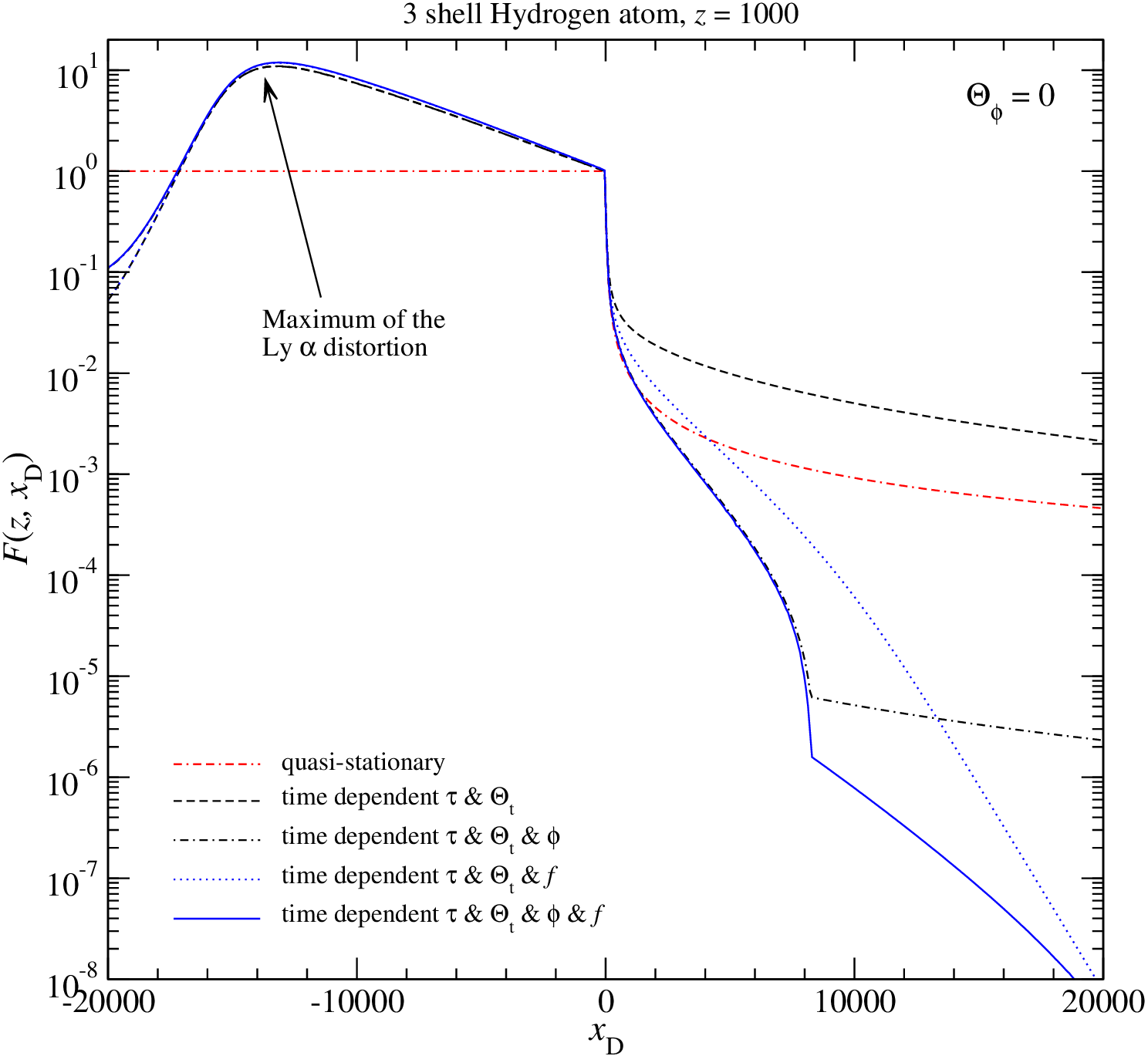}
\\
\includegraphics[width=0.9\columnwidth]{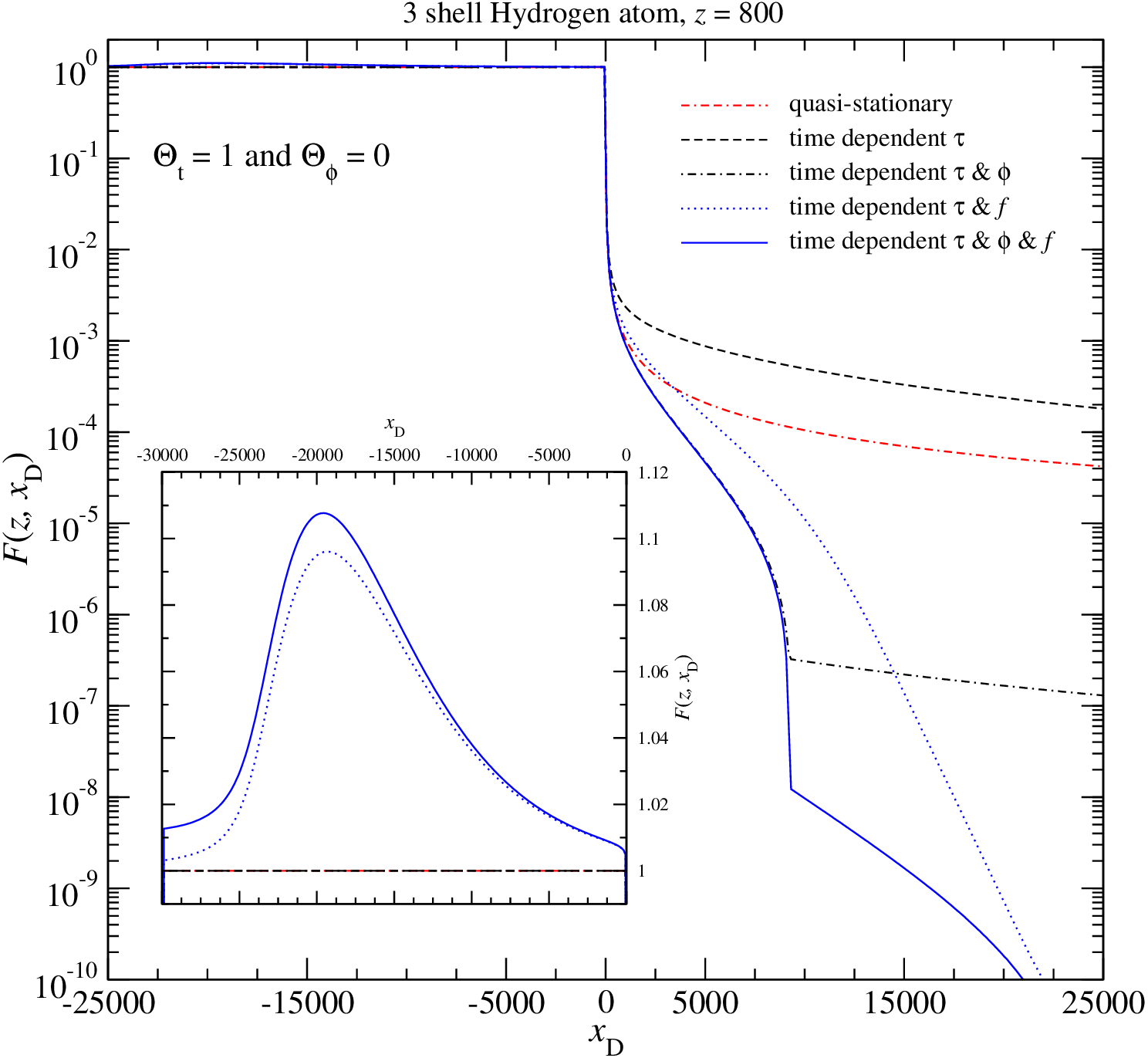}
\hspace{2mm}
\includegraphics[width=0.9\columnwidth]{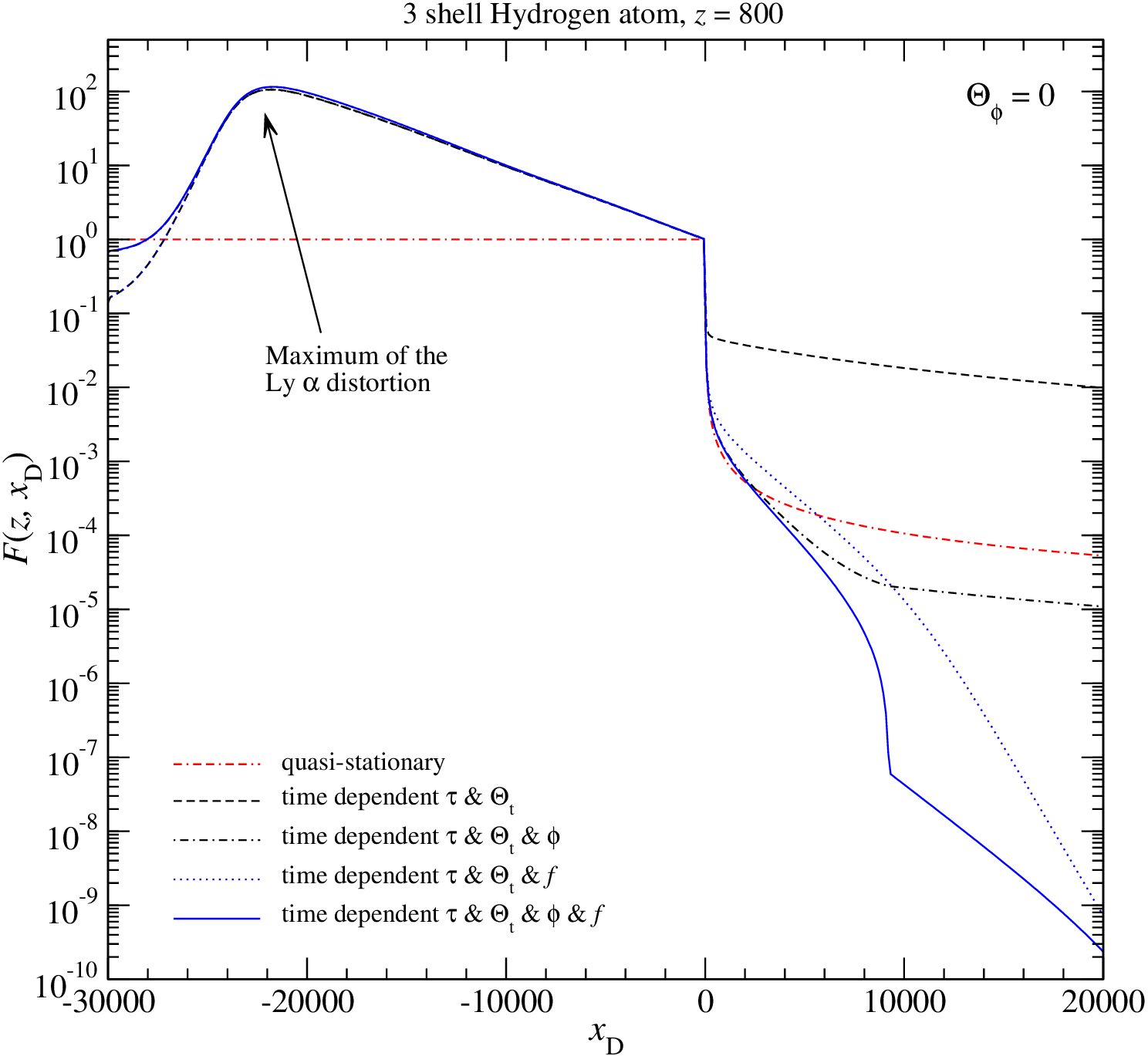}
\caption
{The Lyman $\alpha$ spectral distortion at different redshifts and in a wide range of frequencies around the line center. For the all computations shown in the left column we set $\Theta^{\rm t}=1$ and $\Theta^{\phi} =0$, while in the right we only set $\Theta^{\phi} =0$. The cases labeled with $f$ are computed using $\Theta^{\rm a}=1/f_{\nu'}$ (left column) and $\Theta^{\rm a}=\Theta^{\rm t}/f_{\nu'}$ (right column), while for the others we set $f_{\nu}=1$. In addition the quoted correction factors were included for $\tau_{\rm abs}(\nu, z_{\rm s}, z')$. We assumed a 3 shell hydrogen atom.}
\label{fig:Ff.z}
\label{fig:Ft.z}
\end{figure*}

\subsubsection{Behavior very close to the line center}
\label{sec:f_center}
In Fig.~\ref{fig:DFf.core.1100.z} as an example we show the Lyman $\alpha$ spectral distortion at $z=1100$ in the close vicinity of the line center.
We compare the results with the normal quasi-stationary solution \citep[see][for details]{Chluba2008b}
\beal
\label{eq:FQS}
F^{\rm qs}(\nu, z)=1-e^{-\tau_{\rm d}}\,e^{\tau_{\rm d}\,\chi},
\end{align}
with $\chi=\int_0^\nu\,\varphi_{\rm V}(\nu')\id\nu'$.
We show the result obtained for the pure time-dependent correction to $\tau_{\rm abs}$ (dashed curve), which was already discussed earlier \citep{Chluba2008b}.
At $\xD\lesssim 4$ the distortion is practically identical with the result in the quasi-stationary case, while at $\xD\gtrsim 4$ the time-dependent corrections to $\tau_{\rm abs}$ start to be important. 
One can see that \changeI{there} $F_\nu=1- e^{-\tau_{\rm abs}} \gtrsim F^{\rm qs}_\nu$, which as explained \changeI{in Sect.~\ref{sec:tau_time}} is related to the fact that including the time-dependent correction, at this redshift the value of $\tau_{\rm abs}$ is slightly larger than in the quasi-stationary approximation, \changeI{leading to\footnote{\changeI{The upper indices will henceforth indicate which correction was included. 't' will stand for the time-dependent correction, 'f' for the thermodynamic factor, and '$\phi$' for the profile correction. When all corrections are included simultaneously we will use 'a'.}} $\tau^{\rm t}_{\rm abs}\gtrsim \tau^{\rm qs}_{\rm abs}$}. 

If we also include the correction due to the shape of the absorption profile in the computation of $\tau_{\rm abs}$ (dashed-dotted curve), then we see that at $\xD\lesssim 0$ again the solution is practically identical with the solution in the quasi-stationary case. Although one does expect some corrections to the exact value of $\tau_{\rm abs}$ at different frequencies below the line center\footnote{Since in the red wing $\phi^\ast_{\rm abs}\gtrsim\phi_{\rm V}$ (cf. Fig.~\ref{fig:phi_abs}) one expects $\tau^{\phi}_{\rm abs}\gtrsim \tau^{\rm qs}_{\rm abs}$.}, since $\tau_{\rm abs}\gg 1$ the effect on the shape of $F_\nu \sim 1- e^{-\tau_{\rm abs}}$ will be exponentially small.
\changeI{However, looking at $\xD\gtrsim 4$ we can see that $F_\nu\lesssim F^{\rm qs}_\nu$, implying that $\tau^{\rm t,\phi}_{\rm abs}\lesssim \tau^{\rm qs}_{\rm abs}$. Since blue-ward of the resonance $\phi^\ast_{\rm abs}\lesssim\phi_{\rm V}$ (see right panel of Fig.~\ref{fig:phi_abs}) it is already expected that the curve lies below the one for the pure time-dependent correction (i.e.  $\tau^{\rm t,\phi}_{\rm abs}\lesssim \tau^{\rm t}_{\rm abs}$). But it even turns out that the correction due to the shape of $\phi^\ast_{\rm abs}$ overcompensates the pure time-dependent effect, which alone leads to $\tau^{\rm t}_{\rm abs}\gtrsim \tau^{\rm qs}_{\rm abs}$. 
This shows that at the considered redshift the correction due to the profile is slightly more important than due to time-dependence.}

If we now only include the time-dependent correction and the effect of the thermodynamic correction factor \changeI{(in both $\tau_{\rm abs}$ and $\Theta^{\rm a}$)} then we obtain the dotted line. As expected from Eq.~\eqref{app:F_QS_f}, very close to the Doppler core ($|\xD|\lesssim 4$) one has $F_\nu (z)\sim 1/f_\nu(z)$.
We also found this scaling at other redshifts (marginally visible in Fig.~\ref{fig:DFf.z}), as long as the optical depth across the Doppler core is much larger that unity, implying that the quasi-stationary approximation is valid.
However, outside this region the distortion differs significantly from the previous cases. In particular one finds $F_\nu\gtrsim F_\nu^{\rm qs}$ at $\xD\lesssim -4$, \changeI{which is the result of $\Theta^{\rm a}=1/f_\nu$. 
If we include the thermodynamic correction factor only in the computation of $\tau_{\rm abs}$, i.e. setting $\Theta^{\rm a}=1$, then we find $F_\nu\sim 1$ instead.}

\changeI{It also turns out that $F_\nu\lesssim 1/f_\nu$ at $\xD\lesssim -4$. This is in contrast to the result $F^{\rm qs}_{\rm f}(\nu, z)$, as given by Eq.~\eqref{app:F_QS_f}, for which we assume quasi-stationary conditions. This implies that in the red damping wing deviations from the quasi-stationary assumption become important.}

If we in addition include the correction due to the \changeII{shape of the} absorption profile, then we can see that at $\xD\lesssim 4$ the curve basically coincides with the one from the previous case.
This again is expected since the tiny corrections to the value of a very large optical will not lead to visible changes in $F_\nu$.
In addition, at $\xD\gtrsim 4$ one can see that the difference to the previous case are about the same as for the difference between the first two cases, owing to the fact that the corrections are small \changeI{and hence additive}.

\subsubsection{Behavior at intermediate and large distances from the line center}
We now look at the corrections in a slightly wider range around the line center. In  Fig.~\ref{fig:DFf.z} we show the same cases as above, but now also varying the redshift.
As before one can see that differences due to the shape of the absorption profile are negligible at $\xD\lesssim 0$. Furthermore, on the blue side of the resonance the correction due to the shape of the absorption profile is always negative, as also seen in the previous paragraph.

\changeII{Taking the} differences between the curves for $\Theta^{\rm a}=1$ (first two lines) and those for $\Theta^{\rm a}=1/f_{\nu'}$ (last two lines) one can also see that at  $z=1200$ these are practically the same. However for $z=1000$ and $z=800$ higher order terms already start to become important. For example, at $\xD=100$ the difference of the curves for $\Theta^{\rm a}=1$ is $\sim \pot{2}{-3}$, while it is about  $\pot{8}{-4}$ for \changeI{those with} $\Theta^{\rm a}=1/f_{\nu'}$.

If we consider the distortion in an even wider range of frequencies (Fig.~\ref{fig:Ff.z}), then we can make several important observation. First, as expected from the discussion related to Eq.~\eqref{app:F_QS_f} in the limit $\xD\rightarrow \infty$ for $\Theta^{\rm a}=1/f_{\nu'}$ one finds $F_\nu\sim 1/f_\nu$, regardless if the normal Voigt profile was used or the effective absorption profile, as given by Eq.~\eqref{sec:tau_abs_all_b}.
However, for $\phi^\ast_{\rm abs}=\phi_{\rm V}$ the limit $F_\nu\sim 1/f_\nu$ is reached closer to the line center than for the effective absorption profile. This is expected, since for the 3 shell atom the effective absorption profile only has a small admixture of the Voigt profile (due to the description of routes connecting to the continuum). If $\phi^\ast_{\rm abs}=\phi_{\rm 3d}$ then the limit $F_\nu\sim 1/f_\nu$ would never be reached, simply because at $\nu\gtrsim\nu_{31}$ the contribution to $\tau_{\rm abs}$ would be zero. Reducing the admixture of the pure Voigt profile therefore moves the transition to $F_\nu\sim 1/f_\nu$ towards larger frequencies.

The second important observation is that in the frequency range $\nu_{21}\lesssim \nu\lesssim \nu_{31}$ on the blue side of the resonance the correction due to the shape of the absorption profile is much more important than both the pure time-dependent correction to $\tau_{\rm abs}$ and the correction due to the $1/f_{\nu'}$ factor in $\Theta^{\rm a}$.
And finally, in the red wing the correction \changeI{to the spectral distortion} is dominated by the $1/f_{\nu'}$ scaling of $\Theta^{\rm a}$, however, the correction is very small, in particular in comparison with the one coming from $\Theta^{\rm t}$ (see Sect.~\ref{sec:F_p0}).

\subsection{Corrections related to both $\tau_{\rm abs}$ and $\Theta^{\rm t}$ but for $\Theta^{\phi} =0$}
\label{sec:F_p0}
We now want to understand the effect of changes in the ionization history and death probability on the effective emission rate. We therefore allow $\Theta^{\rm t}\neq 1$ but still set $\Theta^{\phi} =0$. We then again discuss different combinations of corrections, like in the previous section. As we will see the corrections due to $\Theta^{\rm t}\neq 1$ dominate at large distances on the red side of the resonance, while the shape of the profile is most important for the spectral distortion on the blue side of the resonance (see Fig.~\ref{fig:Ft.z}). In the vicinity of the resonance basically {\it all} the correction factors are important.

\subsubsection{Behavior very close to the line center}
In Fig.~\ref{fig:DFt.core.1100.z} we show the Lyman $\alpha$ spectral distortion at $z=1100$ in the very close vicinity of the line center, now also including $\Theta^{\rm t}\neq 1$. 
\changeI{If we first look at the curves for $f_\nu=1$ (dashed and dashed-dotted), then we can see that very close to the resonance ($|\xD| \lesssim 4$) the solution is not affected by the inclusion of $\Theta^{\rm t}\neq 1$. Due to the huge optical depth across the Doppler core (corresponding to $\Delta z/z\sim 10^{-5}$) there practically any time-dependent variation of the effective emission coefficient is erased. 
However, moving towards the wings time-dependent aspects become important and in particular now also $F_\nu\neq F^{\rm qs}_\nu$ at $\xD<0$.
}

Also one can see that at the considered redshift the difference in comparison to the case $\Theta^{\rm t}=1$ (see Fig.~\ref{fig:DFf.core.1100.z}) is very small at $\xD\lesssim -4$. There clearly the correction due to the thermodynamic factor $f_{\nu}$ (dotted and solid lines) is dominant.
However, at $\xD\gtrsim 4$ the time-dependent changes in the effective emission coefficient lead to a correction that is similarly important as the one due to $f_{\nu}$.
One can also see that all the corrections add roughly linearly.
\changeI{Note that at $\xD<4$ the curves are not affected when accounting for the corrections to the shape of the line profile.}

\subsubsection{Behavior at intermediate and large distances from the line center}
Looking at the right column of Fig.~\ref{fig:DFt.z} we can see that at $\xD\lesssim 0$ the correction due to the inclusion of $f_{\nu}$ dominates at high redshifts, while at $z\sim 1000$ the correction due to $\Theta^{\rm t}$ is already comparable, and clearly dominates at low redshifts.
In addition, at $\xD\gtrsim 0$ one can see that at high redshift all the corrections practically add linearly, while for  $z=1000$ and $z=800$ the correction due to the inclusion of $\Theta^{\rm t}$ practically disappears when including the correct shape of the effective absorption profile.
Also when including the thermodynamic correction factor the large excess of photons seen for the case $f_\nu=1$ and $\phi_{\rm abs}=\phi_{\rm V}$ practically vanishes.
This implies that the {\it self-feedback} effect at low redshifts that was reported elsewhere \citep{Chluba2007b} is expected to disappear. 
As explained there, this unphysical aspect of the solution in the '$1+1$' photon pictures is due to the fact that the Voigt profile in principle extends up to arbitrarily large frequencies, so that photon emitted at $z\sim 1400$ in the very distant blue damping wing will \changeI{still} be able to reach the line center at low redshift, strongly enhancing the photon occupation number.
However, when including the thermodynamic correction factor, due to the exponential enhancement of the absorption optical depth at large distances blue-ward of the resonance, such photons disappear much before this.
Similarly, when including the shape of the effective emission profile such excess of photons will never be produced in the first place, so that from this region the residual correction due to $\Theta^{\rm t}\neq 1$ is much smaller.

Looking at the spectral distortion in a very large range of frequencies around the line center (Fig.~\ref{fig:Ft.z}) it is clear that at all redshift the shape of the distortion is dominated by the correction due to $\Theta^{\rm t}\neq 1$ for $\xD\lesssim 0$. Both the thermodynamic factor and the shape of the absorption and emission profile only lead to small \changeI{additional} modifications there. The largest correction is due to the fact that the 3d two-photon emission profile is larger than the Voigt profile at $\nu\rightarrow \nu_{31}/2$, explaining the small addition of photons in comparison to the case $\phi_{\rm abs}^\ast=\phi_{\rm V}$ seen close to the lowest frequencies shown in the figures.
\changeI{On the other hand, at frequencies above the line center clearly the correction due to the shape of the line profile is most important. In the line center all sources of correction contribute to changes of the Lyman $\alpha$ spectral distortion with respect to the quasi-stationary solution.}

\begin{figure}
\centering 
\includegraphics[width=0.95\columnwidth]{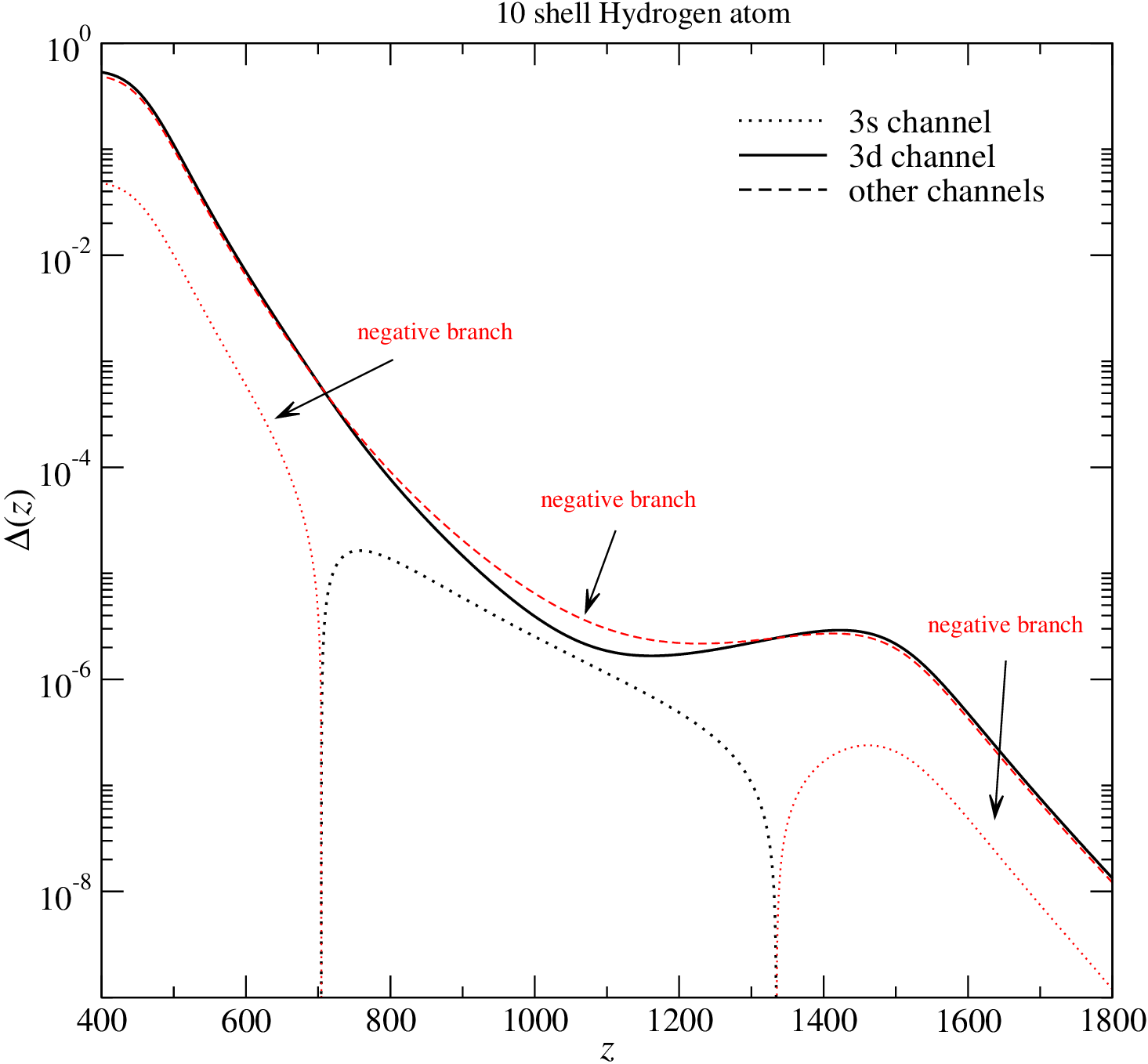}
\caption
{Source of the asymmetry between the absorption and emission profiles. We present $\Delta(z)=R^{i,-}_{\rm 2p}/R^{-}_{\rm 2p}-R^{i,+}_{\rm 2p}/R^{+}_{\rm 2p}$ for a given level as a function of redshift. For all curves we used the solution for the 10 shell hydrogen atom.}
\label{fig:DRm_DRp}
\end{figure}

\subsection{Corrections related to $\tau_{\rm abs}$ and $\Theta^{\rm t}$ including $\Theta^{\phi}\neq 0$}
\label{sec:F_all}
\label{sec:F_phi_corr}
We also ran cases for $\Theta^{\phi}\neq 0$. However, the correction was always extremely small. Therefore we decided to omit the corresponding plots for $F_\nu$.
As mentioned \changeI{in Sect.~\ref{sec:profile_i}}, this is expected since the deviations of $R^{i,-}_{\rm 2p}/R^{-}_{\rm 2p}$ and $R^{i,+}_{\rm 2p}/R^{+}_{\rm 2p}$ from their equilibrium values is always very small in the relevant redshift range, so that $R^{i,+}_{\rm 2p}/R^{+}_{\rm 2p}\approx R^{i,-}_{\rm 2p}/R^{-}_{\rm 2p}$, and hence $\phi^\ast_{\rm abs}\approx \phi_{\rm em}$. In Fig.~\ref{fig:DRm_DRp} we explicitly show this fact. As an example, for the 3d-channel one can see that at $z\sim 1100$ one has $R^{\rm 3d,-}_{\rm 2p}/R^{-}_{\rm 2p}-R^{\rm 3d,+}_{\rm 2p}/R^{+}_{\rm 2p}\sim 10^{-6}-10^{-5}$. Therefore one would expect corrections to the effective escape probability at the level of $\sim 10^{-4}-10^{-3}\%$, which is clearly negligible in the context of the CMB anisotropies. 
We confirmed this statement numerically.

One should also mention that although at low redshifts the expected difference between emission and absorption profile increases, there the value of the effective escape probability is dominated by the contribution from the Doppler core, where details of the profiles will not matter very much. In addition corrections to the escape probability will not propagate very much into the ionization history, so that \changeI{here} we do not discuss this point any further.

\section{Changes in the effective escape probability}
\label{sec:Pesc}
%
%
%
%

\begin{figure*}
\centering 
\includegraphics[width=0.98\columnwidth]{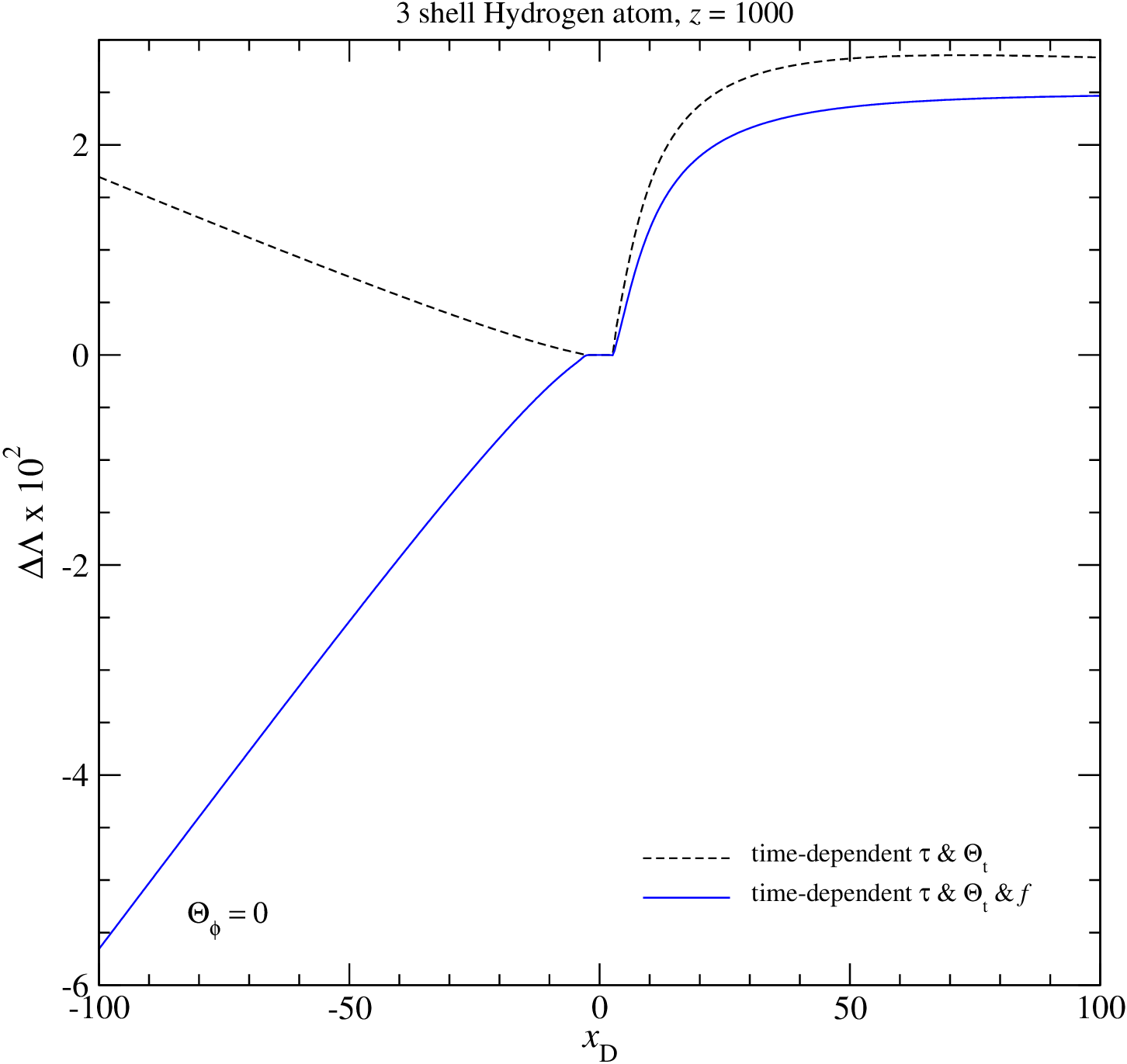}
\includegraphics[width=0.98\columnwidth]{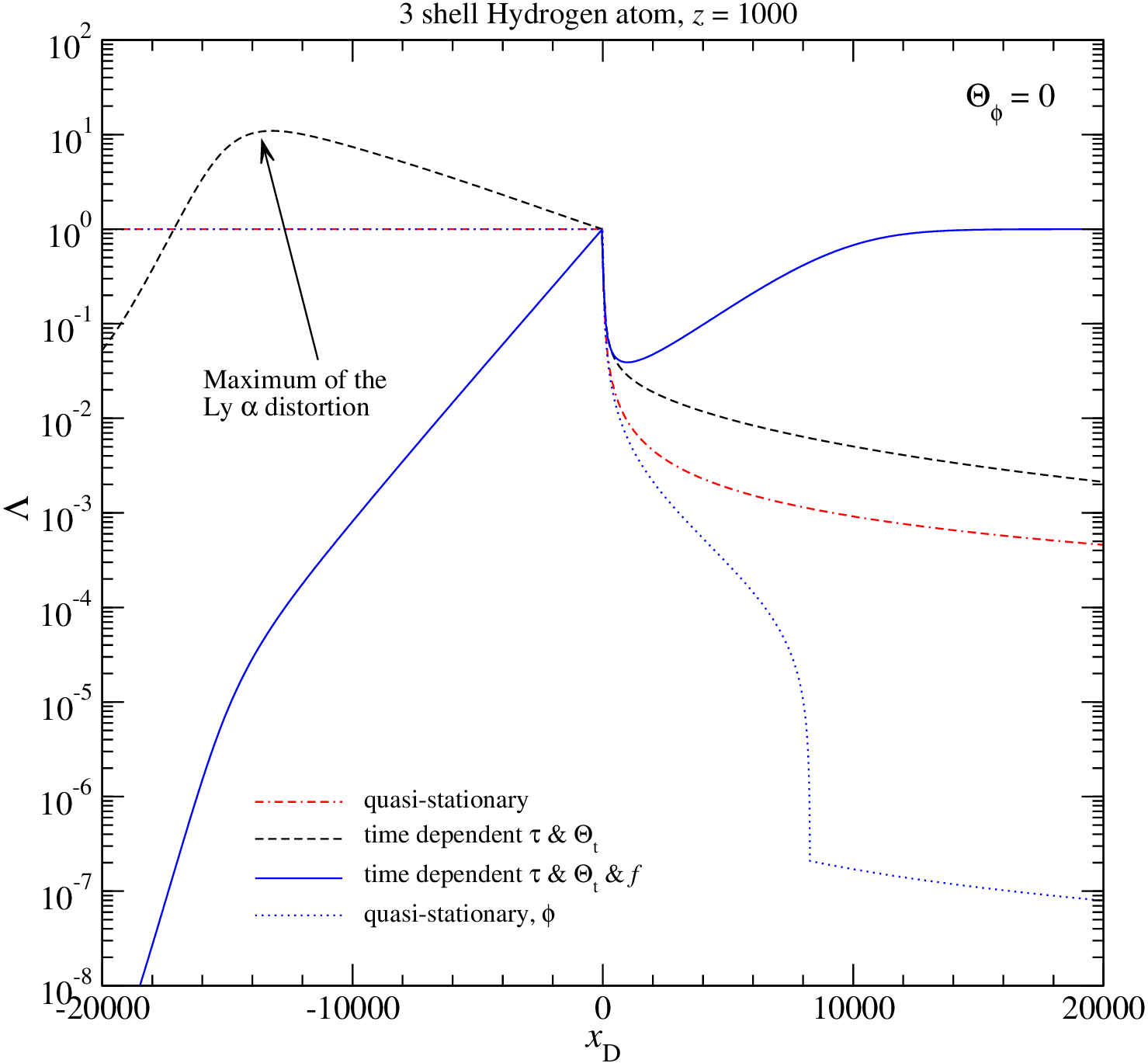}
\caption
{The functions $\mathcal{F}^{\rm f}_\nu=f_\nu\,F^{\rm f}_\nu$ at redshift $z=1000$. In the left panel we show the difference of $\Delta \Lambda=\mathcal{F}^{\rm f}_\nu-F^{\rm qs}_\nu$, while in the right we directly show $\Lambda=\mathcal{F}_\nu$.
The cases labeled with $f$ are computed using $\Theta^{\rm a}=\Theta^{\rm t}/f_{\nu'}$, the others with $\Theta^{\rm a}=\Theta^{\rm t}$. In addition the quoted correction factors were included in the computation of $\tau_{\rm abs}(\nu, z_{\rm s}, z')$. 
For comparison in the right panel we also directly show the normal quasi-stationary solution $F^{\rm qs}_\nu$ (double dash dotted line) and $F^{\phi, \rm qs}_\nu$ (dotted line) that includes the correction due to the profile $\phi_{\rm abs}^\ast\neq \phi_{\rm V}$ (see Sect.~\ref{sec:corr_P}).
We assumed a 3 shell hydrogen atom.}
\label{fig:Ftf.z}
\end{figure*}

\subsection{Effect of the thermodynamic correction factor}
\label{sec:normal}
First we consider the normal Lyman $\alpha$ transfer equation~\eqref{eq:real_em_abs_11}, but including the thermodynamic correction factor $f_{\nu}$, in order to correct for the small imbalance in the emission and absorption process in the line wings coming from the normal '$1+1$' photon formulation (see Sect.~\ref{sec:thermodyn}). 
In this case one has $\phi_{i}(\nu)=\phi_{\rm V}(\nu)$ and hence $\phi_{\rm em}\equiv \phi^\ast_{\rm abs}\equiv \phi_{\rm V}(\nu)$, so that from \eqref{app:kin_abs_em_Sol_phys_asym_F} one can find
\bsub
\label{eq:F_f}
\beal
\label{eq:F_f_a}
F^{\rm f}_\nu
&=
\!\int_{z_{\rm s}}^z \! \Theta^{\rm f}(z, z')\,
\partial_{z'}e^{-\tau_{\rm abs}(\nu, z', z)}\id z'
\\[1mm]
\label{eq:F_f_b}
\Theta^{\rm f}(z, z')&=
\frac{1}{f_{\nu'}(z')}
\frac{
\tilde{N}_{\rm em}(z')-\tilde{N}^{\rm pl}_{x'_{21}} }{\tilde{N}_{\rm em}(z)-\tilde{N}^{\rm pl}_{x_{21}}}
=
\frac{\Theta^{\rm t}(z,z')}{f_{\nu'}(z')},
\end{align}
\esub
with $\tau_{\rm abs}$ as given by Eq.~\eqref{sec:tau_abs_all_a} but for $\phi^\ast_{\rm abs}(\nu)\rightarrow \phi_{\rm V}(\nu)$.

If here one drops the factors due to $f_\nu$ in the definition of $\Theta^{\rm f}$ and also $\tau_{\rm abs}$, one obtains the purely time-dependent case, $F^{\rm t}_\nu$, that has been addressed earlier \citep{Chluba2008b}. However, here the term $\Theta^{\rm t}$ naturally is given by the line center values of $\tilde{N}_{\rm em}(z')-\tilde{N}^{\rm pl}_{x}$, which in the previous work had to be enforced for consistency with the standard approximations \citep[see comments in Sect.~3.4.1 of][]{Chluba2008b}.

To understand the role of $f_\nu$ in the final correction to $P_{\rm eff}$, Eq.~\eqref{eq:DP_eff}, we first look at the term $\Delta P_{\rm ind}$, Eq.~\eqref{eq:DP_ind}. It is clear that for $\phi_{\rm em}=\phi_{\rm abs}^\ast=\phi_{\rm V}$ one has $\Delta P_{\rm ind}\equiv 0$, since in this case $\Delta\bar{\varphi}_{\rm em}\!\equiv\! \Delta\bar{\varphi}^\ast_{\rm abs}\!=\!0$.
As we will show below (Sect.~\ref{sec:corr_DP_ind}), $\Delta P_{\rm ind}\neq 0$ when taking into account the effect of stimulated emission in the blackbody radiation field, however, the contribution to the final correction turns out to be negligible. 

If we now look at the definition of $P$, Eq.~\eqref{eq:def_P}, then for the considered case we can rewrite it as
\beal
\label{eq:P_rew_f}
P^{\rm f}&=\int\varphi_{\rm V}(\nu)\,[1-\mathcal{F}^{\rm f}_\nu]\id\nu
\end{align}
where we introduced the function 
\beal
\label{eq:F_f_cal}
\mathcal{F}^{\rm f}_\nu=f_\nu(z)\, F^{\rm f}_\nu.
\end{align}
This representation allows to directly reveal the expected differences in comparison with the standard quasi-stationary case, for which one has ($\varphi_{\rm em}=\varphi^\ast_{\rm abs}\equiv\varphi_{\rm V}$ and $f_\nu\equiv 1$)
\beal
\label{eq:F_cal_qs}
\mathcal{F}^{\rm qs}_\nu\equiv F^{\rm qs}_\nu=1-e^{-\tau_{\rm d}}\,e^{\tau_{\rm d}\,\chi}.
\end{align}
If instead of $\mathcal{F}^{\rm f}_\nu$ we insert this into Eq.~\eqref{eq:P_rew_f}, we directly obtain $P_{\rm d}=\frac{1-e^{-\tau_{\rm d}}}{\tau_{\rm d}}$.
Therefore we can write the correction with respect to the quasi-stationary solution as
\beal
\label{eq:DP_f_d}
\Delta P^{\rm f}_{\rm d}&=P^{\rm f}-P_{\rm d}=-\int\phi_{\rm V}(\nu)\,[\mathcal{F}^{\rm f}_\nu-\mathcal{F}^{\rm qs}_\nu]\id\nu.
\end{align}
This expression shows that for the correction to the effective escape probability it in fact is not important how $F^{\rm f}_\nu$ behaves, but how $\mathcal{F}^{\rm f}_\nu$ looks. Since $f_\nu$ is a very strong function of frequency this makes a big difference in the conclusions, as we explain below. 
Furthermore, any difference to the value of $\mathcal{F}^{\rm qs}_\nu$ will lead to a contribution to $\Delta P^{\rm f}_{\rm d}$ with {\it opposite} sign.
Also, it is possible to compute the {\it partial contributions} to the total correction in the escape probability by only integrating $\Delta P^{\rm f}_{\rm d}$ as defined by Eq.~\eqref{eq:DP_f_d} over a given range of frequencies.

\subsubsection{Behavior of $\mathcal{F}^{\rm f}_\nu$ and the expected corrections to $P_{\rm eff}$}
\label{sec:Fcal_f_beh}
In Fig.~\ref{fig:Ftf.z}, as an example we show $ \mathcal{F}^{\rm f}_\nu$ at redshift $z=1000$ (dotted lines).
For comparison we also show the pure time-dependent solution (dashed curves) for which $\mathcal{F}^{\rm t}_\nu\equiv F^{\rm t}_\nu$.
Note that we include the correction terms in both $\tau_{\rm abs}$ and $\Theta^{\rm a}$.

If we focus on the behavior at $-100\leq \xD \lesssim -4$, then we can see that although in the considered cases $F^{\rm f}_\nu>F^{\rm t}_\nu>F^{\rm qs}_\nu$ (compare with Fig.~\ref{fig:DFt.z}, right column, case $z=1000$), for $\mathcal{F}^{\rm f}_\nu$ one has $\mathcal{F}^{\rm t}_\nu>\mathcal{F}^{\rm qs}_\nu>\mathcal{F}^{\rm f}_\nu$. This change in the sequence is only due to the factor $f_\nu<1$ in the definition of $\mathcal{F}^{\rm f}_\nu$, which appears due to $\phi_{\rm abs}=f_\nu\phi_{\rm abs}^\ast$ in the escape integral, Eq.~\eqref{eq:def_P}.
Instead of an additional {\it negative} contribution to the escape probability ($\Delta P^{\rm f}<\Delta P^{\rm t}<0$), as it would follow from the differences seen in $F_\nu$, when including the thermodynamic correction factor $f_\nu$ one therefore expects a {\it positive} contribution from the considered spectral region.
This effect becomes even more apparent when looking at the very distant red wing, where clearly $\mathcal{F}^{\rm t}_\nu>\mathcal{F}^{\rm qs}_\nu\gg\mathcal{F}^{\rm f}_\nu$, owing to the exponential cutoff introduced by $f_\nu$.
The behavior shows that in the very distant red wing the excess Lyman $\alpha$ photons no longer supports the flow of electrons toward higher levels and the continuum. These photons only undergo line scattering events, with tiny shifts in the frequency due to the Doppler motion of the atom, but do not die anymore, and therefore have effectively already escaped.
As we will demonstrate below (Sect.~\ref{sec:red_blue_f} and \ref{sec:estimate_red_f}) in fact the main correction due to $f_\nu$ is coming from the change in the absorption cross section in the red wing of the line profile.
The real modifications of the spectral distortion due to $f_\nu$ are not so important.

If we now look at the behavior in the range $4\leq \xD \lesssim 100$ we can see that the sequence $F^{\rm t}_\nu>F^{\rm f}_\nu>F^{\rm qs}_\nu$ does not change when considering $\mathcal{F}^{\rm t}_\nu>\mathcal{F}^{\rm f}_\nu>\mathcal{F}^{\rm qs}_\nu$, but $\mathcal{F}^{\rm f}_\nu$ becomes more similar to $\mathcal{F}^{\rm t}_\nu$. 
From the behavior at $-100\lesssim \xD\lesssim -4$ and the strength of the changes seen there one might have expected that at $4\leq \xD \lesssim 100$ also $\mathcal{F}^{\rm t}_\nu\geq\mathcal{F}^{\rm f}_\nu$, since $|\Delta f_\nu/f_\nu|$ in fact is similar in both regions. However, at $4\lesssim \xD\lesssim 100$ the spectral distortion $F_\nu$ is very steep, so that small changes $\Delta f_\nu/f_\nu$ cannot affect the shape of $\mathcal{F}^{\rm f}_\nu$ so much in comparison to ${F}^{\rm f}_\nu$.
Only when going to much larger distances on the blue side of the resonance, where $f_\nu$ again behaves exponentially, one can see $\mathcal{F}^{\rm f}_\nu\gg\mathcal{F}^{\rm t}_\nu>\mathcal{F}^{\rm qs}_\nu$ although there $F^{\rm t}_\nu\gg F^{\rm f}_\nu$ (compare with Fig.~\ref{fig:Ft.z}, right column, case $z=1000$).
In comparison to the pure time-dependent correction from the range $4\leq \xD \lesssim 100$ one therefore expects a slightly smaller (negative) correction to the total value of $P$, while the contributions from very large distances in the blue wing should be significantly larger than in the pure time-dependent case.
However, here it is important to mention that these very distant wing contributions will always be very minor, since the Voigt profile drops like $\propto 1/x^2_{\rm D}$ (see Sect.~\ref{sec:blue_wing_f}).

Finally, looking at the central region $-4 \lesssim\xD\lesssim 4$ we can see that there now $\mathcal{F}^{\rm t}_\nu\approx\mathcal{F}^{\rm f}_\nu\approx\mathcal{F}^{\rm qs}_\nu$. This is because $F^{\rm f}_\nu\approx 1/ f_\nu$ in that regions (see Sect.~\ref{sec:f_center}), so that $\mathcal{F}^{\rm f}_\nu\approx 1$ with very high accuracy. 
This implies that although the Lyman $\alpha$ spectral distortions in the case $F^{\rm f}_\nu$ and $F^{\rm qs}_\nu$ look rather different the correction factor $f_\nu$ does not lead to any real correction in the escape probability from inside the Doppler core. There everything is well described within the quasi-stationary assumption, for which the whole Doppler core reaches thermodynamic equilibrium with the ambient radiation field, but now also including the small additional variation of the photon distribution over $\nu$.

\begin{figure}
\centering 
\includegraphics[width=0.95\columnwidth]{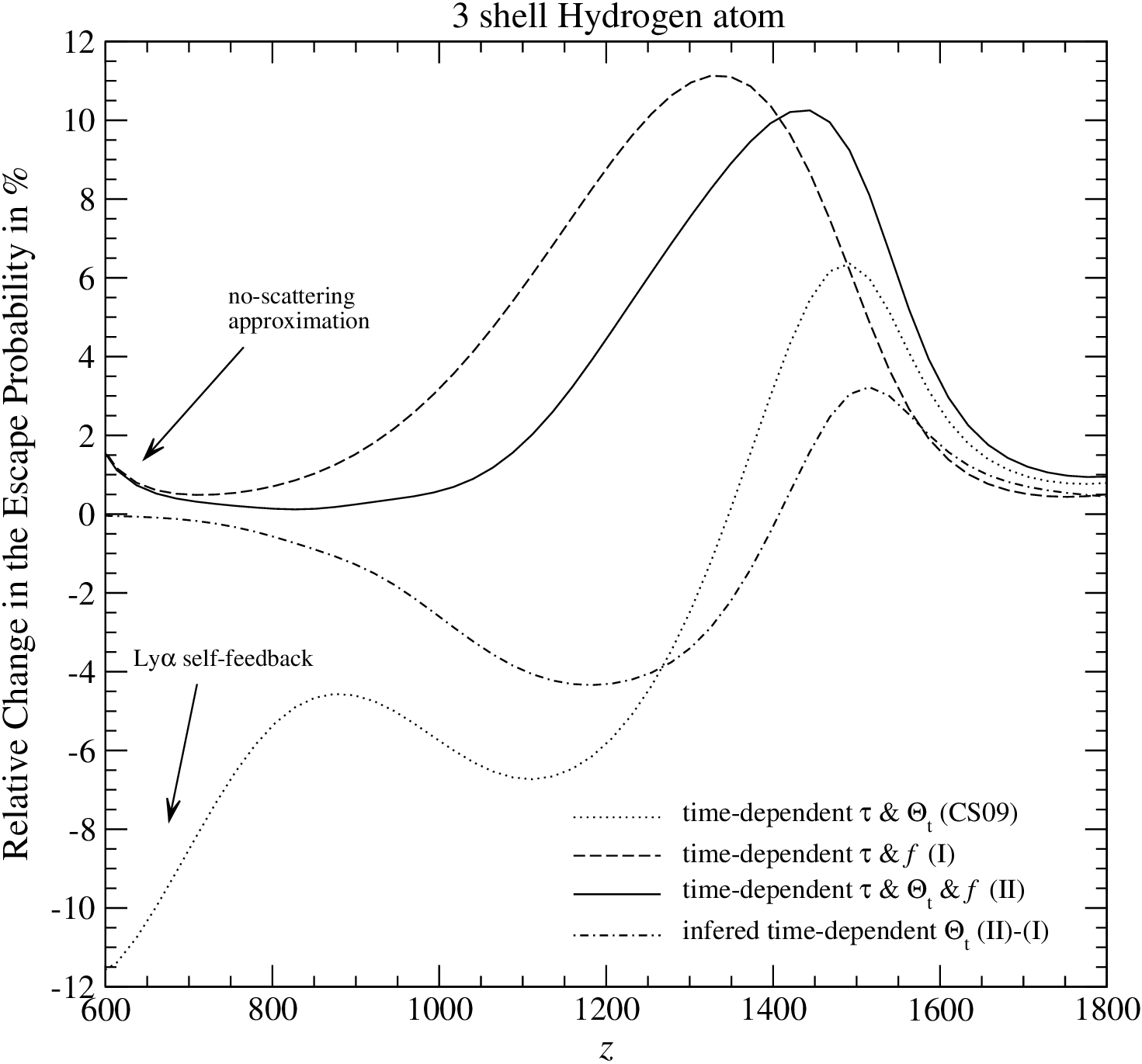}
\caption
{Relative difference in the effective escape probability with respect to the Sobolev escape probability: effect of the thermodynamic correction factor.}
\label{fig:DP_P.f.comp}
\end{figure}
\subsubsection{Correction to the escape probability}
\label{sec:Pesc_f}
In Fig.~\ref{fig:DP_P.f.comp} we show the result for the effective escape probability and the effect of $f_\nu$. For comparison we also show the result for the pure time-dependent correction (dotted line) which was already discussed elsewhere \citep{Chluba2008b}. 
At low redshifts we indicate the rise in the amplitude of the correction, which was attributed to the late self feedback of Lyman $\alpha$ photons for this case.  For the other cases we also point towards the difference which arises due to the no scattering approximation. It is due to the differences in $P_{\rm d}$ and $P_{\rm S}$ itself which close to the maximum of the Thomson visibility function at $z\sim 1100$ are negligible, but become notable both at  very low and very high redshifts  \citep{Chluba2008b}. 
\changeII{However, there the changes have no effect on the free electron fraction.}

When now including the thermodynamic correction factor in the computation of the absorption optical depth and the effective emission rate $\Theta^{\rm a}$, but setting $\Theta^{\rm t}=1$ (dashed line), one can see that the correction to the escape probability becomes positive at all redshifts, with a maximum of $\Delta P/P\sim 11\%$ at $z\sim 1350$.
As we will see in Sect.~\ref{sec:red_blue_f} and \ref{sec:estimate_red_f} bulk of this total correction is coming from changes in the absorption process on the red side of the resonance, where in this case $F_\nu\approx F^{\rm qs}_\nu\approx 1$ (cf. Fig.~\ref{fig:DFt.z} and \ref{fig:Ft.z}, left column).

When also including the variation of $\Theta^{\rm t} \neq 1$ over time (solid line) the result changes significantly, shifting the maximum of the correction $\Delta P/P\sim 10\%$ to $z\sim 1450$, but still leading to $\Delta P/P>0$ everywhere.
However, especially the low redshift part is strongly modified, reducing the total correction by a factor $\sim 2$ at $z\sim 1100$.
We also show the infered correction due to $\Theta^{\rm t}$ alone, which was obtained by taking the difference between the curves labeled (I) and (II).  
The result shows that the final correction close to the maximum of the Thomson visibility function has important contributions from both $f_\nu\neq 1$ and the time-dependence of the problem.

\subsubsection{Effect at large distances blue-ward of the line center and the Lyman $\alpha$ self feedback}
\label{sec:blue_wing_f}
The thermodynamic factor clearly strongly changes the behavior of the correction to the effective escape probability. The purely time-dependent correction is no longer dominant, and in particular the thermodynamic factor removes the {\it self feedback} problem of Lyman $\alpha$ at low redshifts, which was already realized to be an artifact of the standard '$1+1$' photon formulation \citep{Chluba2008b}.
This is due to the fact that when taking $f_\nu$ into account, exponentially fewer photons remain in the photon distribution at large distances on the blue side of the resonance (cf. Fig.~\ref{fig:Ft.z} where $F_\nu\sim 1/f_\nu$ at large $\xD$). As explained in Sect.~\ref{sec:sources_of_corr} every photon emitted at $\xD\gg 1$ will be re-absorbed quasi-instantaneously. This is due to the exponentially larger amount of CMB photons red-ward of the the Balmer $\alpha$ line, than close to the line center, so that line absorption is more effective. 
The main process for the death here is the $\rm 1s\rightarrow 2p \rightarrow 3d$, where the last step is considered to lead to a complete redistribution, so that the absorbed Lyman $\alpha$ photon ($\rm 1s\rightarrow 2p$) will most likely reappear close to the line center. Note that in this Section we still have not included the two-photon corrections to the shape of the absorption profile, but already now the two-photon character of the process leads to this conclusion.

However, one has to mention that basically all the photons present at these large distances contribute to the escape integral. This is because $\mathcal{F}^{\rm f}_\nu\sim 1$ while $\mathcal{F}^{\rm qs}_\nu\ll1$, so that $\mathcal{F}^{\rm f}_\nu-\mathcal{F}^{\rm qs}_\nu\sim 1$ (see Sect.~\ref{sec:Fcal_f_beh} and Fig.~\ref{fig:Ftf.z}).
Therefore in this case the very distant blue wing contribution to the value of $\Delta P^{\rm f}_{\rm bw}$ behaves like
\beal
\label{eq:DP_wing}
\frac{\Delta P^{\rm f}_{\rm bw}}{P_{\rm d}}\approx -\int_{\nu_{\rm bw}}^\infty \phi_{\rm V}\id\nu' \approx 
-\frac{a}{\pi}\frac{\tau_{\rm d}}{x_{\rm D}^{\rm bw}},
\end{align}
at $z\sim 1100$ implying an additional $\lesssim 0.1\%$ correction from $x_{\rm D}^{\rm bw}\gtrsim 10^4$, showing that the absolute contribution becomes negligible far beyond that point.
In numerical computations we therefore typically integrated the profile up to $\xD\sim \pot{1.5}{4}$, introducing a very small error for cases with $f_\nu\neq 1$ and practically no error when also including the correct shape of the emission profile, since then naturally very few photons appear at larger distances blue-ward of the resonance (e.g. see Fig.~\ref{fig:Ftf.z}, solid line).

\begin{figure}
\centering 
\includegraphics[width=0.95\columnwidth]{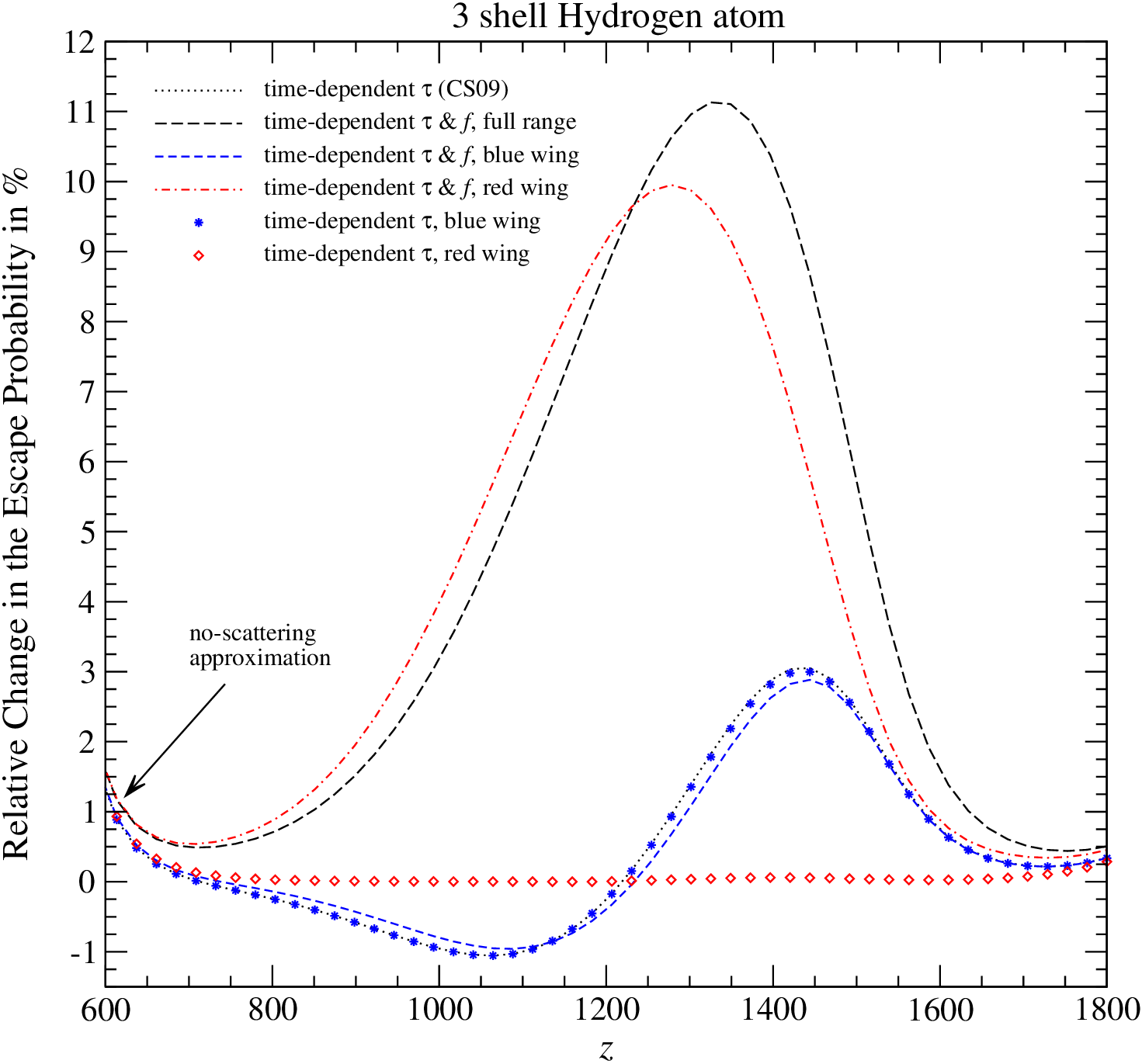}
\\[1mm]
\includegraphics[width=0.95\columnwidth]{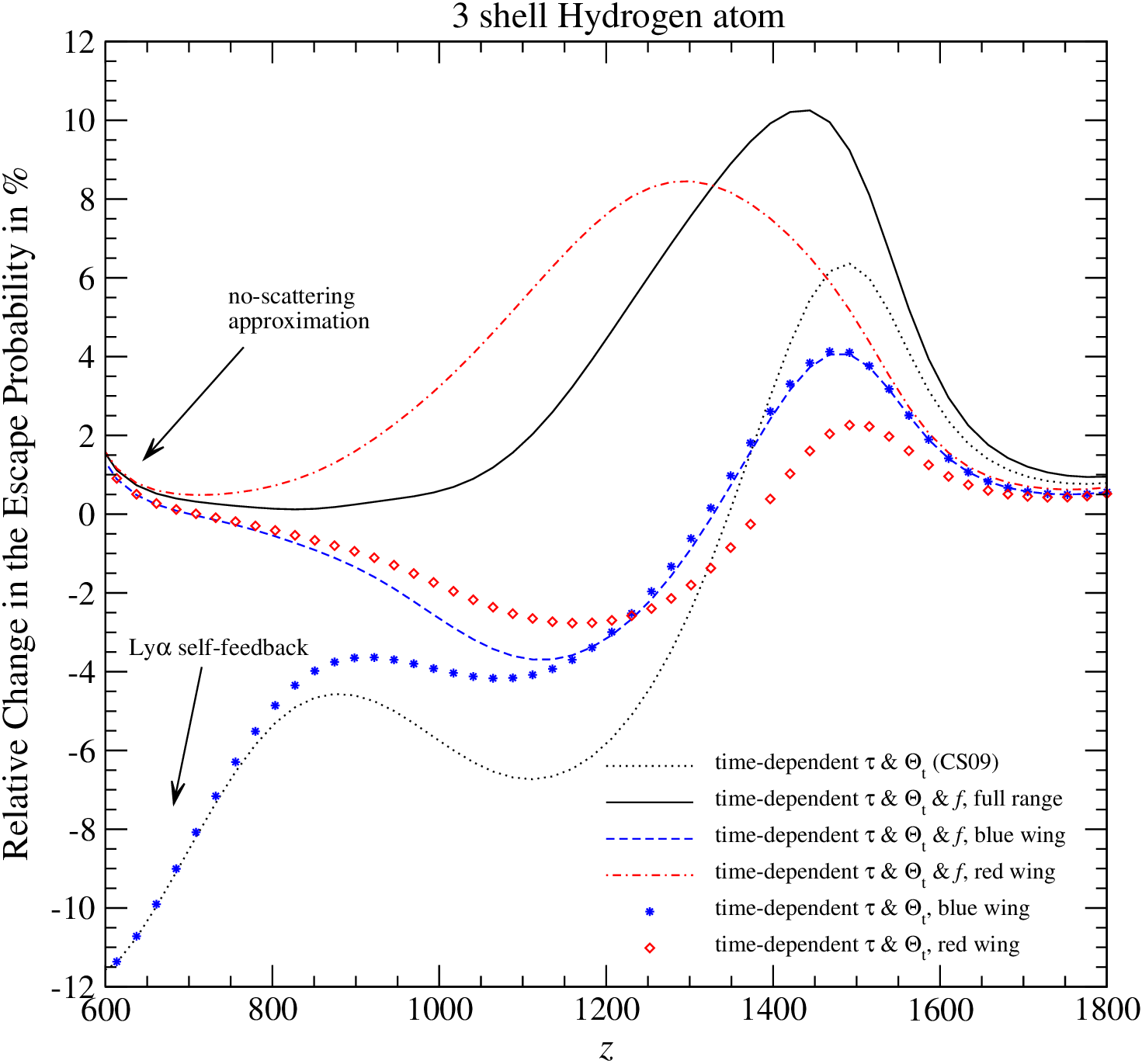}
\caption
{Relative difference in the effective escape probability with respect to the Sobolev escape probability: effect of the thermodynamic correction factor in the blue and red wing of the resonance. For the upper panel we set $\Theta^{\rm t}=1$ and $\Theta^{\phi}=0$. The lower panel was computed also including $\Theta^{\rm t}\neq 1$.}
\label{fig:DP_P.f.comp.rb}
\end{figure}
\subsubsection{Contributions from the blue and red wing}
\label{sec:red_blue_f}
Returning to the correction at higher redshifts ($z\gtrsim 800-900$), it was already shown earlier, that there the Lyman $\alpha$ self feedback is not important \citep{Chluba2008b}.
From Fig.~\ref{fig:DP_P.f.comp}, it is clear that both considered cases for the effect of $f_\nu$ imply that at a given redshift effectively fewer photons support the flow of electrons towards higher levels and the continuum than in the quasi-stationary case, albeit the fact the more photons are produced. 
The latter statement can also be confirmed by looking at Fig.~\ref{fig:Ft.z} and the amplitude of the Lyman $\alpha$ distortion in the distant red wing around its maximum. Note that these photons have already been emitted at $z\sim 1400$.

However, where does the main correction to the escape probability come from at these redshifts? Looking at Fig.~\ref{fig:Ftf.z}, we can see that at $0\lesssim \xD \lesssim 100$ the function $\mathcal{F}^{\rm f}_\nu$ is very similar to $\mathcal{F}^{\rm t}_\nu$, which results in the pure time-dependent correction. Also at slightly larger distances ($100\lesssim \xD \lesssim 10^3$) one still has $\mathcal{F}^{\rm f}_\nu\sim\mathcal{F}^{\rm t}_\nu$. Therefore, one does not expect very dramatic changes in the contribution to the effective escape from this part of the Lyman $\alpha$ distortion in comparison to the purely time-dependent case. 

On the other hand, in the red wing one finds $\mathcal{F}^{\rm f}_\nu\lesssim\mathcal{F}^{\rm t}_\nu$, and at very large distances one even has $\mathcal{F}^{\rm f}_\nu\ll\mathcal{F}^{\rm t}_\nu$. 
Physically this reflects the fact that due to $f_\nu$ the re-absorption process in the distant red wing is  exponentially suppressed, so that there photons can escape more directly, than in the normal '$1+1$' photon formulation.  This is now related to the exponentially smaller amount of CMB photons blue-ward of the Balmer $\alpha$ line, so that the main absorption channel $\rm 1s\rightarrow 2p \rightarrow 3d$ becomes practically inactive at $\xD\lesssim -10^3$.
It therefore is the modifications in the red wing absorption process from which one expects the largest effect in connection with $f_\nu$ at high redshifts, before the appearance of the self feedback problem.

Numerically we studied this statement by simply assuming that at $\xD\leq 0$ the solution is given by the quasi-stationary result (implying $\mathcal{F}^{\rm f}_\nu\equiv \mathcal{F}^{\rm qs}_\nu\approx 1$), while at $0\leq \xD$ we used the real solution for $\mathcal{F}_\nu$ in the considered case.
In this way it is possible to separate the 'blue wing' contribution to the total correction in the effective escape probability, and similarly one can obtain the 'red wing' contribution.
In Fig.~\ref{fig:DP_P.f.comp.rb} we show the comparison of these computations for the cases $\Theta^{\rm t}=1$ and $\Theta^{\phi}=0$ (upper panel; only the corrections to $\tau$ are included and $\Theta^{\rm a}=1/f_{\nu'}$), and $\Theta^{\rm t}\neq 1$ and $\Theta^{\phi}=0$ (lower panel; $\Theta^{\rm a}=\Theta^{\rm t}/f_{\nu'}$).
For comparison we also show the results obtained for the purely time-dependent correction in the considered cases \citep[cf.][]{Chluba2008b}.
In the first case (upper panel), one can clearly see that the blue wing contribution from $\mathcal{F}^{\rm f}_\nu$  is very close to the purely time-dependent result (dotted curve), which itself in fact has no significant contribution \changeII{from} the red wing in the first place (diamonds). This shows that for this case the effect of $f_\nu$ is not important at $\xD\geq 0$.
One can see that indeed the main correction due to the effect of $f_\nu$ arises from the red wing, and that this correction is significantly larger than the time-dependent case alone.

If we look at the comparison in the full time-dependent case (lower panel), one can see that when including the correction factor $f_\nu$, at high redshifts the blue wing contribution (dashed curve) is about 50\% of the total result presented in \citet{Chluba2008b}. At high redshifts the blue-wing contribution in the pure time-dependent case (stars) practically coincides with the one that includes $f_\nu$, implying again that the blue wing contribution is not affected much by the thermodynamic correction factor. 
However, one can see that at low redshifts $f_\nu$ is very important for avoiding the self feedback problem, as explained in Sect.~\ref{sec:blue_wing_f}.
Note that in contrast to the curve quoted '$\Delta P^{\rm t}$ (only blue)' in Fig.~8 of \citet{Chluba2008b}, here the blue wing contribution takes into account the time-dependent correction on $\tau$ and $\Theta^{\rm t}$ simultaneously.
%

\begin{figure}
\centering 
\includegraphics[width=0.95\columnwidth]{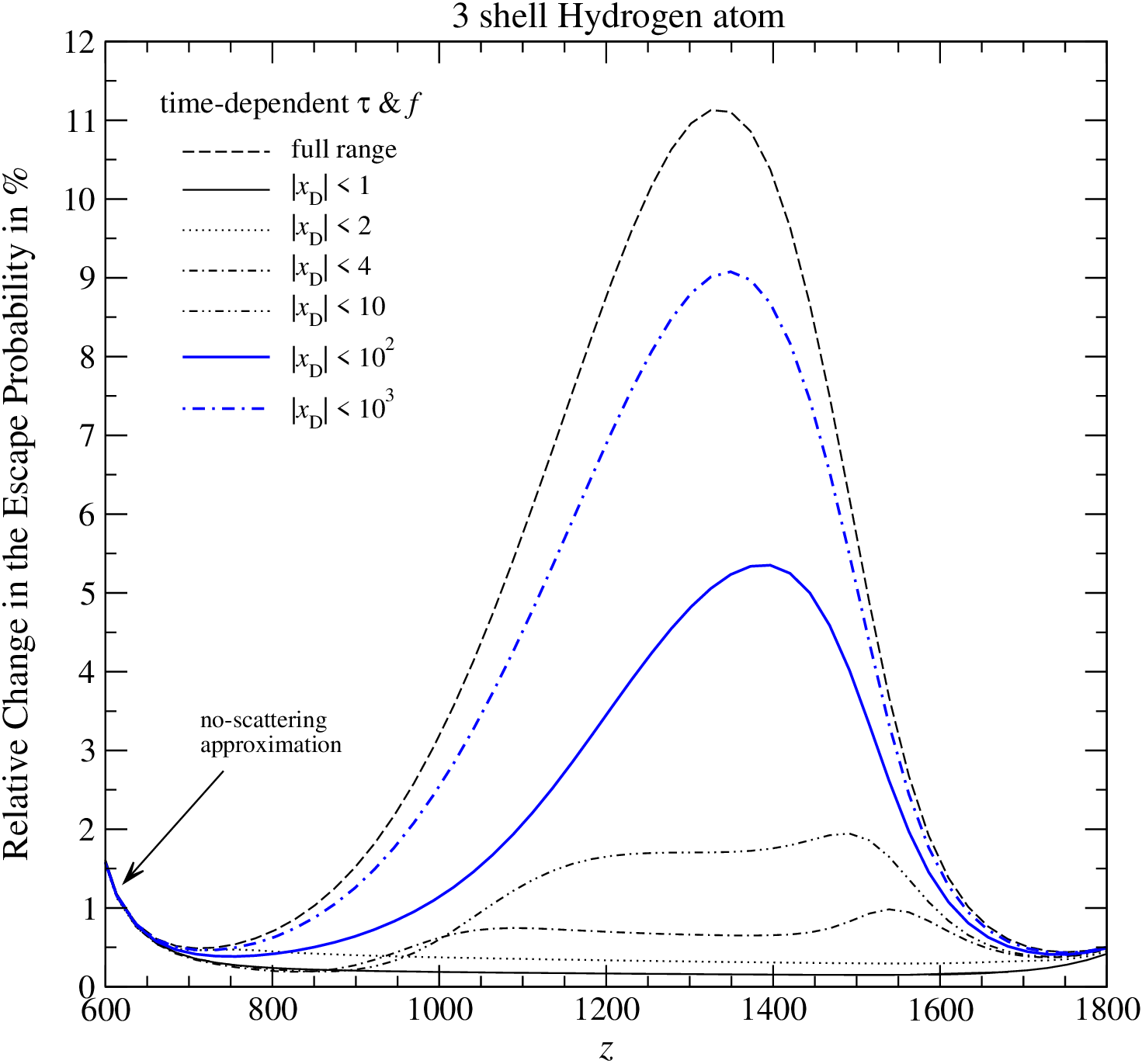}
\\[1mm]
\includegraphics[width=0.95\columnwidth]{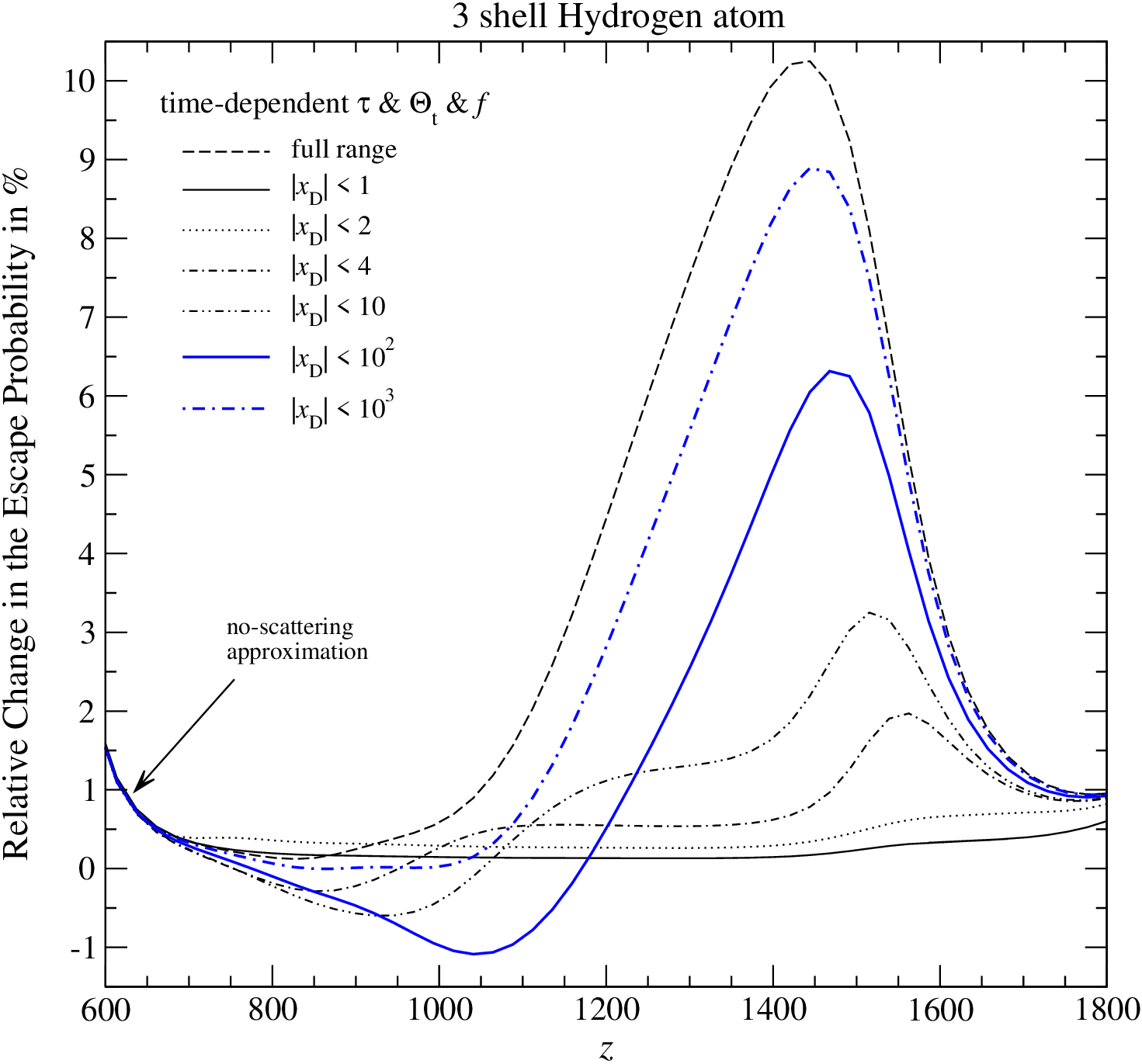}
\caption
{Relative difference in the effective escape probability with respect to the Sobolev escape probability: effect of the thermodynamic correction factor at different distances to the line center. For the upper panel we set $\Theta^{\rm t}=1$ and $\Theta^{\phi}=0$. The lower panel was computed also including $\Theta^{\rm t}\neq 1$.}
\label{fig:DP_P.f.comp.xD}
\end{figure}
Looking at the red wing contributions for this case one can see that for $f_\nu=1$ (diamonds) now the contribution is non zero. This was also already seen earlier \citep{Chluba2008b} and is due to the fact that $F_\nu^{\rm t}\neq 1$. However, the contribution from the red wing is much larger when including $f_{\nu}\neq 1$, and in particular it is only positive due to the fact that $\mathcal{F}^{\rm f}_\nu\lesssim 1$ at all frequencies, so that $\Delta P>0$.
The conclusion clearly is that the dominant correction due to the inclusion of $f_\nu$ is coming from the red wing of the Lyman $\alpha$ resonance. 

%
%

\subsubsection{Simple estimate for the red wing correction}
\label{sec:estimate_red_f}
We can also perform another rough estimate for the expected correction, assuming that in the red wing $\mathcal{F}^{\rm f}_\nu\sim f_\nu$ as suggested by Fig.~\ref{fig:Ftf.z}. This will overestimate the result, since with the inclusion of $f_\nu$ alone one already obtains $\mathcal{F}^{\rm f}_\nu\gtrsim f_\nu$.
In comparison with the quasi-stationary case ($\mathcal{F}^{\rm qs}_\nu\approx 1$) we then have
\beal
\label{eq:DP_redwing}
\frac{\Delta P^{\rm f}_{\rm rw}}{P_{\rm d}}
\! \approx \!-\tau_{\rm d}\!\int_0^{\nu_{\rm core}}\!\!\!\!\! [f_\nu-1] \phi_{\rm V}\id\nu' 
\!\approx \!
\tau_{\rm d}\,\frac{a}{\pi}\! \int_{-\infty}^{-4}\!
 \frac{1-e^{\frac{h\Delta\nu_{\rm D}}{k\Tg}\xD}}{x_{\rm D}^2}\id\xD.
\end{align}
%
Since $\frac{h\Delta\nu_{\rm D}}{k\Tg}\sim 10^{-3}\left[\frac{1+z}{1100}\right]^{-1/2}\ll 1$ one can show
\beal
\label{eq:DP_redwing_exp}
\frac{\Delta P^{\rm f}_{\rm rw}}{P_{\rm d}}
\approx 
\pot{1.6}{-6} \,\tau_{\rm d}\,\left[\frac{1+z}{1100}\right]^{-1}
\end{align}
At redshift $z\sim 1100$ one has $\tau_{\rm d}\sim \pot{6.8}{4}$, so that only from the red wing correction one expects $\frac{\Delta P^{\rm f}_{\rm rw}}{P_{\rm d}}\sim 11\%$, while at redshift $z\sim 1350$ one finds $\frac{\Delta P^{\rm f}_{\rm rw}}{P_{\rm d}}\sim 16\%$. 
Looking at the upper panel in Fig.~\ref{fig:DP_P.f.comp.rb} shows that this is the right order of magnitude, although the final correction is about $1.5$ times smaller than given by this simple estimate. 

\subsubsection{Dependence on the distance to the line center}
Finally we want to look at the dependence of the correction on the distance to the line center. For this we computed the results including the deviation for the quasi-stationary case in a given range around the line center. 
The results of these computations are shown in Fig.~\ref{fig:DP_P.f.comp.xD}. Clearly a large fraction of the total correction is coming from large distances ($10\lesssim |\xD|\lesssim 100-1000$) from the line center, while the contributions from within the Doppler core ($|\xD|\lesssim 4$)  are very small.
The latter result again reflects the fact that there neither $f_\nu$ deviates very strongly from unity, nor does any time-dependent effect become important. The Doppler core can be considered quasi-stationary \citep[for more explanation see][]{Chluba2008b} and in full equilibrium with the line center value.

\subsection{Dependence on the shape of the absorption profile}
\label{sec:phi_corr}
As a next step we want to understand how the two-photon corrections to the shape of the effective line profile affect the escape probability. For this we ran computations only including the fact that $\phi^\ast_{\rm abs}\neq \phi_{\rm V}$, but for the moment we shall neglect the correction due to $f_\nu$ and also assume $\phi^\ast_{\rm abs}=\phi_{\rm em}$ inside $\Theta^{\rm a}$. As explained in  Sect.~\ref{sec:F_phi_corr} the latter correction for our purpose is negligible during cosmological recombination, but the inclusion of $f_\nu\neq 1$ is still expected to be very important, as we will discuss in Sect.~\ref{sec:total_corr} for the combined effect.

\begin{figure}
\centering 
\includegraphics[width=0.95\columnwidth]{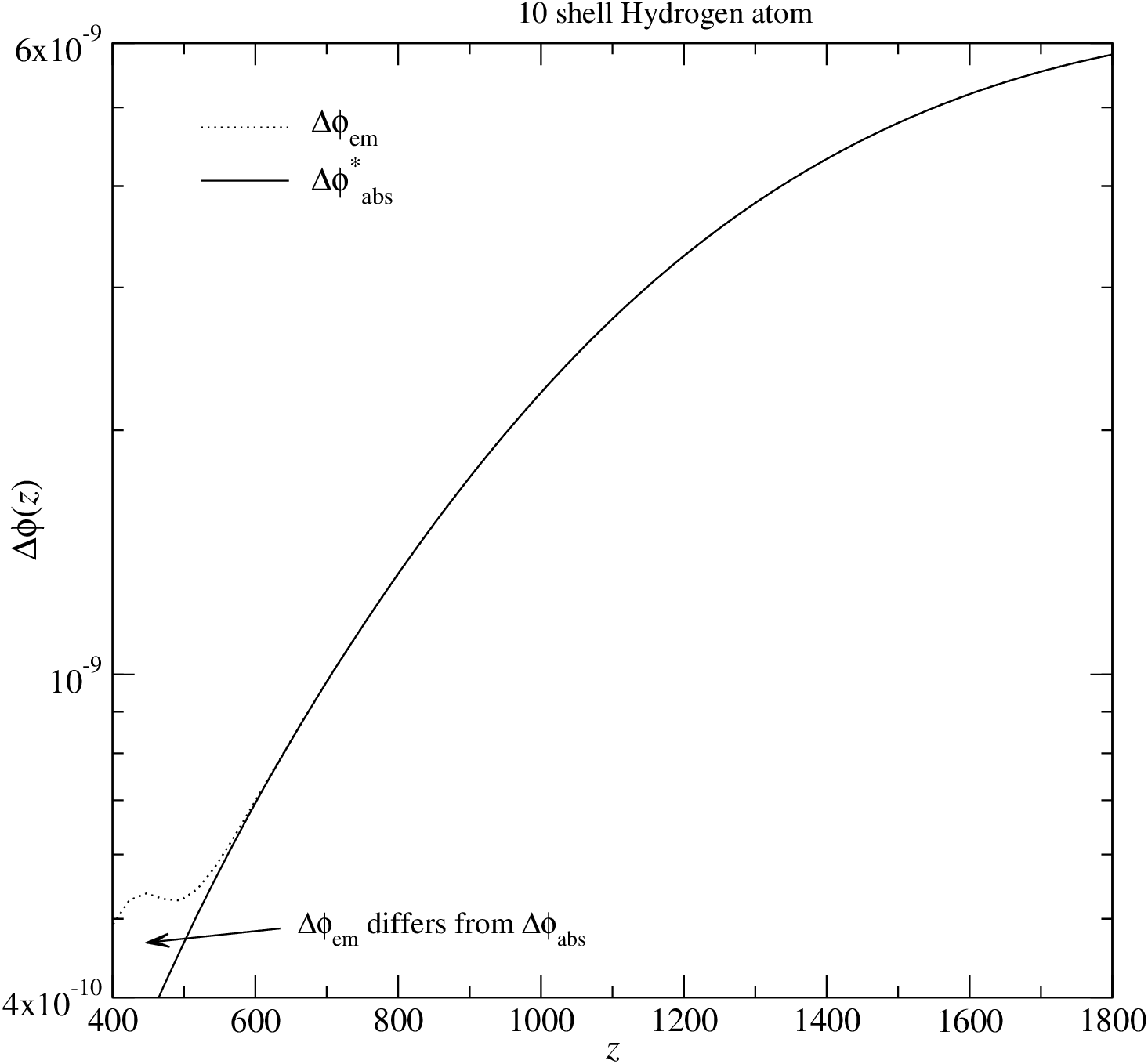}
\caption
{Deviation of the normalization of the different line-profiles from unity. The effect of stimulated emission in the ambient CMB blackbody radiation field was included.}
\label{fig:Dphi_int}
\end{figure}

\subsubsection{Correction due to $\Delta P_{\rm ind}$}
\label{sec:corr_DP_ind}
%
Looking at Eq.~\eqref{eq:DP_ind}, it is clear that the contribution $\Delta P_{\rm ind}$ to $P_{\rm eff}$ is purely due to induced effects, since in vacuum one would find $\Delta\bar{\varphi}_{\rm em}\!\equiv\! \Delta\bar{\varphi}^\ast_{\rm abs}\!=\!0$, and hence $\Delta P_{\rm ind}\equiv 0$.
In Fig.~\ref{fig:Dphi_int} we present the deviation in the normalization of the emission and absorption profile from unity, which have been computed using Eq.~\eqref{sec:phi_em} and \eqref{sec:tau_abs_all_b}. 
First one can see that at  basically all redshifts of interest in the recombination problem $\Delta\bar{\varphi}_{\rm em}\approx \Delta\bar{\varphi}^\ast_{\rm abs}$, implying that $\Delta P_{\rm ind}\approx \Delta\bar{\varphi}_{\rm em}$.
Since $\Delta\bar{\varphi}_{\rm em}\lesssim \pot{6}{-9}$ at all shown redshifts, comparing $\Delta P_{\rm ind}$ with $P_{\rm d}$ implies that the associated correction $\Delta P_{\rm S}$ should never exceed the level of $0.1\%$. In a more detailed computation we find a maximal correction of $\Delta P_{\rm S}/P_{\rm S}\sim 0.06\%$ at $z\sim 1300$.
%
In addition, this correction is practically canceled by another contribution, $\Delta P_{\rm ind, norm}=-\Delta \bar{\varphi}^\ast_{\rm abs}$, which appears as a result of stimulated emission on the overall normalization of the effective absorption profile (see Sect.~\ref{sec:corr_P}).
We therefore can neglect this term in the following.


We would like to mention that the main contribution to $\bar{\varphi}_{\rm em}\neq 1$ is coming from the region $\nu\sim \nu_{31}$. This can be seen in Fig.~\ref{fig:phi_abs}, where only there the effective two-photon emission profile differs significantly from the vacuum profile, $\varphi^\ast_{\rm em, vac}$. 
At this large distance from the line center the overall profile already dropped by a factor of $10^{11}-10^{12}$ relative to the line center (the value there is $\sim 1/\sqrt{\pi}\sim0.56$).
From Fig.~\ref{fig:phi_abs} one can see that there $\varphi^\ast_{\rm em}-\varphi^\ast_{\rm em, vac}\sim 10^{-12}$ over a region $\Delta\xD\sim 10^3$, so that one expects $\Delta P_{\rm ind}\sim 10^{-9}$ at $z\sim 1300$, which is in good agreement with the full numerical result.

\subsubsection{Expression for $P_{\rm eff}$ and its connection to $F_\nu$.}
\label{sec:corr_P}
In order to understand the corrections in the effective escape probability and its relation to the differences in the spectral distortion we again look at the definition of $P$, Eq.~\eqref{eq:def_P}, with $f_\nu\equiv 1$ and rewrite it as
\beal
\label{eq:P_rew_I}
P^{\rm t,\phi}&=\int [\varphi_{\rm V}-\varphi^\ast_{\rm abs}]\id\nu
+
\int\varphi^\ast_{\rm abs}\,[1-F^{\rm t, \phi}_\nu]\id\nu.
\end{align}
The first integral is given by
\beal
\label{eq:P_indnorm}
\Delta P_{\rm ind,norm}
&=\int [\varphi_{\rm V}-\varphi^\ast_{\rm abs, vac}]\id\nu
+\int [\varphi^\ast_{\rm abs, vac}-\varphi^\ast_{\rm abs}]\id\nu
\nonumber\\[1mm]
&=-\Delta \bar{\varphi}^\ast_{\rm abs}.
\end{align}
Note that $\int [\varphi_{\rm V}-\varphi^\ast_{\rm abs, vac}]\id\nu\equiv 0$ even though the partial contributions from the red and blue wing are non zero.
As mentioned in Sect.~\ref{sec:corr_DP_ind}, $\Delta P_{\rm ind,norm}$ cancels with the correction due to $\Delta P_{\rm ind}$, so that we finally have 
\beal
\label{eq:P_eff_rew_I_a}
P^{\rm t, \phi}_{\rm eff}&\approx \int\varphi^\ast_{\rm abs}\,[1-F^{\rm t, \phi}_\nu]\id\nu
\end{align}
This expression now allows to compute the correction to the escape probability.

In order to find out how the shape of the profile enters into the problem, it is illustrative to look at the result for $P^{\rm t,\phi}_{\rm eff}$ when assuming quasi-stationary conditions, but including the correction due to the profile.
In this case one has $\tau^{\phi,\rm qs}_{\rm abs}=\tau_{\rm d}(z)\int^{\infty}_\nu \varphi^\ast_{\rm abs} \id\nu'$ and with $F^{\phi,\rm qs}_\nu(\nu)\equiv 1-e^{-\tau^{\phi,\rm qs}_{\rm abs}}$ resulting from Eq.~\eqref{eq:F0} one finds
\beal
\label{eq:P_eff_phi_qs}
P^{\phi,\rm qs}_{\rm eff}
&= \int\varphi^\ast_{\rm abs}\,e^{-\tau^{\phi,\rm qs }_{\rm abs}} \id\nu 
= \int_0^{\chi^{\ast}_{\rm a, \infty}} e^{-\tau_{\rm d}[\chi^{\ast}_{\rm a, \infty}-\chi^{\ast}_{\rm a}]} 
\id\chi^\ast_{\rm a} 
\approx P_{\rm d}.
\end{align}
Here we have introduced the variable $\chi^{\ast}_{\rm a}(\nu)=\int^{\nu}_0 \varphi^\ast_{\rm abs} \id\nu'$ and $\chi^{\ast}_{\rm a, \infty}=\chi^{\ast}_{\rm a}(\infty)$.
This expression shows that in the quasi-stationary case the shape of the profile does not matter. The result will still be extremely close to\footnote{There is a tiny difference due to the fact that $\chi^{\ast}_{\rm a, \infty}\gtrsim \chi_{\infty}$.} $P_{\rm d}$.
This also implies that the shape of the profile is only entering as correction to correction, i.e. combination of time-dependence and profile. Therefore the changes due to the exact shape of the profile are expected to be smaller than the corrections due to $f_\nu$ and the time-dependence alone.
As we will see below they still are not negligible, in particular when taking all corrections into account simultaneously (Sect.~\ref{sec:total_corr}).

For the correction to the effective escape probability we can therefore finally write
\beal
\label{eq:DP_eff_phi}
\Delta P^{\rm t, \phi}_{\rm d}&=P^{\rm t, \phi}_{\rm eff}-P_{\rm d}
\approx -\int\varphi^\ast_{\rm abs}\,[F^{\rm t,\phi}_\nu-F^{\phi,\rm qs}_\nu]\id\nu.
\end{align}
Now it is clear that for the contributions of the total correction it is important how $F^{\rm t, \phi}_\nu$ deviates from $F^{\phi, \rm qs}_\nu$. For $F^{\rm t, \phi}_\nu<F^{\phi,\rm qs}_\nu$ one will have a {\it positive} contribution and for $F^{\rm t, \phi}_\nu>F^{\phi,\rm qs}_\nu$ a {\it negative}.
In addition due to the appearance of $\phi^\ast_{\rm abs}$ in the outer integral, the red wing contribution ($\phi^\ast_{\rm abs}\gtrsim\phi_{\rm V}$) will be slightly {\it overweighted}, while the blue wing ($\phi^\ast_{\rm abs}\lesssim\phi_{\rm V}$) will be {\it underweighted}. 
%
%

We would like to mention that $F^{\phi,\rm qs}_\nu(\nu)$ behaves very similar to $F^{\rm qs}_\nu(\nu)$. In particular at frequencies $\xD\lesssim 4$ it \changeII{also} becomes very close to unity, so that there $F^{\phi,\rm qs}_\nu(\nu)\approx F^{\rm qs}_\nu(\nu)$. The main difference appears in the blue wing of the line, where $F^{\phi,\rm qs}_\nu(\nu)$ depends strongly on the differences between $\phi_{\rm abs}^\ast$ and $\phi_{\rm V}$. Both aspects can be seen in the right panel of Fig.~\ref{fig:Ftf.z}. 
%

\begin{figure}
\centering 
\includegraphics[width=0.95\columnwidth]{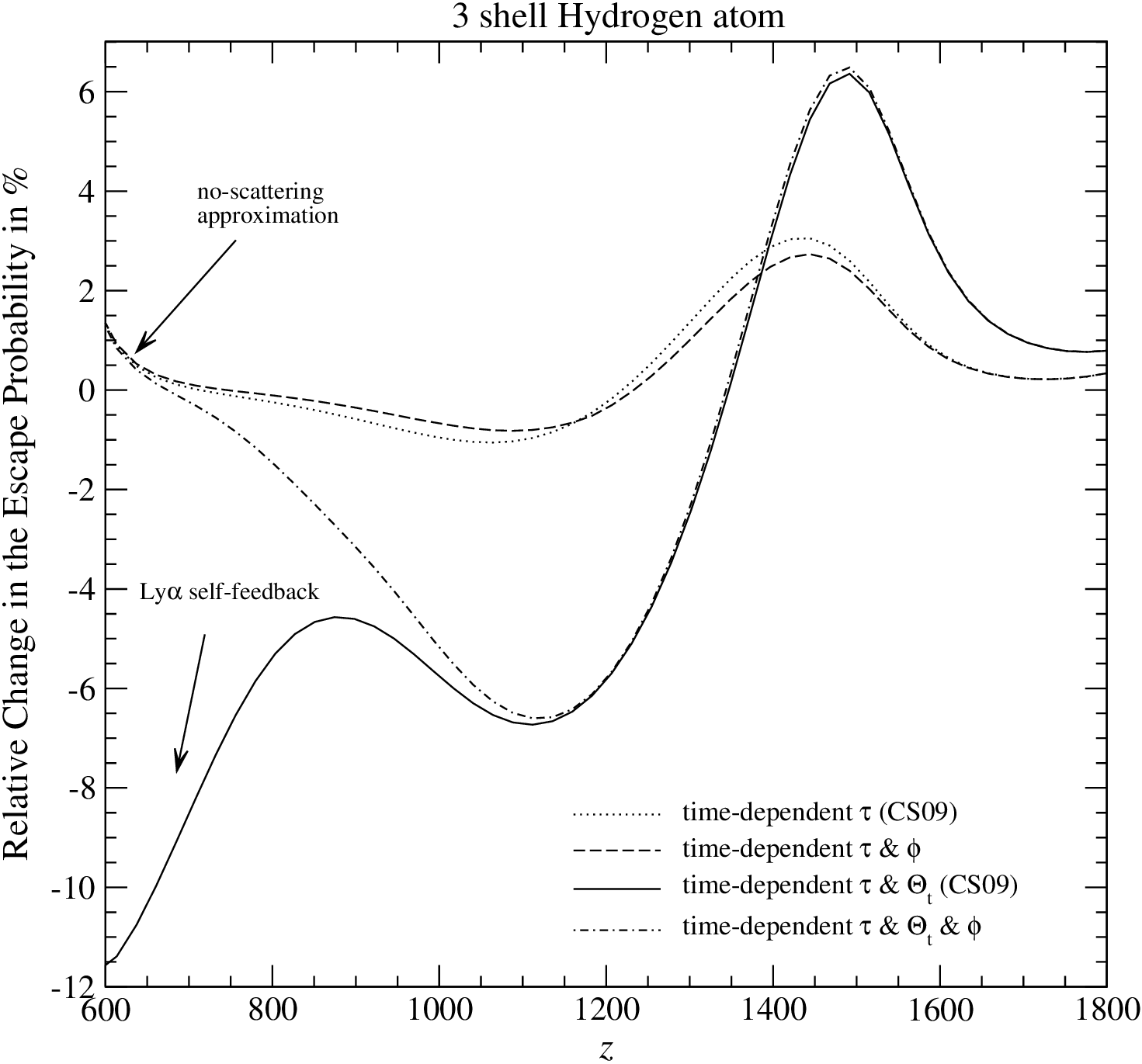}
\caption
{Relative difference in the effective escape probability with respect to the Sobolev escape probability: effect of the line profile. For the curves labeled with 'CS09' \citep{Chluba2008b} we used the standard Voigt profile, while for the others we included the two-photon corrections for the 3s and 3d channels.}
\label{fig:DP_P.phi.comp}
\end{figure}
\subsubsection{Total corrections and the blue and red wing contributions}
\label{sec:P_phi_b_r}
In Fig.~\ref{fig:DP_P.phi.comp} we show the results for the total correction to the effective escape probability when including the modifications in the shape of the absorption profile. 
We used the expression Eq.~\eqref{eq:P_eff_rew_I_a} to compute the different curves.
Again for comparison we also give the results for the time-dependent corrections only. 
As one can see the main effect of the profile is the removal of the self feedback at low redshift. 
%
Other than that in the considered cases the modifications in comparison with the time-dependent result are rather small (less than $\sim10\%$ of $\Delta P/P$ for the cases with $\Theta^{\rm t}=1$ and less than $\sim1\%$ of $\Delta P/P$ for those with $\Theta^{\rm t}\neq 1$). 
%

\begin{figure}
\centering 
\includegraphics[width=0.95\columnwidth]{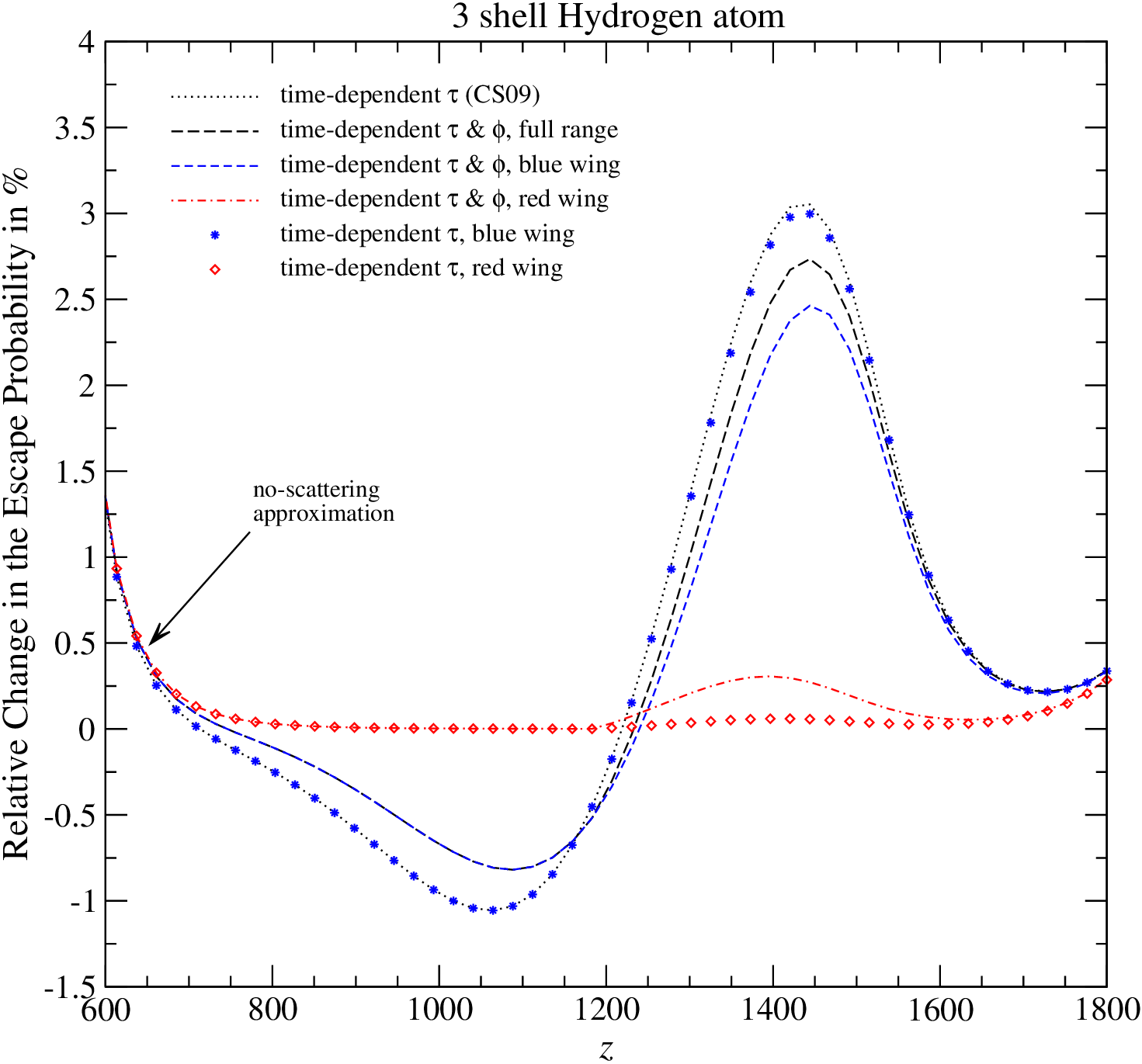}
\\[1mm]
\includegraphics[width=0.95\columnwidth]{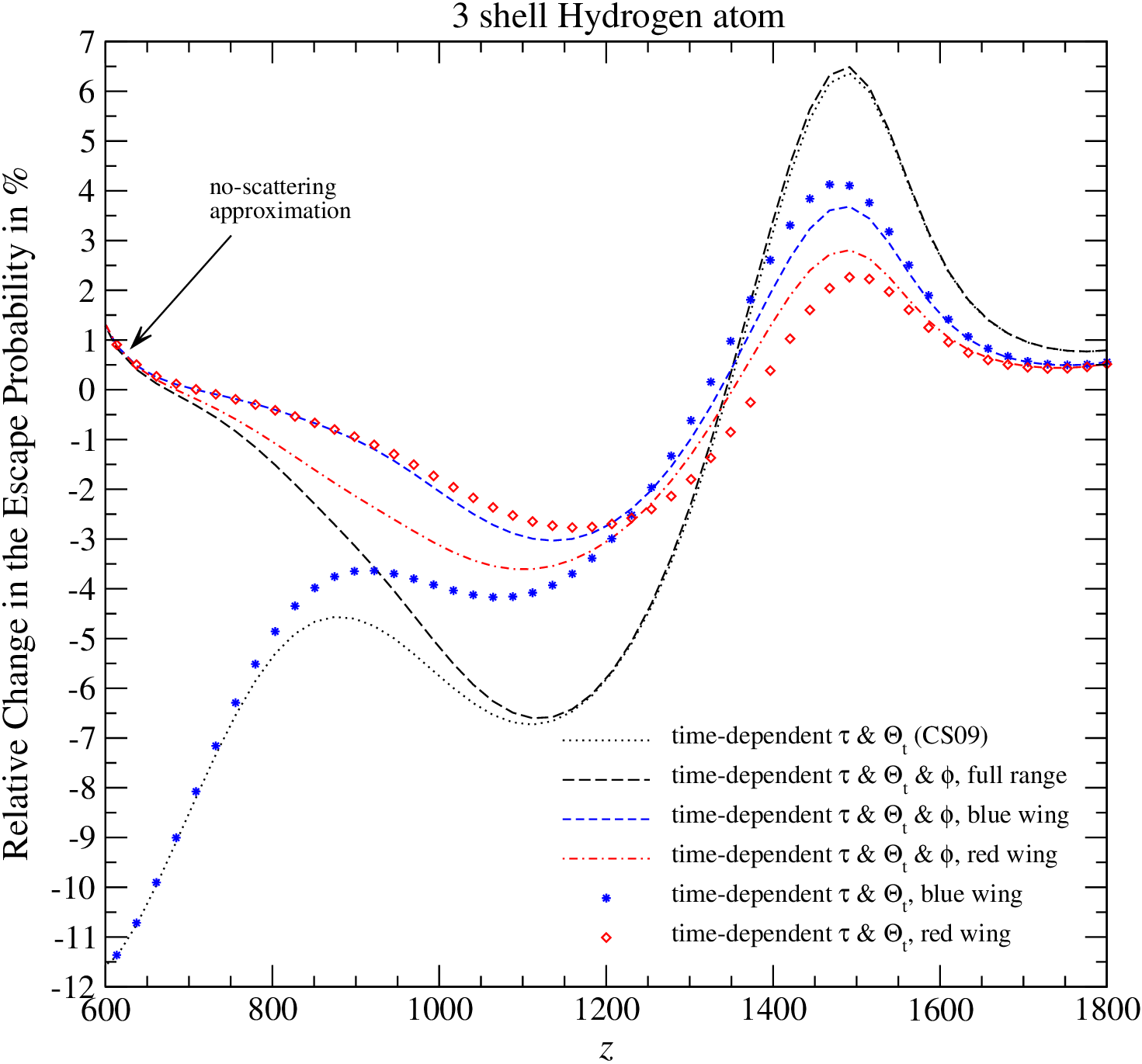}
\caption
{Relative difference in the effective escape probability with respect to the Sobolev escape probability: correction due to the shape of the absorption profile in the blue and red wing of the resonance. For the upper panel we set $\Theta^{\rm t}=1$ and $\Theta^{\phi}=0$. The lower panel was computed also including $\Theta^{\rm t}\neq 1$. In all cases we used $f_\nu=1$. Those curves labeled with $\phi$ were computed including the 3s and 3d two-photon corrections.}
\label{fig:DP_P.phi.comp.rb}
\end{figure}

This shows that there is a cancelation of corrections from the red and the blue wing, since it is clear that already the modulation of these partial contributions due to the presence of $\phi_{\rm abs}^\ast$ in the outer integral of Eq.~\eqref{eq:P_eff_rew_I_a} should have some effect, even if it is of higher order in the correction.
To understand the results we therefore look at the differential contribution from the red and blue wing separately. For Fig.~\ref{fig:DP_P.phi.comp.rb} we ran computations including the corrections to the quasi-stationary result either on the red or blue side of the resonance. We compare the total and partial corrections in both the pure time-dependent case and when also including the shape of the absorption profile. 

\subsubsection*{Corrections in the case of $\Theta^{\rm t}=1$}
When only including the profile corrections to $\tau_{\rm abs}$ (upper panel, $\Theta^{\rm t}=1$ and $\Theta^{\phi}=0$), we can see that the effect of $\phi_{\rm abs}^\ast$ is not leading to any important correction from the red wing. 
Looking at Fig.~\ref{fig:DFf.z} and \ref{fig:Ff.z} (left column), it is clear that in there $F^{\rm t, \phi}_\nu\approx F^{\rm qs}_\nu~\approx~1$. 
Since also $F^{\rm qs}_\nu\approx F^{\phi, \rm qs}_\nu$ (see Sect.~\ref{sec:corr_P}) one has $F^{\rm t, \phi}_\nu-F^{\phi, \rm qs}_\nu\approx 0$ and hence with Eq.~\eqref{eq:DP_eff_phi} one expects a partial contribution of $\Delta P^{\rm t, \phi}_{\rm d} \approx 0$, confirming the above behavior.
The small positive bump seen at $z\sim 1400$ is mainly due to the fact that we started our computation of the spectral distortion at some particular time in the past ($z_{\rm s}=2000$), so that in the considered case the numerical solution for the spectral distortion, as computed using Eq.~\eqref{app:kin_abs_em_Sol_phys_asym_F}, drops towards zero below some distance $\xD\lesssim -10^4$ \changeII{instead of staying close to unity}.
Therefore we have \changeII{$F^{\rm t, \phi}_\nu-F^{\phi, \rm qs}_\nu\approx -1$ and hence}
$\Delta P^{\phi}_{\rm d}(\xD<-10^4)\approx \int \phi_{\rm abs}^\ast \id\xD \sim \text{few}\times 10^{-3}$ from that region, explaining this small excess with respect to $P_{\rm d}$. 
We also checked this statement numerically by increasing $z_{\rm s}$ as expected finding that the bump became smaller.
%
%
When also including the time-dependence of the emission coefficient ($\Theta^{\rm t}\neq 1$) this small inconsistency of our computation is no longer important, since the spectral distortion  by itself drops very fast  toward zero (cf. Fig.~\ref{fig:DFf.z} and \ref{fig:Ff.z}, right column).

Again looking at the upper panel in Fig.~\ref{fig:DP_P.phi.comp.rb} we can also see that the largest contribution to the total correction is coming from the blue wing, and that the difference to the time-dependent case is rather small, with $\Delta P^{\rm t, \phi}_{\rm d}$ being slightly smaller.
This can be understood when writing $\tau^{\rm t,\phi}_{\rm abs}=\tau^{\phi, \rm qs}_{\rm abs}+\Delta\tau^{\rm t, \phi}$ with 
\beal
\label{eq:Dtau_phi_t}
\Delta\tau^{\rm t,\phi}
&\approx\!\! \int_\nu^\infty \Delta\tau_{\rm d}(z')\,\varphi^\ast_{\rm abs}(\nu', z') \id \nu'
\nonumber
\\
&=
\!\!\int_\nu^\infty \!\!\!\Delta\tau_{\rm d}(z')\,\varphi_{\rm V}(\nu', z) \id \nu' 
+ \!\!\int_\nu^\infty\!\! \!\Delta\tau_{\rm d}(z') \,\Delta\varphi^\ast_{\rm abs}(\nu', z') \id \nu'  .
\end{align}
with  the abbreviations $1+z'=\frac{\nu'}{\nu}(1+z)$, $\Delta\tau_{\rm d}(z')=\tau_{\rm d}(z')-\tau_{\rm d}(z)$, and $\Delta\varphi^\ast_{\rm abs}(\nu', z')=\varphi^\ast_{\rm abs}(\nu', z')-\varphi_{\rm V}(\nu', z)$.
This approximation shows that one has $\Delta\tau^{\rm t,\phi}\approx \int_\nu^\infty \!\!\!\Delta\tau_{\rm d}(z')\,\varphi_{\rm V}(\nu', z) \id \nu' \equiv \Delta\tau^{\rm t} $, since the profile correction with respect to $\tau^{\phi,\rm qs}_{\rm abs}$ is already of higher order.
For $\Theta^{\rm a}=1$ one therefore expects 
\beal
\label{eq:DP_eff_phi_appr}
\Delta P^{\rm t,\phi}_{\rm d}
&\approx \int\varphi^\ast_{\rm abs}\,e^{-\tau^{\phi,\rm qs}_{\rm abs}}[e^{-\Delta\tau^{\rm t,\phi}}-1]\id\nu
\nonumber
\\
&\approx \int\varphi_{\rm V}\,e^{-\tau^{\rm qs}_{\rm d}}[e^{-\Delta\tau^{\rm t}}-1]\id\nu 
=\Delta P^{\rm t}_{\rm d},
\end{align}
confirming the result seen in Fig.~\ref{fig:DP_P.phi.comp.rb} in lowest order.

The remaining difference is mainly due to the second order term in Eq.~\eqref{eq:Dtau_phi_t} $\Delta\tau^{\rm t,\phi}-\Delta\tau^{\rm t}=\int_\nu^\infty\!\! \!\Delta\tau_{\rm d}(z') \,\Delta\varphi^\ast_{\rm abs}(\nu', z') \id \nu'$, which we neglected in Eq.~\eqref{eq:DP_eff_phi_appr}.
The factor $\varphi^\ast_{\rm abs}\,e^{-\tau^{\phi,\rm qs}_{\rm abs}}$ plays a minor role, since the function $e^{-\Delta\tau^{\rm t,\phi}}-1$ is varying much faster. We confirmed these statements numerically, finding that for the considered case the modulation of the blue wing correction due to $\phi^\ast_{\rm abs}$ in the outer integral can be neglected.

\subsubsection*{Corrections in the case of $\Theta^{\rm t}\neq 1$}
To understand the result in the case when also including the change in the effective emission coefficient $\Theta^{\rm t}\neq 1$, we again look at the red and blue wing contribution separately. Since in the red wing $F^{\phi, \rm qs}_\nu\approx 1$ and because we already saw in Sect.~\ref{sec:F_p0} that there the solution for the spectral distortion is dominated by the correction due to $\Theta^{\rm t}\neq 1$ only (cf. Fig.~\ref{fig:DFf.z} and \ref{fig:Ff.z} right column), we expect that the partial contribution form the red wing will be very close to
\beal
\label{eq:DP_eff_phi_Theta}
\Delta P^\phi_{\rm rw}&
\approx \int \phi^\ast_{\rm abs}\,[1-F^{\rm t}_\nu]\id\nu,
\end{align}
where $F^{\rm t}_\nu$ is the spectral distortion for the purely time-dependent case.
Therefore to lowest order again one will have $\Delta P^\phi_{\rm rw}\approx \Delta P^{\rm t}_{\rm rw}$, but in next order this correction will in addition be slightly larger in amplitude due to the fact that for $\xD\leq 0$ one has $\phi^\ast_{\rm abs}\gtrsim \phi_{\rm V}$. We can see that this indeed is true comparing the diamonds with the dash-dotted curve in the lower panel of Fig.~\ref{fig:DP_P.phi.comp.rb}.

For the blue wing one can argue in a very similar way as above. We know that for $\Theta^{\rm t}=1$ the correction to the escape probability is basically given by the time-dependent correction in the value of $\tau_{\rm abs}$, but only to higher order due to $\phi_{\rm abs}^\ast\neq \phi_{\rm V}$. 
If now including $\Theta_{\rm t}\neq 1$ the lowest order correction will still be given by the purely time-dependent case. The additional modulation of the resulting spectral distortion by $\phi^\ast_{\rm abs}\lesssim \phi_{\rm V}$ will in addition lead to a small decrease in the total amplitude of the contribution to the correction. 
Again this can be seen in Fig.~\ref{fig:DP_P.phi.comp.rb} when comparing the stars with the short-dashed curve.
Only at low redshifts the shape of the profile determines the amplitude of the correction, removing the self-feedback problem. This is because unlike in the case of $\phi_{\rm V}$ photons are emitted only in a limited range of frequencies. This avoids that photons which are released at $z\sim 1400$ and $\xD\gg 10^4$ will redshift into the Lyman $\alpha$ line at $z\lesssim 1000$, as seen in the normal '$1+1$' formulation of the problem \citep{Chluba2008b}.

Furthermore, it is clear that the sum of both the red and blue wing contribution should again be very close to the purely time-dependent case, since the modulation of the contributions form the red (enhancement) and blue wing (suppression) in lowest order will cancel, due to the symmetry  around the line center. 
%

\begin{figure}
\centering 
\includegraphics[width=0.95\columnwidth]{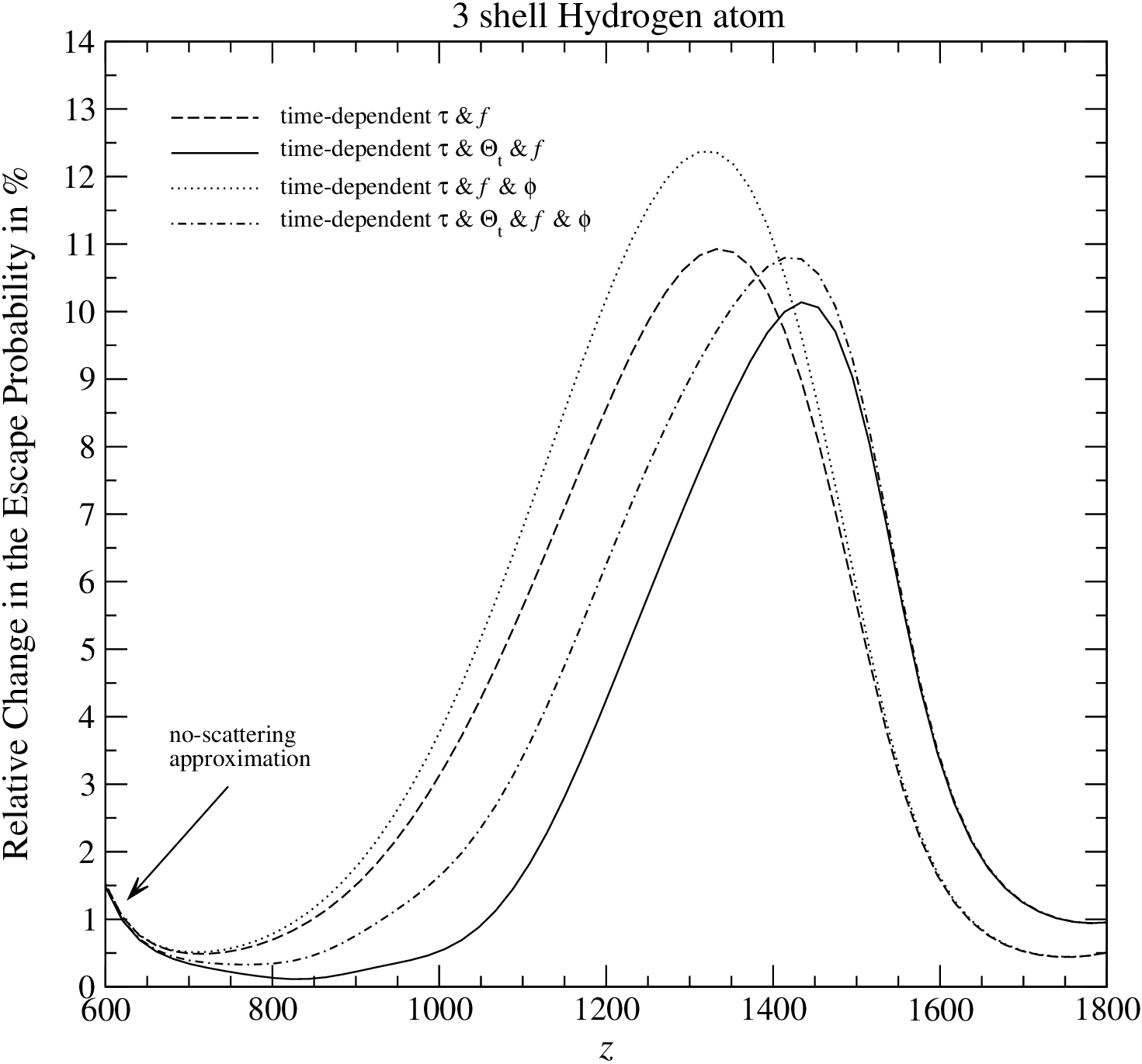}
\caption
{Relative difference in the effective escape probability with respect to the Sobolev escape probability: combined effect of the thermodynamic correction factor and the shape of the line profile.}
\label{fig:DP_P.phi.f.comp}
\end{figure}

\subsection{Combined effect of $f_\nu$ and $\phi^\ast_{\rm abs}\neq \phi_{\rm V}$}
\label{sec:total_corr}
With the derivations in the previous Sections it is now straightforward to understand the results for the combined effect of all corrections.
Following the same line of thoughts we obtain
\beal
\label{eq:DP_eff_all}
\Delta P^{\rm a}_{\rm d}
&=-\int\phi^\ast_{\rm abs}\,[f_\nu F^{\rm a}_\nu-F^{\phi,\rm qs}_\nu]\id\nu.
\end{align}
where in $F^{\rm a}_\nu$ we include all the corrections simultaneously.

As in the previous Section it is now clear that on the red side of the resonance the profile will enter the computation mainly due to its presence in the outer integral. For $F^{\rm a}_\nu$ and $\xD\leq 0$ it only leads to  a very small correction (cf. Fig.~\ref{fig:DFf.z} and \ref{fig:Ff.z} right column).
However, on the blue side of the resonance the profile correction again can be neglected in the outer integral, but should be taken into account when computing the difference $\Delta_\nu=f_\nu F^{\rm a}_\nu-F^{\phi,\rm qs}_\nu$.
Also one can conclude that the shape of the profile plays the key role in removing the low redshift self-feedback problem. The latter statement can be confirmed when looking at the shape of the Lyman $\alpha$ distortion at intermediate to high frequencies blue-ward of the resonance (cf. Fig.~\ref{fig:DFf.z} and \ref{fig:Ff.z} right column), which is clearly dominated by the profile rather than $f_\nu$.

Therefore in lowest order one expects the total correction to be the superposition of the time-dependent correction and the one from the thermodynamic correction factor, where on the red side of the resonance each of them is modulated by the shape of the profile in the outer integral of Eq.~\eqref{eq:DP_eff_all} in addition, while on the blue side the contribution is slightly suppressed due the profile corrections to $\Delta_\nu$.
Here it is important that because $f_\nu$ strongly changes the symmetry of the problem (the main correction is coming from the red wing as shown in Sect.~\ref{sec:normal}), it is clear that the main effect of $\phi_{\rm abs}^\ast\neq \phi_{\rm V}$ will be an enhancement of the final correction.

In Fig.~\ref{fig:DP_P.phi.f.comp} we present the results from our numerical calculation for different cases. Indeed we find that when including the shape of the profile the corresponding correction is  slightly increased as explained above.

\begin{figure}
\centering 
\includegraphics[width=0.95\columnwidth]{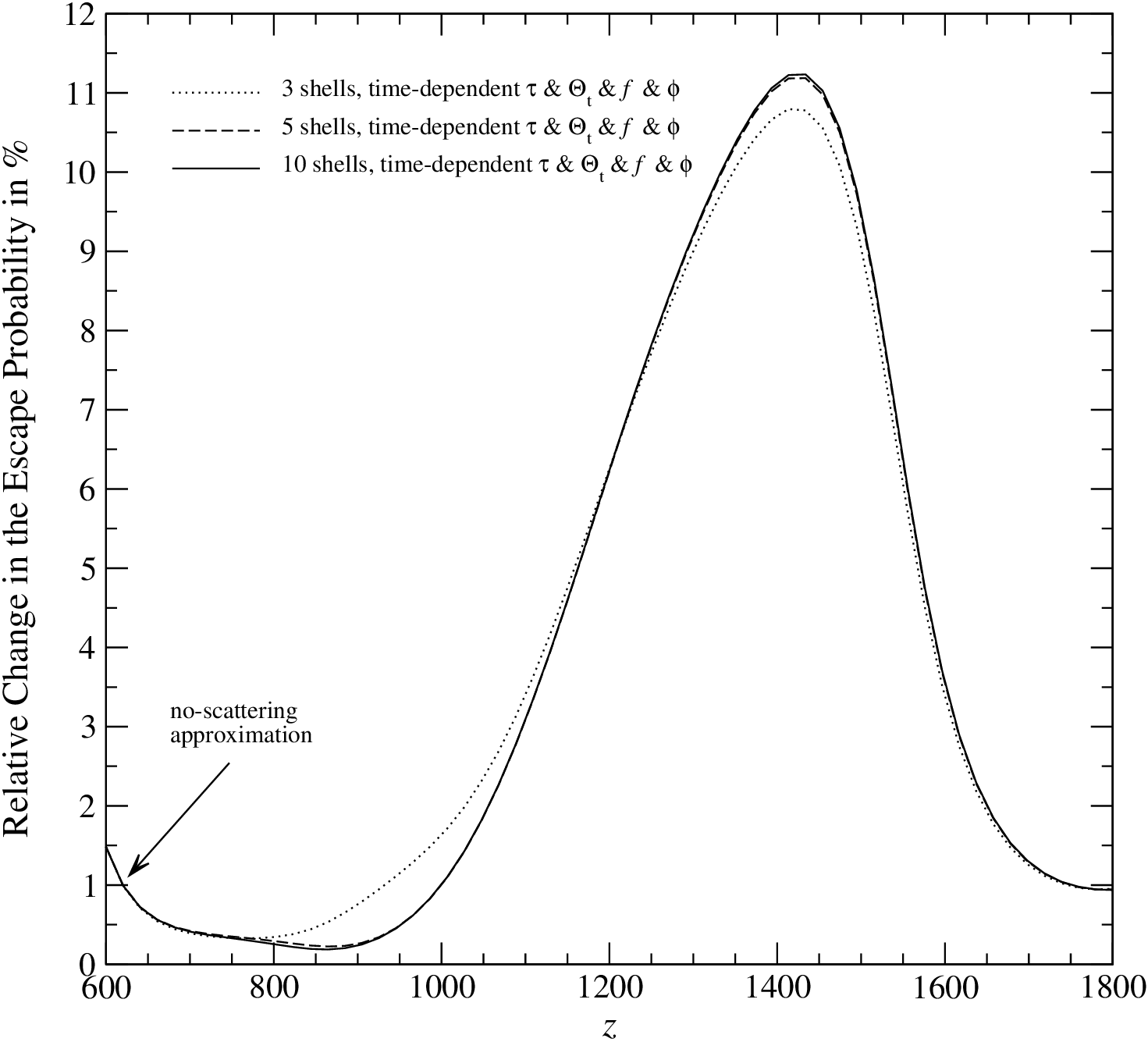}
\caption
{Total relative difference in the effective escape probability with respect to the Sobolev escape probability: dependence on the total number of shells. Note that the curves for the 5 shell and 10 shell cases practically coincide.}
\label{fig:DP_P.all.nS.comp}
\end{figure}
\subsubsection{Dependence on the included number of shells}
\label{sec:total_corr_nS}
For the purely time-dependent correction it has been shown that in particular at low \changeII{redshifts} the result depends strongly on the total number of shells that were included in the computation. 
Since there the correction was very strongly dominated by the self-feedback of Lyman $\alpha$ photons, here we do expect this dependence on the number of shells to be more mild.

In Fig.~\ref{fig:DP_P.all.nS.comp} we show the results of our computations for 3, 5 and 10 shells. The changes between the 3 and 5 shell cases is still rather significant, but the difference between the 5 and 10 shell case is already very minor. This show that in our description the total correction is already converged when including $\sim 5$ shells into the computation. 
%

\begin{figure}
\centering 
\includegraphics[width=0.95\columnwidth]{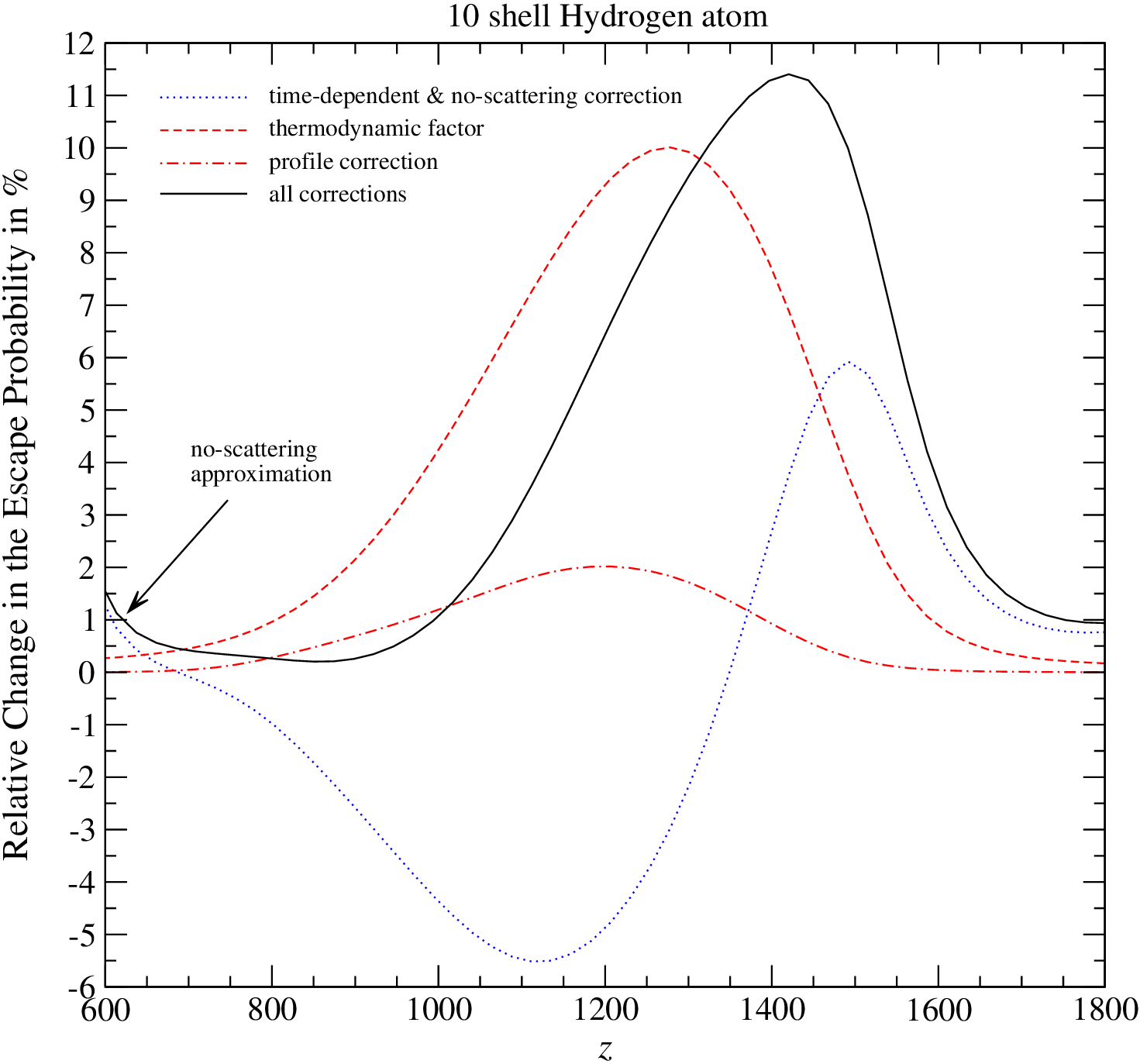}
\caption
{Relative difference in the effective escape probability with respect to the Sobolev escape probability: separate contributions due to the time-dependent correction, the thermodynamic factor and the shape of the profile.}
\label{fig:DP_P.final}
\end{figure}

\begin{figure}
\centering 
\includegraphics[width=0.95\columnwidth]{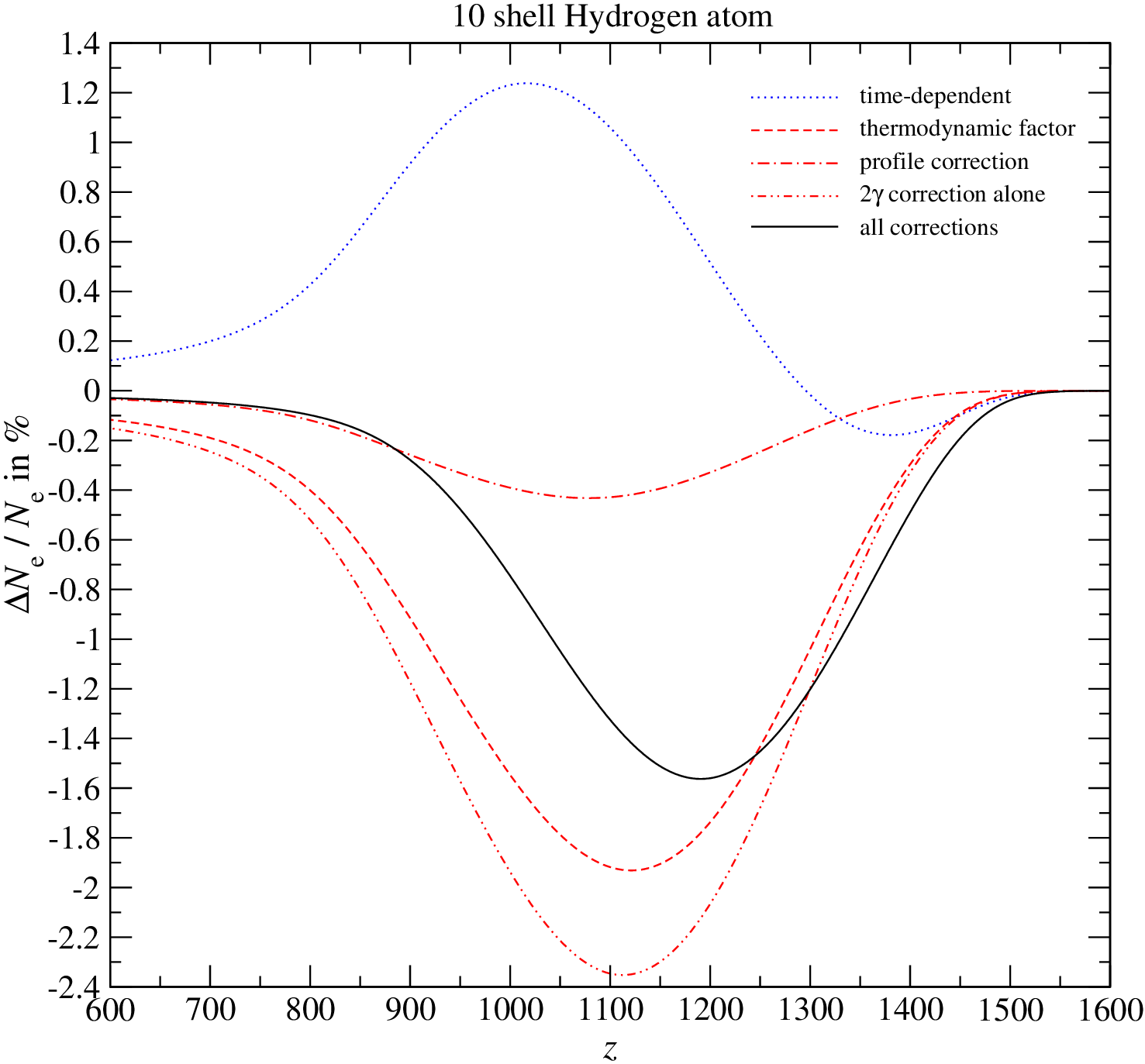}
\caption
{Changes in the free electron fraction: separate contributions due to the time-dependent correction, the thermodynamic factor and the shape of the profile.}
\label{fig:DN_N.final}
\end{figure}

\section{Effect on the ionization history and the CMB power spectra}
\label{sec:Ne}
In this Section we now give the expected correction to the ionization history when including the processes discussed in this paper.
For this we modified the {\sc Recfast} code \citep{SeagerRecfast1999}, so that we can load the pre-computed change in the Sobolev escape probability studied here.

In Fig.~\ref{fig:DP_P.final} we present the final curves for $\Delta P/P$ as obtained for the different processes discussed in this paper.
In Fig.~\ref{fig:DN_N.final} we show the corresponding correction in the free electron fraction computed with the modified version of {\sc Recfast}.
One can clearly see that the dominant correction is due to the thermodynamic factor, resulting in $\Delta P/P\sim+10\%$ at $z\sim 1280$ and $\Delta N_{\rm e}/N_{\rm e}\sim-1.9\%$ at $z\sim 1120$. 
The next largest correction is due to the time-dependent aspects of the problem, leading to $\Delta P/P\sim-5.6\%$ at $z\sim 1120$ and $\Delta P/P\sim+5.9\%$ at $z\sim 1490$. The associated correction in the free electron fraction has a maximum of $\Delta N_{\rm e}/N_{\rm e}\sim+1.2\%$ at $z\sim 1020$.
The smallest correction is due to the exact shape of the effective line profile, resulting in $\Delta P/P\sim+2.0\%$ at $z\sim 1200$ and $\Delta N_{\rm e}/N_{\rm e}\sim-0.4\%$ at $z\sim 1080$.
The total correction then is $\Delta P/P\sim+11\%$ at $z\sim 1420$ and $\Delta N_{\rm e}/N_{\rm e}\sim-1.6\%$ at $z\sim 1190$. 
This is an important speed up of hydrogen recombination, although at $z\sim 1100$ a large part of the correction due to $f_\nu$ alone is canceled by the time-dependent correction.
At the maximum of the Thomson visibility function $z\sim 1100$ we find $\Delta N_{\rm e}/N_{\rm e}\sim -1.3\%$, where about $\Delta N_{\rm e}/N_{\rm e}\sim -0.4\%$ is coming from the shape of the profile alone.  

For completeness we also show the correction that is due to the two-photon formulation alone, i.e. where we subtracted the time-dependent contribution from the total correction.
It leads to $\Delta N_{\rm e}/N_{\rm e}\sim-2.4\%$ at $z\sim 1110$. 

\begin{figure}
\centering 
\includegraphics[width=0.95\columnwidth]{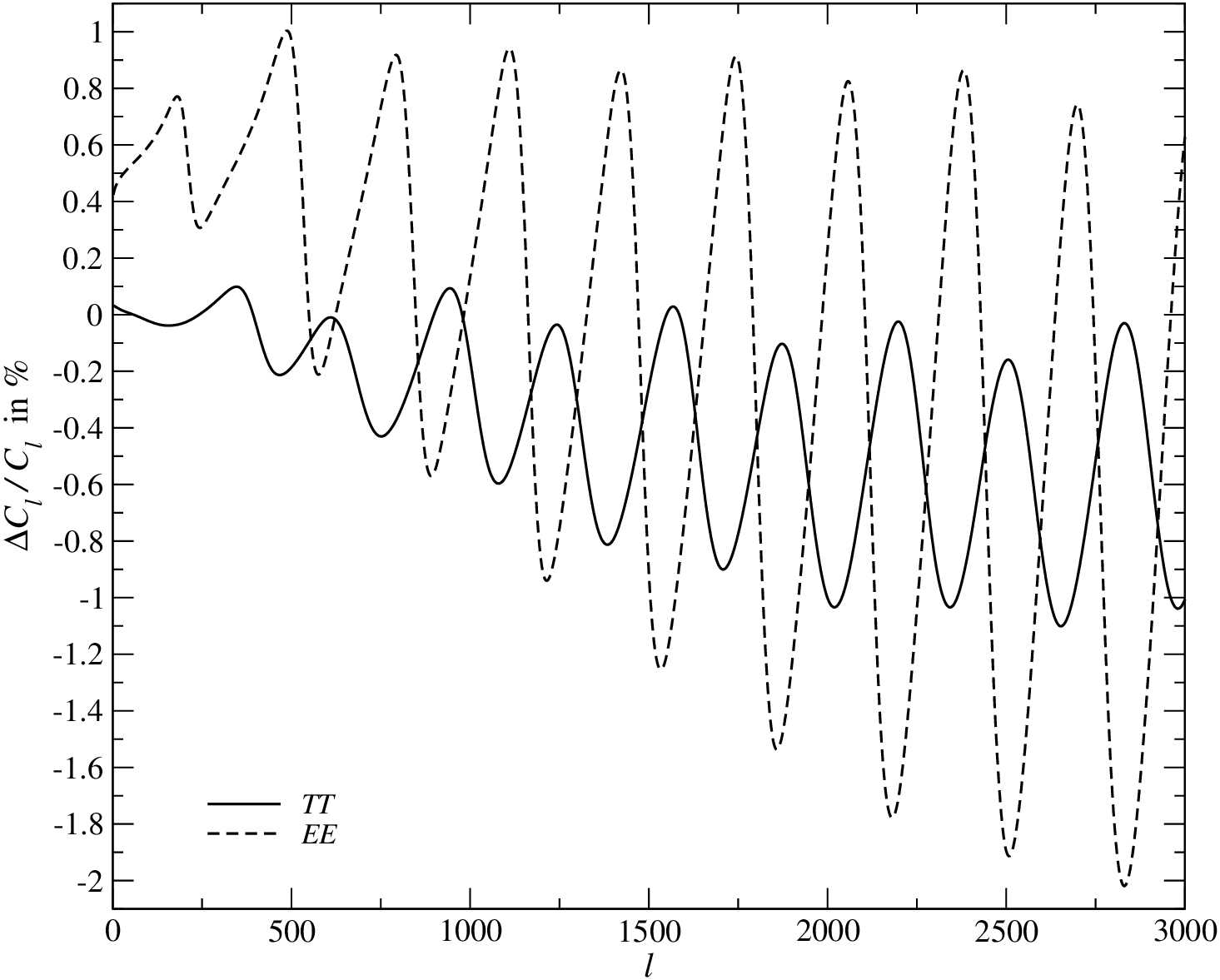}
\caption
{Changes in the CMB temperature and polarization power spectra. We included effect of the time-dependent correction, the thermodynamic correction factor and the profile correction, simultaneously. We used the result for the 10 shell hydrogen atom.}
\label{fig:DCl}
\end{figure}
In Fig.~\ref{fig:DCl} we finally show the changes in the CMB temperature and polarization power spectra coming from the total correction $\Delta N_{\rm e}/N_{\rm e}$ as given by the solid line in Fig.~\ref{fig:DN_N.final}.
In particular the changes in the EE power spectrum are impressive, with peak to peak amplitude $\sim 2\%-3\%$ at $l\gtrsim 1500$.
One can also see that the modifications in the $C_l$'s \changeII{correspond more to} a shift in the positions of the peaks rather than changes in the absolute amplitude.
This is connected to the fact that the correction in the free electron fraction leads to a small shift in the maximum of the Thomson visibility, but does not affect the Silk damping length \citep{Silk1968} as much.
It will be very important to take these changes into account in the analysis of future CMB data.

\section{Discussion and conclusions}
\label{sec:conc}

In this paper we gave a formulation of the Lyman $\alpha$ transfer equation which allows us to include the two-photon corrections for the 3s-1s and 3d-1s channels.
We then solved this transfer equation and presented the results for the Lyman $\alpha$ distortion at different redshifts (Sect.~\ref{sec:F}). From this we computed the effective Lyman $\alpha$ escape probability and derived the corresponding corrections to the Sobolev escape probability (Sect.~\ref{sec:Pesc}). 
We used these results to compute the corrections in the free electron fraction during hydrogen recombination and the associated changes in the CMB power spectra (Sect.~\ref{sec:Ne}).
Below we now shortly summarize the results of each of these Sections and also give a short discussion for future work and improvements.

\subsection{The resulting Lyman $\alpha$ spectral distortion}
In Sect.~\ref{sec:F} we discussed the influence of the different processes on the Lyman $\alpha$ distortion at different redshifts.
Including all the corrections considered here simultaneously one can conclude that at $\xD\lesssim 0$ the shape of the distortion is mainly determined by the time-dependence of the effective emission rate (cf. Fig.~\ref{fig:DFf.z} and \ref{fig:Ff.z}, right column). 
In the close vicinity of the resonance {\it all} sources of corrections under discussion here are important for the exact shape of the Lyman $\alpha$ spectral distortion at difference redshifts (cf. Fig.~\ref{fig:DFf.core.1100.z}).
In particular inside the Doppler core the spectral distortion will be very close the value at the line center multiplied by $1/f_\nu$.
On the blue wing the distortion is mainly determined by the shape of the line profile.

With the method given here we can in principle also compute the Lyman $\alpha$ distortion, as it would be observable today.
Since all the processes discussed here affect the exact shape of this distortion, one could in principle probe our understanding of the recombination dynamics by observing it.
As mentioned earlier \citep{Chluba2008b}, similarly one expects additional changes in the detailed shape of the Lyman $\alpha$ distortion due to partial frequency redistribution and electron scattering. All these processes therefore not only affect the dynamics of cosmological hydrogen recombination, but in principle should leave observable traces in cosmological recombination spectrum \citep[e.g. see][and references therein]{Sunyaev2007} \changeII{until} today. 
Measuring the exact shape of the Lyman $\alpha$ distortion and the other recombinational lines would in principle allow us to reveal these differences, and therefore directly probe our understanding of the recombination problem. 
Also if something non-standard happens (e.g. due to early energy release by decaying or annihilating dark matter), this will affect the exact shape of the cosmological recombination radiation \citep{Chluba2008c}.
Therefore, by observing the recombinational radiation one in principle can directly uncover potential unknowns in the cosmological recombination problem, a task that may not be completed otherwise.

\subsection{Corrections to the escape probability}
We have discussed the corrections to the effective Lyman $\alpha$ escape probability, showing that the largest contribution is coming from the thermodynamic factor $f_\nu$ (see Fig.~\ref{fig:DP_P.final}). 
The next largest correction is due to the time-dependent aspects of the recombination problem, where in the formulation given here the self-feedback problem \citep{Chluba2008b} appearing at low redshifts when using the '$1+1$' photon picture disappears (cf. Fig.~\ref{fig:DP_P.final}).
Furthermore the time-dependent correction partially cancels the correction for the thermodynamic factor at low redshift ($z\lesssim 1300-1400$), leaving a significanly smaller change in the escape probability at $z\sim 1100$.
As we explained here, these two corrections can be also obtained in the normal '$1+1$' photon picture, but for the thermodynamic correction factor a natural interpretation can only be given within the two-photon picture (see Sect.~\ref{sec:thermodyn}).
Here the crucial ingredient is that the spectrum in the vicinity of the second, low frequency photon $\gamma'$ is given by the CMB blackbody photon field, which then allows us to write the radiative transfer equation for the Lyman $\alpha$ photon as equation for one photon only.
A very similar formulation should be applicable in the case of expanding envelopes of planetary nebulae and stars, where the photon field in the vicinity of $\gamma'$ will be given by a weakly diluted blackbody spectrum.
However, when the photon distribution in the vicinity of {\it both} photons involved in the absorption process are far from their equilibrium values the derived formulation does not work.

We also showed that the correction coming from the exact shape of the line profile in the vicinity of the Lyman $\alpha$ resonance leads to the \changeII{smallest} separate correction under discussion here.
Only this part of the correction is really related to quantum mechanical modification of the transfer problem. Still the final contribution \changeII{related to this modification} is significant \changeII{at the required level of accuracy} (cf. Fig.~\ref{fig:DP_P.final}).

\subsection{Implications for the ionization history and the CMB power spectra and critical remarks}
The results for the changes in the free electron fraction and CMB power spectra are shown in Fig.~\ref{fig:DN_N.final} and \ref{fig:DCl}, respectively.
The main effect on $N_{\rm e}$ due to the processes discussed here is a net speed up of hydrogen recombination by $\Delta N_{\rm e}/N_{\rm e}\sim -1.3\%$ at $z\sim 1100$. 
About $\Delta N_{\rm e}/N_{\rm e}\sim -0.4\%$ of this correction is coming from the shape of the line profile alone, while the rest is due to the thermodynamic factor and the time-dependent aspects of the recombination problem.
Here we would like to emphasis again, that the latter two corrections can actually also be obtained in the standard '$1+1$' photon picture, when introducing the thermodynamic correction factor using the detailed balance principle.
We would also like to mention that our results for the changes in the free electron fraction seem to be rather similar to those of \citet{Hirata2008} for the contributions from high level two-photon decays alone.
However, we obtained these with a completely independent method. For the future it will be very important to perform a more detailed comparison once all \changeII{relevant additions} are identified.

Regarding the CMB power spectra, in particular the associated changes in the EE power spectrum are impressive, reaching peak to peak amplitude $\sim 2\%-3\%$ at $l\gtrsim 1500$ (see Fig.~\ref{fig:DCl}).
It will be important to take these corrections into account for the analysis of future CMB data.

However, it is also clear that several steps still have to be taken: (i) one still has to study more processes; and (ii) one has to device a sufficiently fast method to simultaneously incorporate all the corrections discussed in the literature so far into the computations of the CMB power spectra, in order to be ready for real parameter estimations using the CMB data.

Regarding the first point, for example, \changeII{the correction due to Raman processes (as explained in \citet{Hirata2008} mostly due to 2s-1s Raman scattering) leads to an} additional delay of recombination at low redshifts with a maximum of $\Delta N_{\rm e}/N_{\rm e}\sim +0.94\%$ at $z\sim 920$.
We did not include this process here, but it certainly is very important for accurate predictions of the CMB power spectra and should be cross-validated in the future.
It is clear that one should also include the effect of higher level two-photon decays (e.g. from the 4d-level), since they are expected to speed up hydrogen recombination in addition, likely affecting the result obtained here by another $\sim 10\%-20\%$ (i.e. $\Delta N_{\rm e}/N_{\rm e}\sim 0.1-0.3\%$).
And finally, the effects of partial frequency redistribution, line recoil and electron scattering should be studied. Here in particular the effect of line recoil will be important, leading to a systematic drift of photons towards lower frequencies which again accelerates hydrogen recombination by $\Delta N_{\rm e}/N_{\rm e}\sim -1.2\%$ at $z\sim 900$ \citep{Grachev2008, Chlubaprep}.

Regarding the second point, fairly recently \citet{Fendt2008} proposed a new approach called {\sc
  Rico}\footnote{http://cosmos.astro.uiuc.edu/rico/}, which uses
multi-dimensional polynomial regression to accurately represent the dependence
of the free electron fraction on redshift and the cosmological parameters.
Instead of running the full (slow) recombination code, one trains {\sc Rico} with a set of models, so that the interpolation between them will be very fast afterwards.
This approach should allow to propagate all the corrections in the ionization
history that are included in the full recombination code to the CMB power
spectra, without using any fudging like in {\sc Recfast}
\citep{SeagerRecfast1999, Wong2008}.
In the future, we plan to provide an updated training set for {\sc Rico},
including the corrections discussed here. This should also make
it easier for other groups to cross-validate our results \changeII{and} will allow us to focus the effort on the physics rather than on approximating it.

\begin{appendix}

\section{Different channels for the emission and death of Lyman $\alpha$ photon involving a sequence of two transitions}
\label{sec:channels}
If we restrict ourselves to the main channels that can lead to the
emission or absorption of photons in the vicinity of the
Lyman $\alpha$ resonance and involve two photons then we are left
with: (i) $n{\rm s}\rightarrow {\rm 1s}$ two-photon decay and
absorption; (ii) $n{\rm d}\rightarrow {\rm 1s}$ two-photon decay and
absorption; (iii) ${\rm c}\rightarrow {\rm 1s}$ two-photon
recombination and photoionization.

The problem is now to compute the emission and absorption profiles
connected with these processes and describe their relative
contributions or probabilities at a given frequency.
This in principle can be done for all possible routes. However, here we
will focus on formulating the problem for the $3{\rm
  s}\rightarrow {\rm 1s}$ and $3{\rm d}\rightarrow {\rm 1s}$
two-photon channels, not altering all other channels.
This is a reasonable first approximation, since \changeI{as we show here} it is already clear
from the '$1+1$' photon picture that the main contribution to the death of
photons is coming from the 3d channel \changeI{(see Fig.~\ref{fig:p_d})}. We only also add the 3s channel, since it is conceptually very similar. 

\subsection{The 3s-1s two-photon channel}
\label{sec:3s-channel}
In this section we would like to derive the rate equation that describes the evolution of the population, $N_{\rm 3s}$, in the 3s level, but where we take the two-photon aspect of the 3s-1s transition into account. In the normal '$1+1$' photon picture this transition is described by the sequence $\rm 3s\rightarrow 2p \rightarrow 1s$ and its inverse process $\rm 1s\rightarrow 2p \rightarrow 3s$. 
For the 3s rate equation it is therefore clear that the two-photon correction due to the 3s-1s channel should affect part of the $\rm 3s\rightarrow 2p$ and $\rm 2p\rightarrow 3s$ rate. Once this part is known one can in principle replace it in the rate equations using a more complete two-photon description.

Here we just give the formulation of this problem, also because it allows to understand the most important aspects of the two-photon picture. 
However, to compute the corrections to the escape probability we directly use the (pre-computed) solution for the populations given in the standard approach, and only solve for the presumably small correction in the evolution of the photon field around the Lyman $\alpha$ resonance.
Solving the complete set of modified rate equations simultaneously will be the final goal, for which one will need the results of the derivation presented here. We leave this problem for a future work.

\subsubsection{Isolating the different '$1+1$' photon routes}
In order to isolate the contribution from the 3s-1s two-photon channel we start by writing down the 3s rate equation in the '$1+1$' photon picture, including all possible ways for electrons in and out of the 3s-level
\beal
\label{eq:dN3sdt}
\Abl{N_{\rm 3s}}{t}=\left.\Abl{N_{\rm 3s}}{t}\right|_{\rm 3s2p}+ R^{+}_{\rm 3s}-R^{-}_{\rm 3s}N_{\rm 3s}.
\end{align}
Here we have directly separated the part due to the Balmer $\alpha$
transition
\beal
\label{eq:R3s2p}
\left.\Abl{N_{\rm 3s}}{t}\right|_{\rm 3s2p}&=R_{\rm 2p3s}-R_{\rm 3s2p}\,N_{\rm 3s}
\nonumber
\\
&\equiv
N_{\rm 2p}\,A_{\rm 3s2p}\,\frac{g_{\rm 3s}}{g_{\rm 2p}} \,\nbb(\nu_{32})-N_{\rm 3s}\,A_{\rm 3s2p}\,[1+\nbb(\nu_{32 })]
\end{align}
which below we want to \changeI{discuss} in more detail. 
\change{Here $N_i$ and $g_i$ denote the population and statistical weight of level $i$, $\nbb(\nu)$ is the CMB blackbody occupation number, and $A_{ij}$ and $\nu_{ij}$ are the transition rate and frequency between level $i$ and $j$.}
All the other possible channels in and out of the 3s-level lead to the terms
\beal
\label{eq:R3s}
R^{-}_{\rm 3s}&=R_{\rm 3s c}+\sum_{i>3\rm s}\frac{g_i}{g_{\rm 3s}} A_{\it i \rm 3s}\,\nbb(\nu_{\it i \rm 3s})\\
R^{+}_{\rm 3s}&=N_{\rm e}N_{\rm p} R_{\rm c 3s}+\sum_{i>3\rm s} N_i\,A_{\it i \rm 3s}\,
[1+\nbb(\nu_{\it i \rm 3s })],
\end{align}
\change{where $N_{\rm e}$ and $N_{\rm p}$ are the free electron and proton number densities, and $R_{{\rm c} i}$ and $R_{i{\rm c}}$ are the recombination and photoionization rates to the level $i$.
Note that, since} at frequencies below the Lyman $\alpha$ line the spectral
distortions during recombination are small \citep[e.g. cf.][]{Chluba2006b}, above we simply replaced
$n(\nu)\rightarrow \nbb(\nu)$ everywhere.

Now we are interested in refining the term connected with the Balmer $\alpha$ channel.
Since any two-photon or multi-photon process only leads to tiny corrections in the total decay rates, it is possible to use the one photon rates to compute the relative contributions of different transition sequences.
It is clear that the term $N_{\rm 3s}\,A^\ast_{\rm 3s2p}\equiv N_{\rm 3s}\,A_{\rm 3s2p}\,[1+\nbb(\nu_{32 })]$ describes the total flow of electrons in the direction of the 2p-state. Once the electron
reached there it can go back to the 3s level with the probability
\beal
\label{eq:p_2p3s}
p_{\rm 2p3s}&=\frac{A_{\rm 3s2p}\,\frac{g_{\rm 3s}}{g_{\rm 2p}}\,\nbb(\nu_{32})}{A^\ast_{\rm 2p1s}+R^{-}_{\rm 2p}}.
\end{align}
Here $A^{\ast}_{\rm 2p1s}=A_{\rm 2p1s}[1+\nbb(\nu_{21})]$ is the
stimulated Lyman $\alpha$ emission rate in the ambient CMB blackbody
field, and $R^{-}_{\rm 2p}$ is the total rate at which electrons can
leave the 2p-level, but excluding the Lyman $\alpha$ line. It is given by 
\beal
\label{eq:Rm2p}
R^{-}_{\rm 2p}&=R_{\rm 2p c}+\sum_{i>2\rm p}\frac{g_i}{g_{\rm 2p}} A_{\it i \rm 2p}\,\nbb(\nu_{\it i \rm 2p}).
\end{align}
Similarly, the electron can take the route $2{\rm p}\rightarrow 1{\rm
  s}$ with the probability
\beal
\label{eq:p_2p1s}
p_{\rm 2p1s}&=\frac{A^{\ast}_{\rm 2p1s}}{A^{\ast}_{\rm 2p1s}+R^{-}_{\rm 2p}} \equiv p^{1\gamma}_{\rm em}, 
\end{align}
or any of the other channels ($2{\rm p}\rightarrow n{\rm s}$ $(n>2)$,
$2{\rm p}\rightarrow n{\rm d}$ or $2{\rm p}\rightarrow {\rm c}$) with
probability $\bar{p}_{\rm 2p1s}=1-p_{\rm 2p1s}$.
Note that here we have neglected the deviations from a blackbody in
the stimulated Lyman $\alpha$ emission rate, which in any case is tiny. We just wanted to maintain the structure of the equations.
Then it is clear that the term $N_{\rm 3s}\,A^{\ast}_{\rm 3s2p}$ can
be interpreted as
\beal
\label{eq:R3s2p_parts}
N_{\rm 3s}\,A^{\ast}_{\rm 3s2p}&\equiv N_{\rm 3s}\,A^{\ast}_{\rm 3s2p}
\times[p_{\rm 2p1s}+\bar{p}_{\rm 2p1s}],
\end{align}
because the total flow of electrons should split up into those electrons that go to the 1s-level (probability $p_{\rm 2p1s}$) and those that don't (probability $\bar{p}_{\rm 2p1s}$).

From the physical point of view it is now clear that the {\it partial} flow connected with $p_{\rm 2p1s}N_{\rm
  3s}\,A^{\ast}_{\rm 3s2p}$ should be interpreted as 3s-1s two-photon emission in
the '$1+1$'-photon picture, which we will later replace with the more proper two-photon formulae. 
The rest ($\bar{p}_{\rm 2p1s}N_{\rm
  3s}\,A^{\ast}_{\rm 3s2p}$) describes the contributions
of all the other possible channels, e.g. also including the Balmer $\alpha$
scattering transition $3{\rm s}\rightarrow 2{\rm p}\rightarrow 3{\rm
  s}$. We will continue to describe all these in the '$1+1$' photon picture.

In order to understand the term connected with the total flow of
electrons from the 2p-level towards the 3s-state, given by $N_{\rm 2p}\,A_{\rm 3s2p}\,\frac{g_{\rm 3s}}{g_{\rm 2p}} \,\nbb(\nu_{32})$, we have to think
about an electron that is added to the 2p-state.
It will take the route $2{\rm p}\rightarrow 3{\rm s}$ with probability
$p_{\rm 2p3s}$ as given by Eq.~\eqref{eq:p_2p3s}. If we consider all
possible routes into the 2p-state, and write the corresponding total rate as
$R^{+,t}_{\rm 2p}=R^{+}_{\rm 2p}+N_{\rm 1s}\frac{g_{\rm 2p}}{g_{\rm
    1s}}\,A_{\rm 2p1s}\,n(\nu_{21})$, then one has the identity
\bsub
\label{eq:R2p3s_parts}
\beal
N_{\rm 2p}\,A_{\rm 3s2p}\,\frac{g_{\rm 3s}}{g_{\rm 2p}} \,\nbb(\nu_{32})
&\equiv p_{\rm 2p3s}\,[R^{+}_{\rm 2p}+N_{\rm 1s}\frac{g_{\rm 2p}}{g_{\rm
    1s}}\,A_{\rm 2p1s}\,n(\nu_{21})]
\end{align}
where 
\beal
\label{eq:Rp2p}
R^{+}_{\rm 2p}&=N_{\rm e}N_{\rm p} R_{\rm c 2p}+\sum_{i>2\rm p} N_i\,A_{\it i \rm 2p}\,[1+\nbb(\nu_{\it i \rm 2p })].
\end{align}
\esub
Now one can write
\bsub
\label{eq:R3s2p_allpart}
\beal
\label{eq:R3s2p_allpart_a}
\left.\Abl{N_{\rm 3s}}{t}\right|_{\rm 3s2p}
&=p_{\rm 2p3s}\,R^{+}_{\rm 2p}-\bar{p}_{\rm 2p1s}\,N_{\rm 3s}\,A^{\ast}_{\rm 3s2p}
+\left.\Abl{N_{\rm 3s}}{t}\right|_{\rm 3s2p1s}^{1+1}
\end{align}
where
\beal
\label{eq:R3s2p_2g_channel}
\left.\Abl{N_{\rm 3s}}{t}\right|_{\rm 3s2p1s}^{1+1}= 
p_{\rm 2p3s}N_{\rm 1s}\frac{g_{\rm 2p}}{g_{\rm 1s}}\,A_{\rm 2p1s}\,n(\nu_{21})
-p_{\rm 2p1s}\,N_{\rm 3s}\,A^{\ast}_{\rm 3s2p}.
\end{align}
\esub
The first two terms in Eq.~\eqref{eq:R3s2p_allpart_a} describe the partial flow of electrons towards the 3s-state, but where it is certain that the electron did not pass through the Lyman $\alpha$ line before.
The last term is the fractional contribution of the $3{\rm s}\leftrightarrow 2{\rm
  p}\leftrightarrow 1{\rm s}$-channel in the 3s rate equation, but described in the '$1+1$' photon picture.
  This is the term which in the end we will want to replace with the two-photon formulae.

If we now identify 
\beal
\label{eq:A2gamma_3s2p}
A^{2\gamma}_{\rm 3s1s}=\frac{A_{\rm 3s2p}\,A_{\rm 2p1s}}{A^\ast_{\rm 2p1s}+R^{-}_{\rm 2p}}, 
\end{align}
then we can finally rewrite Eq.~\eqref{eq:R3s2p_2g_channel} as
\beal
\label{eq:R3s2p_2g_channel_tilde}
\left.\Abl{N_{\rm 3s}}{t}\right|_{\rm 3s2p1s}^{1+1}
&= 
A^{2\gamma}_{\rm 3s1s}\,N_{\rm 1s}\frac{g_{\rm 3s}}{g_{\rm 1s}}\,n(\nu_{21})\,\nbb(\nu_{32})
\nonumber
\\
&\qquad
-A^{2\gamma}_{\rm 3s1s}\,N_{\rm 3s}[1+\nbb(\nu_{21})][1+\nbb(\nu_{32})].
\end{align}
Note that in vacuum one would have $A^{2\gamma}_{\rm 3s1s}\equiv
A_{\rm 3s2p}$, as it should be, since the electron will only have one way to leave the 2p-state.
However, within an intense CMB background field, also the other channels will become active (e.g. $\rm 3s\leftrightarrow 2p \leftrightarrow c$), so that part of the $\rm 3s\leftrightarrow 2p$ flow will go through them.
\change{This reduces the effective decay rate $A^{2\gamma}_{\rm 3s1s}$.}

\subsubsection{Replacing the 3s-1s channel in the '$1+1$' photon formulation with the two-photon expression}
We now want to replace the part due to $\left.\Abl{N_{\rm 3s}}{t}\right|_{\rm 3s2p1s}^{1+1}$ with the more proper two-photon terms. For this we have to ask the question how the 3s-1s two-photon term actually looks like.
If one considers an electron that is initially in the 3s-state, then
one can use the vacuum 3s-1s two-photon decay profile in order to
derive the emission profile needed to describe the injection of
Lyman $\alpha$ photons for the escape problem.
Simple formulae for the necessary vacuum two-photon decay profiles can be found in the literature
\citep{Chluba2008a}. 
We shall normalize these profiles like
$\int_0^\infty \frac{\phi^{2\gamma}_i(\nu)}{4\pi\,\Delta\nu_{\rm
    D}}\id\nu\id\Omega\equiv 1$, where $\phi^{2\gamma}_i(\nu)$ already includes the motion of the
atoms in the same way as for the normal Lorentzian lines, usually leading to the Voigt-profiles
\citep[e.g. see][]{Mihalas1978}.
For convenience we chose the Lyman $\alpha$ Doppler-width,
$\Delta\nu_{\rm D}$, in the normalization.

With this the net change of the number density of electrons in the 3s
level via the 3s-1s two-photon channel is given by
\bsub
\beal
\label{eq:dN3sdt_2g_a}
\left.\Abl{N_{\rm 3s}}{t}\right|^{2\gamma}_{\rm 3s1s}
&=A^{2\gamma}_{\rm 3s1s}N_{\rm 1s}\int \varphi^{2\gamma}_{\rm 3s}(\nu) n(\nu)\,n(\nu_{31}-\nu)\id\nu
\nonumber
\\
&\quad
-A^{2\gamma}_{\rm 3s1s}N_{\rm 3s}\int \varphi^{2\gamma}_{\rm 3s}(\nu) [1+n(\nu)] [1+n(\nu_{31}-\nu)]\id\nu
\\
\label{eq:dN3sdt_2g_b}
&\approx
A^{2\gamma}_{\rm 3s1s} N_{\rm 1s} n(\nu_{21})\,\nbb(\nu_{32}) 
\nonumber
\\
&\qquad
- A^{2\gamma}_{\rm 3s1s} N_{\rm 3s} [1+\nbb(\nu_{21})] [1+\nbb(\nu_{32})],
\end{align}
\esub
where $A^{2\gamma}_{\rm 3s1s}$ is the effective 3s-1s
two-photon decay rate, which in vacuum is\footnote{Here the approximate sign is due to the fact that the rate coefficient in the two-photon formulation should contain some small ($\sim 10^{-6}-10^{-5}$) quantum mechanical correction to the one photon rate. This will not lead to any significant correction in the escape probability.} $A^{2\gamma}_{\rm 3s1s}\approx A_{\rm 3s2p}$, but within an ambient blackbody radiation field should take the value given by Eq.~\eqref{eq:A2gamma_3s2p}.
Furthermore, $\varphi^{2\gamma}_{\rm
  3s}=\phi^{2\gamma}_{\rm
  3s}(\nu)/\Delta\nu_{\rm D}$ denotes the 3s-1s
two-photon decay profile, and $\nu_{31}$ is the 3s-1s transition
frequency.

For the approximation Eq.~\eqref{eq:dN3sdt_2g_b} three comments should be made: first  we have assumed that the main contributions
to the integrals over the two-photon line profiles are coming from the
poles close to $\nu\sim\nu_{21}$ and $\nu\sim\nu_{32}$.  Second, we
have used the fact that the CMB spectral distortion around the
Balmer $\alpha$ line are tiny. Also the stimulated term in the vicinty
of the Lyman $\alpha$ resonance is completely negligable, so that we
just can use $1+n(\nu_{21})\approx1+\nbb(\nu_{21})$ instead, without
changing anything.
And finally, we assumed that only for the 1s-3s two-photon absorption rate the
deviations of the CMB spectrum from a blackbody in the vicinity of the
Lyman $\alpha$ resonance will matter.

The result Eq.~\eqref{eq:dN3sdt_2g_b} is identical with the term given by the '$1+1$' photon picture, Eq.~\eqref{eq:R3s2p_2g_channel_tilde}. This is not surprising, since with the above approximations we have simply turned from the two-photon to the '$1+1$' photon picture.
In order to include the effect of two-photon transitions in to the
rate equation of the 3s-level, we should therefore replace
$\left.\Abl{N_{\rm 3s}}{t}\right|_{\rm 3s2p1s}^{1+1}$ with the full
integral given by Eq.~\eqref{eq:dN3sdt_2g_a}.

\subsubsection{Term in the Lyman $\alpha$ radiative transfer equation}
%
In order to use the integral \eqref{eq:dN3sdt_2g_a} in the computations of the ionization history, we also have to give the solution of the CMB spectral distortion in the vicinity of the Lyman $\alpha$ resonance. We therefore have to explicitly write the 3s-1s two-photon emission and absorption terms for the evolution of the photon field and solve the corresponding transfer equation.
In particular we want to bring the transfer equation into the form Eq.~\eqref{eq:real_em_abs_11_gen}.

From Eq.~\eqref{eq:dN3sdt_2g_a} it directly follows
\beal
\label{eq:dNdt_3s1s_emabs}
\left.\frac{1}{c}\,\frac{\partial N_{\nu}}{\partial t}\right|^{2\gamma}_{\rm 3s1s}
&=\frac{2A^{2\gamma}_{\rm 3s1s}}{4\pi} N_{\rm 3s} \varphi^{2\gamma}_{\rm 3s}(\nu) [1+n(\nu)] [1+n(\nu_{31}-\nu)]
\nonumber
\\
&\qquad
-\frac{2A^{2\gamma}_{\rm 3s1s}}{4\pi} N_{\rm 1s} \varphi^{2\gamma}_{\rm 3s}(\nu)\, 
n(\nu)\,n(\nu_{31}-\nu).
\end{align}
Here the factor of 2 is due to the fact that per electron two photons are involved, and the factor of $4\pi$ converts to per steradian, \changeI{having in mind that the medium is isotropic}.
Furthermore in this form it is assumed that every two-photon interaction in the 3s-1s channel leads to a {\it complete redistribution} of the photons over the whole two-photon profile. This also means that we have not distinguished two-photon emission and absorption from two-photon scattering events.
However, this should be a very good approximation, since the scattering event involves two photons. This means that the total energy of the incoming photons will be split up such that in most cases the scattered photons will have energy $\nu\sim \nu_{21}$ and $\nu'\sim \nu_{32}$.
Note that this does not imply that we are using a complete redistribution approximation for the Lyman $\alpha$ resonance scattering itself, since only about $\sim 10^{-4}-10^{-3}$ of all interactions will lead to the 3s- and 3d-state via two-photon interactions \citep{Chluba2008b}.

Neglecting the deviations from the blackbody spectrum in the emission term and comparing with Eq.~\eqref{eq:real_em_abs_11_gen} we can identify
\bsub
\label{eq:phi_Rp_3s}
\beal
\phi_{\rm 3s\leftrightarrow 1s}(\nu)&=2\,\phi^{2\gamma}_{\rm 3s}(\nu) 
\frac{1+\nbb(\nu)}{1+\nbb(\nu_{21})} \frac{1+\nbb(\nu_{31}-\nu)}{1+\nbb(\nu_{32})}
\\
&\stackrel{\stackrel{\nu\gtrsim\nu_{31}/2}{\downarrow}}{\approx}
2\,\phi^{2\gamma}_{\rm 3s}(\nu) 
\frac{1+\nbb(\nu_{31}-\nu)}{1+\nbb(\nu_{32})}
\\[1mm]
p^{1\gamma}_{\rm em}&=\frac{A^\ast_{\rm 2p1s}}{A^\ast_{\rm 2p1s}+R^{-}_{\rm 2p}}
\\[1mm]
R^{\rm 3s, +}_{\rm 2p}&=A^\ast_{\rm 3s2p}\,N_{\rm 3s}. 
\end{align}
\esub
This result shows that the effective profile of the process as expected is given by the two-photon profile for the 3s-1s channel including the induced terms relative to the values at the Lyman and Balmer $\alpha$ resonance. 
Also the emission probability is exactly the Lyman $\alpha$ emission probability including the induced emission for the central frequency of the Lyman $\alpha$ line. And the last term simply represents the number density of 3s-electrons that reach the 2p-state per second in the '$1+1$' photon picture, where again the stimulated emission due to CMB photons close to the Balmer $\alpha$ frequency was included.

Note that $\phi_{\rm 3s\leftrightarrow 1s}(\nu)$ is no longer normalized to unity. In vacuum one would find $\int_0^\infty \frac{\phi^{2\gamma}_{\rm 3s\leftrightarrow 1s}(\nu)}{4\pi\,\Delta\nu_{\rm
    D}}\id\nu\id\Omega\equiv 2$, while within the CMB blackbody field $\int_0^\infty \frac{\phi^{2\gamma}_{\rm 3s\leftrightarrow 1s}(\nu)}{4\pi\,\Delta\nu_{\rm
    D}}\id\nu\id\Omega\gtrsim 2$. 
However the relative correction to the overall normalization of the profile due to stimulated emission is of the order of $\sim 10^{-9}-10^{-8}$ (see Sect.~\ref{sec:corr_DP_ind}).
Also one should mention that due to the symmetry of the profile around $\nu=\nu_{31}/2$, by restricting the range of integration to $\nu_{31}/2\leq \nu\leq \nu_{31}$ one can avoid counting both the Lyman $\alpha$ and Balmer $\alpha$ photons.
We will use this fact to simplify the numerical integration (see Sect.~\ref{sec:integration}).

If we now look at the absorption term in Eq.~\eqref{eq:dNdt_3s1s_emabs}, using the definitions \eqref{eq:phi_Rp_3s} we can directly write
\beal
\left.\frac{1}{c}\,\frac{\partial N_{\nu}}{\partial t}\right|^{2\gamma}_{\rm 3s1s, abs}
&=\frac{\phi_{\rm 3s\leftrightarrow 1s}(\nu)}{4\pi\,\Delta\nu_{\rm D}}\,
A^{2\gamma}_{\rm 3s1s}\,N_{\rm 1s}
\,n(\nu)\,\nbb(\nu_{31}-\nu)
\nonumber\\
&\qquad\times
\frac{1+\nbb(\nu_{21})}{1+\nbb(\nu)} \frac{1+\nbb(\nu_{32})}{1+\nbb(\nu_{31}-\nu)}.
\end{align}
Here we have already assumed that the important part for our problem is the region $\nu\geq \nu_{31}/2$. This implies that $\nu_{31}-\nu\leq \nu_{31}/2$, so that the deviations from the CMB blackbody can be neglected leading to $n(\nu)\,\nbb(\nu_{31}-\nu)$ instead of $n(\nu)\,n(\nu_{31}-\nu)$.
Since according to Eq.~\eqref{eq:p_2p3s} and \eqref{eq:A2gamma_3s2p} $A^{2\gamma}_{\rm 3s1s}\equiv p_{\rm 2p3s}\frac{g_{\rm 2p}\, A_{\rm 2p1s} }{g_{\rm 3s} \nbb(\nu_{32})}$, and because $[1+\nbb(\nu)]/\nbb(\nu)=e^{h\nu/k\Tg}$, we finally find
\bsub
\beal
\label{eq:sss_1}
\left.\frac{1}{c}\,\frac{\partial N_{\nu}}{\partial t}\right|^{2\gamma}_{\rm 3s1s, abs}
&\!\!\!\!=\frac{\phi_{\rm 3s\leftrightarrow 1s}(\nu)}{4\pi\,\Delta\nu_{\rm D}}
\frac{A_{\rm 2p1s}\,p_{\rm 2p3s}}{g_{\rm 3s}/g_{\rm 2p}}
e^{\frac{h[\nu-\nu_{21}]}{k\Tg}} \frac{1+\nbb(\nu_{21})}{1+\nbb(\nu)} \,{n(\nu)}N_{\rm 1s}
\\
&\!\!\!\!\!\!\!\stackrel{\stackrel{\nu\gtrsim\nu_{31}/2}{\downarrow}}{\approx}
\!\!\!\frac{\phi_{\rm 3s\leftrightarrow 1s}(\nu)}{4\pi\,\Delta\nu_{\rm D}}
\frac{g_{\rm 2p}}{g_{\rm 1s}} A_{\rm 2p1s} p_{\rm 2p3s}\,
e^{\frac{h[\nu-\nu_{21}]}{k\Tg}}\!n(\nu)N_{\rm 1s}.
\end{align}
\esub
Note that $g_{\rm 3s}/g_{\rm 2p}\equiv g_{\rm 1s}/g_{\rm 2p}$. 

With the Einstein relations it is then easy to show that 
$\frac{g_{\rm 2p}}{g_{\rm 1s}}\,A_{\rm 2p1s}\,n(\nu)=h\nu_{21}\,B_{12} \frac{\nu^2_{21}}{\nu^2}N_\nu$, so that we directly verify the thermodynamic correction factor\footnote{From Eq.~\eqref{eq:sss_1} bycomparing with Eq.~\eqref{eq:real_em_abs_11_gen} we can see that with the choice of coefficients and variables we more rigorously infer $f_\nu=\frac{\nu_{21}^2}{\nu^2}e^{\frac{h[\nu-\nu_{21}]}{k\Tg}} \frac{1+\nbb(\nu_{21})}{1+\nbb(\nu)}\equiv
\frac{\nu_{21}^2}{\nu^2}\,\frac{\nbb(\nu_{21})}{\nbb(\nu)}\approx \frac{\nu_{21}^2}{\nu^2}\,e^{h[\nu-\nu_{21}]/k\Tg} $ for $h\nu\gg k\Tg$, a condition that is fulfilled during cosmological recombination in the vicinity ($\nu\gtrsim\nu_{31}/2$) of the Lyman $\alpha$ resonance.} $f_\nu=\frac{\nu_{21}^2}{\nu^2}\,e^{h[\nu-\nu_{21}]/k\Tg} $ and find $p_{\rm d}^{\rm 1s3s}\equiv p_{\rm 2p3s}$.
We therefore have confirmed the completeness of the form of the Eq.~\eqref{eq:real_em_abs_11_gen} for the 3s-1s two-photon channel.

\subsection{The 3d-1s two-photon channel}
After going through the argument for the 3s-1s channel it is easy to do the same for the 3d-1s channel.
For the rate equation analog to  Eq.~\eqref{eq:dN3sdt}, \eqref{eq:R3s}, and \eqref{eq:R3s2p_allpart} one has
\bsub
\label{eq:R3d2p_allpart}
\beal
\label{eq:R3d2p_allpart_a}
\Abl{N_{\rm 3d}}{t}
&=\left.\Abl{N_{\rm 3d}}{t}\right|_{\rm 3d2p}+ R^{+}_{\rm 3d}-R^{-}_{\rm 3d}N_{\rm 3d}
\\
R^{-}_{\rm 3d}&=R_{\rm 3d c}+\sum_{i>3\rm d}\frac{g_i}{g_{\rm 3d}} A_{\it i \rm 3d}\,\nbb(\nu_{\it i \rm 3d})
\\
R^{+}_{\rm 3d}&=N_{\rm e}N_{\rm p} R_{\rm c 3d}+\sum_{i>3\rm d} N_i\,A_{\it i \rm 3s}\,
[1+\nbb(\nu_{\it i\rm 3d})] 
\end{align}
where the Balmer $\alpha$ channel is defined by
\beal
\label{eq:R3d2p_allpart_b}
\left.\Abl{N_{\rm 3d}}{t}\right|_{\rm 3d2p}
&=p_{\rm 2p3d}\,R^{+}_{\rm 2p}-\bar{p}_{\rm 2p1s}\,N_{\rm 3d}\,A^{\ast}_{\rm 3d2p}
\!+\!\!\left.\Abl{N_{\rm 3d}}{t}\right|_{\rm 3d2p1s}^{1+1}
\\
\label{eq:p_2p3d}
p_{\rm 2p3d}&=\frac{A_{\rm 3d2p}\,\frac{g_{\rm 3d}}{g_{\rm 2p}}\,\nbb(\nu_{32})}{A^\ast_{\rm 2p1s}+R^{-}_{\rm 2p}}
\\
\label{eq:R3d2p_2g_channel_tilde}
\left.\Abl{N_{\rm 3d}}{t}\right|_{\rm 3d2p1s}^{1+1}
&= 
A^{2\gamma}_{\rm 3d1s}\,N_{\rm 1s}\frac{g_{\rm 3d}}{g_{\rm 1s}}\,n(\nu_{21})\,\nbb(\nu_{32})
\nonumber
\\
&\quad
-A^{2\gamma}_{\rm 3d1s}\,N_{\rm 3d}[1+\nbb(\nu_{21})][1+\nbb(\nu_{32})].
\end{align}
\esub
As before one should now replace Eq.~\eqref{eq:R3d2p_2g_channel_tilde} with
\bsub
\beal
\label{eq:dN3ddt_2g_a}
\left.\Abl{N_{\rm 3d}}{t}\right|^{2\gamma}_{\rm 3d1s}
&=A^{2\gamma}_{\rm 3d1s}\frac{g_{\rm 3d}}{g_{\rm 1s}}\,N_{\rm 1s}\int \varphi^{2\gamma}_{\rm 3d}(\nu) n(\nu)\,n(\nu_{31}-\nu)\id\nu
\nonumber
\\
&\quad
-A^{2\gamma}_{\rm 3d1s}N_{\rm 3d}
\int  \varphi^{2\gamma}_{\rm 3d}(\nu) [1+n(\nu)] [1+n(\nu_{31}-\nu)]\id\nu
\\
\label{eq:A2gamma_3d2p}
A^{2\gamma}_{\rm 3d1s}
&=\frac{A_{\rm 3d2p}\,A_{\rm 2p1s}}{A^\ast_{\rm 2p1s}+R^{-}_{\rm 2p}}
\end{align}
\esub
if one is interested in the 3d-1s two-photon correction to the 3d-rate equation. Note that here the ratio of the statistical weights is not unity like in the case of the 3s-1s channel.

The terms for the transfer equation can also be cast into the form \eqref{eq:real_em_abs_11_gen} where the important coefficients are given by
\bsub
\label{eq:phi_Rp_3d}
\beal
\phi_{\rm 3d\leftrightarrow 1s}(\nu)&=2\,\phi^{2\gamma}_{\rm 3d}(\nu) 
\frac{1+\nbb(\nu)}{1+\nbb(\nu_{21})} \frac{1+\nbb(\nu_{31}-\nu)}{1+\nbb(\nu_{32})}
\\
&\approx
2\,\phi^{2\gamma}_{\rm 3d}(\nu) 
\frac{1+\nbb(\nu_{31}-\nu)}{1+\nbb(\nu_{32})}
\\[1mm]
R^{\rm 3s, +}_{\rm 2p}&=A^\ast_{\rm 3d2p}\,N_{\rm 3d}
\\[1mm]
p^{\rm 1s3d}_{\rm d}&=p_{\rm 2p3d}. 
\end{align}
\esub

\subsection{The other channels}
For the other channels in and out of the 2p-state we can also derive
the corresponding partial rates in a similar way as for the 3s and 3d state. However, since the main
correction is expected to come from the 3s and 3d two-photon channels for these we will
simply use the '$1+1$' photon picture. This means that we will not replace the corresponding $i\leftrightarrow \rm 2p$ rates with the two-photon description.
For all the $n$s and $n$d-states with $n>3$ the rate equations therefore will be similar to Eq.~\eqref{eq:R3d2p_allpart},
and for electrons in the continuum one will have
\bsub
\label{eq:Rc2p_allpart}
\beal
\label{eq:Rc2p_allpart_a}
\Abl{N_{\rm e}}{t}
&=\left.\Abl{N_{\rm e}}{t}\right|_{\rm c2p}
+
 \sum_{i>2} \left[R_{i \rm c} N_{i} - N_{\rm e} N_{\rm p} R_{\rm c \it i}
 \right]
\end{align}
with
\beal
\label{eq:Rc2p_allpart_b}
\left.\Abl{N_{\rm e}}{t}\right|_{\rm c2p}
&=p_{\rm 2pc}\,R^{+}_{\rm 2p}-\bar{p}_{\rm 2p1s}\,N_{\rm e}\,N_{\rm p}\,R_{\rm c2p}
+\left.\Abl{N_{\rm e}}{t}\right|_{\rm c2p1s}^{1+1}
\\
\label{eq:p_2pc}
p_{\rm 2pc}&=\frac{R_{\rm 2pc}}{A^\ast_{\rm 2p1s}+R^{-}_{\rm 2p}}
\\
\label{eq:Rc2p_2g_channel_tilde}
\left.\Abl{N_{\rm e}}{t}\right|_{\rm c2p1s}^{1+1}
&= 
N_{\rm 1s}\frac{g_{\rm 2p}}{g_{\rm 1s}}\,A_{\rm 2p1s}\,n(\nu_{21})\,p_{\rm 2p c}
-p_{\rm 2p1s}\,N_{\rm e}\,N_{\rm p}\,R_{\rm c2p}.
\end{align}
\esub
%
Still there is a small difference to the normal rate equations. In the formulation given above the population of the 2p-state has vanished from {\it all} the rate equations, and in particular from those for the 3s and 3d-state. For the 3s and 3d-state physically this is expected, since in the two-photon picture on the way to the 1s-level the electron is not really passing through the 2p-state. In the full two-photon picture the electron reaches the 1s level via {\it all} intermediate p-states, including those in the continuum. 
For the other levels the above formulation would have also been obtained by simply replacing the solution of the 2p-state with the quasi-stationary value in the '$1+1$' photon approach. In this way one again has a closed system of rate equations, which avoids the difficulty in attaching a population to the 2p-state.

\section{Derivation of the thermodynamic factor using the '$1+1$'  photon picture.}
\label{app:DB_f}
As mentioned in the introduction and also earlier \citep{Chluba2008b}, in the normal '$1+1$' photon approximation the term describing the emission and absorption of Lyman $\alpha$ photons in full thermodynamic equilibrium is {\it not} exactly conserving a blackbody spectrum at all frequencies.
This can be directly seen from Eq.~\eqref{eq:real_em_abs_11}, since in full thermodynamic equilibrium one should have 
$\left(p_{\rm em}^{\rm 1\gamma}\,R^{+}_{\rm 2p}\right)^{\rm eq}
\equiv
\left( p^{\rm 1\gamma}_{\rm d}\,h\nu_{\rm 21}\,B_{12}\, N_{1\rm s}\,N_{\nu} \right)^{\rm eq}$. 
 Using the definitions of the previous Section, in equilibrium one expects $\left(R^{+}_{\rm 2p}\right)^{\rm eq}\equiv  \left( R^{-}_{\rm 2p} N_{\rm 2p}\right)^{\rm eq}$, 
$N_{\rm 2p}^{\rm eq}\equiv \frac{g_{\rm 2p}}{g_{\rm 1s}} N_{\rm 1s}^{\rm eq} \, e^{-\frac{h\nu_{21}}{k\Tg}}$, 
$h\nu_{\rm 21}B_{12}\equiv \frac{g_{\rm 2p}}{g_{\rm 1s}}\,\frac{c^2 A_{21}}{2\nu_{21}^2}$
and 
$p_{\rm em}^{\rm 1\gamma}\equiv A_{21}\,[1+\nbb(\nu_{21})]\,p^{\rm 1\gamma}_{\rm d}/R^{-}_{\rm 2p}$, so that one finds
\beal
\label{eq:rEQ_f}
\left( 
\frac{p^{\rm 1\gamma}_{\rm d}\,h\nu_{\rm 21}\,B_{12}\, N_{1\rm s}\,N_{\nu}}
{p_{\rm em}^{\rm 1\gamma}\,R^{+}_{\rm 2p}} \right)^{\rm eq}
\equiv 
\frac{c^2}{2\nu_{21}^2}\frac{N^{\rm pl}_{\nu}}{1+\nbb(\nu_{21})}\,e^{\frac{h\nu_{21}}{k\Tg}}
\end{align}
Here $N^{\rm pl}_{\nu}=\frac{2\nu^2}{c^2}\,\nbb(\nu)$ defines the blackbody spectrum. 
With the identity $[1+\nbb(\nu)]/\nbb(\nu)\equiv e^{h\nu/k\Tg}$ one therefore has 
\beal
\label{eq:rEQ_f_result}
\left( 
\frac{p^{\rm 1\gamma}_{\rm d}\,h\nu_{\rm 21}\,B_{12}\, N_{1\rm s}\,N_{\nu}}
{p_{\rm em}^{\rm 1\gamma}\,R^{+}_{\rm 2p}} \right)^{\rm eq}
\equiv 
\frac{\nu^2}{\nu_{21}^2}\frac{1+\nbb(\nu)}{1+\nbb(\nu_{21})}\,e^{\frac{h[\nu_{21}-\nu]}{k\Tg}}
\approx 1/f_{\nu}
\end{align}
The in the last step we assumed that $h\nu\gg k\Tg$ and $h\nu_{21}\gg k\Tg$, so that the factors $[1+n]$ could be neglected, an approximation that is certainly possible during cosmological recombination. However it in fact is only due to the used definition of $p_{\rm em}^{\rm 1\gamma}$, in which we evaluated the stimulated emission factors at the Lyman $\alpha$ line center only and in addition directly assumed a blackbody spectrum.
More consistently one should have used $p_{\rm em}^{\rm 1\gamma}=p_{\rm em}^{\rm 1\gamma, \ast}[1+n_{\nu}]$, where we have
\beal
\label{eq:p_em_star}
p^{1\gamma, \ast}_{\rm em}&=\frac{A_{\rm 2p1s}}{A_{\rm 2p1s}[1+\left<n_{\nu}\right>_{\rm em}]+R^{-}_{\rm 2p}}.
\end{align}
Here $\left<n_{\nu}\right>_{\rm em}$ denotes the average of the photon occupation number across the Lyman $\alpha$ emission profile.
%
With this definition in Eq.~\eqref{eq:rEQ_f_result} one would have obtained the factor $1/f_\nu$ directly, without any additional comment.

We would like to mention another way to obtain the thermodynamic factor in the absorption process, which just uses the term in the Lyman $\alpha$ rate equation.
In the standard textbooks \citep[e.g.][]{Mihalas1978} one finds:
\beal
\label{eq:dN2pdt}
\left.\Abl{N_{\rm 2p}}{t}\right|_{\rm 2p1s}^{1+1}= 
N_{\rm 1s}\frac{g_{\rm 2p}}{g_{\rm 1s}}\,A_{\rm 2p1s}\,n(\nu_{21})
-N_{\rm 2p}\,A_{\rm 2p1s}[1+n(\nu_{21})].
\end{align}
Here two comments should be made: (i) more rigorously one should replace $n(\nu_{21})$ with the average occupation number over the line profile, i.e. $n(\nu_{21})\rightarrow \left<n_{\nu}\right>$ and (ii) in addition one should distinguish between the emission and absorption process, implying that $[1+n(\nu_{21})]\rightarrow [1+\left<n_{\nu}\right>_{\rm em}]$, while in the first term one has $n(\nu_{21})\rightarrow\left<n_{\nu}\right>_{\rm abs}$.
In full equilibrium one should then find
$\left(N_{\rm 1s}\frac{g_{\rm 2p}}{g_{\rm 1s}}\,A_{\rm 2p1s}\,\left<n_{\nu}\right>_{\rm abs}\right)^{\rm eq}
=\left(N_{\rm 2p}\,A_{\rm 2p1s}[1+\left<n_{\nu}\right>_{\rm em}]\right)^{\rm eq}$, so that
\beal
\label{eq:rEQ_f_alt}
\left( 
\frac{N_{\rm 1s}\frac{g_{\rm 2p}}{g_{\rm 1s}}\,A_{\rm 2p1s}\,\left<n_{\nu}\right>_{\rm abs}}{N_{\rm 2p}\,A_{\rm 2p1s}[1+\left<n_{\nu}\right>_{\rm em}]}\right)^{\rm eq}
&=\frac{\left<n^{\rm pl}_{\nu}\right>_{\rm abs}\,e^{h\nu_{21}/k\Tg}}
{1+\left<n^{\rm pl}_{\nu}\right>_{\rm em}}\equiv 1
\end{align}
It is now easy to show that
\beal
\label{eq:small_derivation}
\left<n^{\rm pl}_{\nu}\right>_{\rm abs}\,e^{h\nu_{21}/k\Tg}
&\equiv 
\left<\frac{n^{\rm pl}_{\nu}}{1+n^{\rm pl}_{\nu}}\,[1+n^{\rm pl}_{\nu}]\right>_{\rm abs}\,e^{h\nu_{21}/k\Tg}
\nonumber\\
&\equiv \left<e^{h[\nu_{21}-\nu]/k\Tg}\,[1+n^{\rm pl}_{\nu}]\right>_{\rm abs}.
\end{align}
Since according to Eq.~\eqref{eq:rEQ_f_alt} $\left<n^{\rm pl}_{\nu}\right>_{\rm abs}\,e^{h\nu_{21}/k\Tg}\equiv \left<1+n^{\rm pl}_{\nu}\right>_{\rm em}$
one should conclude that
$\left< n_\nu \right>_{\rm abs}\equiv \left< e^{h[\nu-\nu_{21}]/k\Tg} n_\nu \right>_{\rm em}$, where $n_\nu$ now is an arbitrary photon occupation number.
In terms of $N_\nu=\frac{h\nu^2}{c^2}\,n_\nu$ one therefore has $\left< N_\nu \right>_{\rm abs}
\equiv \frac{2\nu_{21}^2}{c^2}\,\left< n_\nu \right>_{\rm abs}\equiv \left< f_{\nu} N_\nu \right>_{\rm em}$.
With this we obtained the thermodynamic correction factor $f_\nu$ in the absorption profile for $N_\nu=\frac{h\nu^2}{c^2}\,n_\nu$, since $\left< N_\nu \right>_{\rm abs}\equiv \left< f_{\nu} N_\nu \right>_{\rm em}$ automatically implies $\phi_{\rm abs}\equiv f_{\nu} \phi_{\rm em}$.

\section{Computation of two-photon profiles}
\label{app:2gamma_profiles}
We compute the two-photon decay profiles according to the work of  \citep{Chluba2008b}. There in particular the infinite sum over intermediate states was split up into those states with principal quantum numbers $n>n_i$ and $n\leq n_i$, where $n_i$ is the initial state principal quantum number. This makes the sum over the resonances (in the case of 3s and 3d only one) finite and allows to give fitting formulae for the remaining contribution to the total matrix element coming from the infinite sum. This procedure is very convenient for numerical evaluations.

Here we would like to mention that the two-photon decay profiles behave like $\phi\propto \nu\,(\nu_{i\rm 1s}-\nu)$ in the limits $\nu\rightarrow 0$ or  $\nu\rightarrow \nu_{i\rm 1s}$. This is because in this limit the main term in the infinte sum is coming from the Matrix element $n_i{\rm s/d} \rightarrow n_i\rm p$, which in the non-relativistic formulation has zero transition frequency. This implies that for $\nu\rightarrow 0$ or  $\nu\rightarrow \nu_{i\rm 1s}$
\beal
\label{eq:phi_0}
\phi^{2\gamma}_{n_i \rm s/d\rightarrow \rm 1s}
&\approx G_{n_il_i}\,y\,(1-y) \left|\left< R_{\rm 1s}\, |\, r \,| \,R_{n_i\rm p} \right>\left< R_{n_i\rm p} \,|\, r\, |\, R_{n_i I_i} \right>\right|^2
\nonumber
\\
&\approx G_{n_il_i}\,y\,(1-y)\, 2^6\,3^2\,n_i^9\frac{(n_i-1)^{2n_i-5}}{(n_i+1)^{2n_i+5}}\,(n_i^2-1|4).
\end{align}
Here $1|4$ means $1$ for the s-states or $4$ for the d-states.
Inserting numbers \citep[for definitions see][]{Chluba2008a} one finds
\bsub
\label{eq:phi_0}
\beal
\phi^{2\gamma}_{n_i \rm s\rightarrow \rm 1s}
&\approx
\pot{1.0598}{4}\,y\,(1-y)\,\frac{(n_i-1)^{2n_i}}{(n_i+1)^{2n_i}}\,\frac{n_i^2-1}{n_i}\,{\rm s^{-1}}
\\
\phi^{2\gamma}_{n_i \rm d\rightarrow \rm 1s}
&\approx
\pot{4.2393}{3}\,y\,(1-y)\,\frac{(n_i-1)^{2n_i}}{(n_i+1)^{2n_i}}\,\frac{n_i^2-4}{n_i}\,{\rm s^{-1}}.
\end{align}
\esub
For the 3s and 3d profiles one therefore has
\bsub
\label{eq:phi_3s2d_0}
\beal
\phi^{2\gamma}_{\rm 3s\rightarrow \rm 1s}(y)
&\approx  441.6\,y\,(1-y)\,{\rm s^{-1}}
\\
\phi^{2\gamma}_{\rm 3d\rightarrow \rm 1s}(y)
&\approx  110.4\,y\,(1-y)\,{\rm s^{-1}}
\end{align}
\esub
We will use these simple formulae to compute the two-photon spectra at $0 \leq y \leq 0.001$ and $0.999\leq y \leq 1$.

The most important consequence of this limiting behavior with frequency is that due to stimulated emission in the ambient CMB blackbody radiation field the two-photon profiles no longer vanish at $y\sim 0$ and $y\sim 1$, since $\nbb(\nu)\sim 1/y$ for $\nu\sim 0$. For the 2s-1s two-photon process this behavior was also seen earlier \citep{Chluba2006}. In the case of 3s and 3d two-photon decays this enhances the emission of photons close to the Lyman $\beta$ resonance (cf. Fig.~\ref{fig:phi_abs}). 
%
\changeI{However, we find that the corrections due to stimulated two-photon emission are not important for the cosmological recombination problem.}

\section{Small corrections due to the motion of the atom}
\label{app:profiles}
To account for the motion of the atoms in the computations of the emission profile one has to compute the following integral \citep[see Sect. 9.2 in][]{Mihalas1978}
\beal
\label{app:therm_av}
\phi_{\rm m}(\nu)=\int_{-\infty}^{\infty} \phi_{\rm r}(\nu'(t))\,e^{-t^2}\id t
\end{align}
over the rest frame emission profile, $\phi_{\rm r}(\nu)$, which for a given frequency $\nu$ due to the Doppler effect has to be evaluated at $\nu'(t)=\nu[1-\frac{\xi_0}{c}\,t]$, where $\xi_0=c\,\sqrt{2k T/m_{\rm H}\,c^2}$. The exponential factor arises from the Maxwell-Boltzmann velocity distribution for the neutral hydrogen atoms.

For the Voigt-profile one normally uses the approximation $\nu'(t)\approx \nu-\nu_{21}\frac{\xi_0}{c}\,t$, so that the emission profile can be written in terms of the Voigt-function
\beal
\label{app:H_V}
H(\xD, a)=\frac{a}{\pi}\int_{-\infty}^{\infty}\frac{e^{-t^2}\id t}{(\xD-t)^2+a^2}
\end{align}
for which simple approximation in terms of the Dawson-integral exist \citep[see Sect. 9.2 in][]{Mihalas1978}.
Here $a$ is the  normal Voigt-parameter, and $\xD$ is the frequency distance from the line center in Doppler units of the Lyman $\alpha$ resonance.

\begin{figure}
\centering 
\includegraphics[width=0.90\columnwidth]
{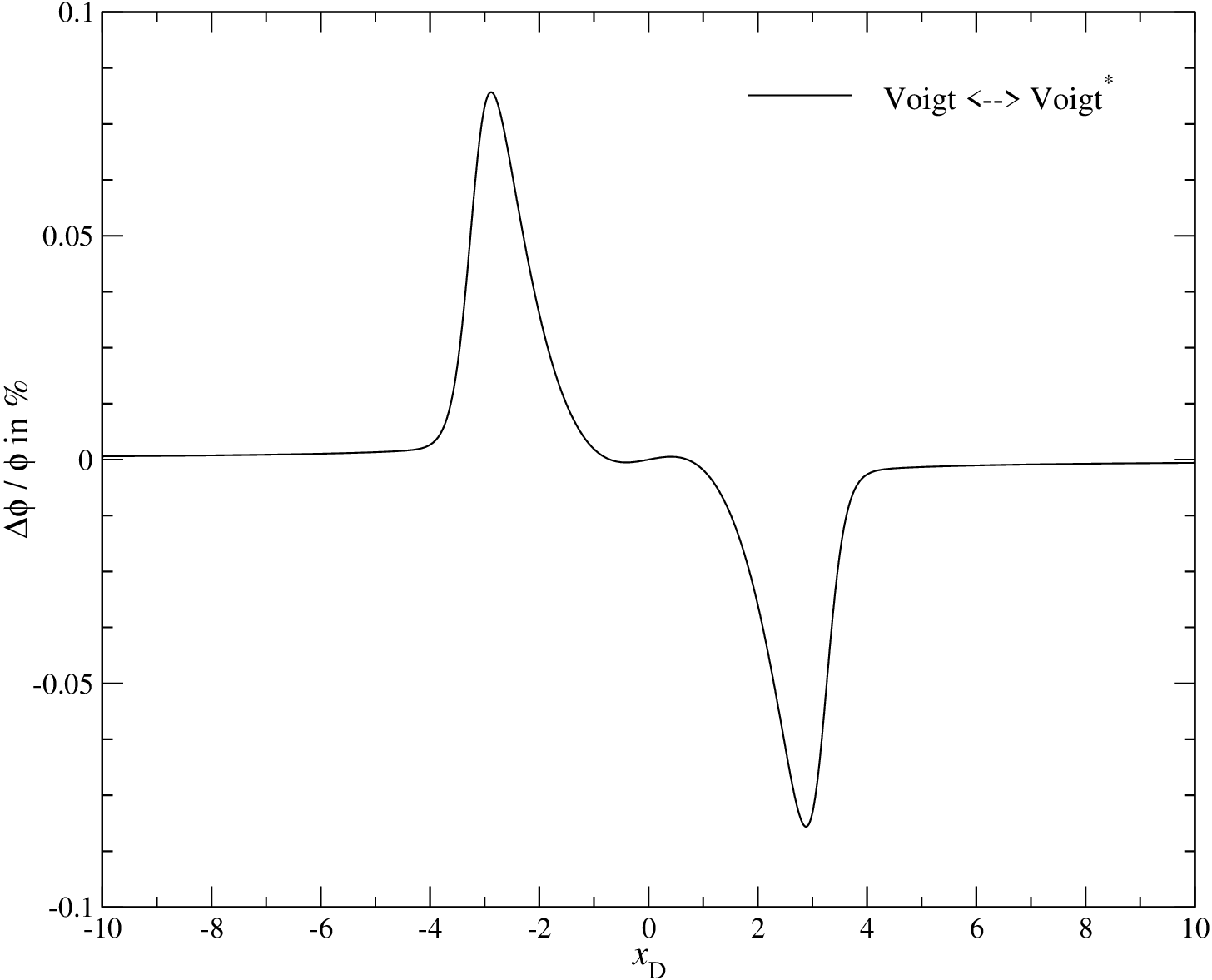}
\caption
{Small correction to the normal Voigt-profile. We show the relative difference of the standard Voigt profile $\phi\propto H(\xD, a)$ in comparison with the modified expression based on $\tilde{H}(\xD, a)$ at $z=1100$. }
\label{fig:Dphi_Voigt}
\end{figure}
%
However, due to the steepness of the Lorentzian close to the line center this approximation leads to a small inaccuracy ($\Delta\phi/\phi\sim 10^{-4}-10^{-3}$), which actually is not necessary.
To avoid this one should simply replace $\xD$ and $a$ in the Voigt-integral with $\tilde{x}_{\rm D}=\xD/[1+\xD\frac{\Delta\nu_{\rm D}}{\nu_{21}}]$ and $\tilde{a}=a/[1+\xD\frac{\Delta\nu_{\rm D}}{\nu_{21}}]$, and in addition multiply $H$ with $1/[1+\xD\frac{\Delta\nu_{\rm D}}{\nu_{21}}]$, yielding
\beal
\label{app:H_V_corr}
\tilde{H}(\xD, a)&=
\frac{H(\tilde{x}_{\rm D}, \tilde{a})}{[1+\xD\frac{\Delta\nu_{\rm D}}{\nu_{21}}]}.
\end{align}
As one can see that then the Voigt profile, $\phi_{\rm V}=\tilde{H}(\xD, a)/\sqrt{\pi}$, effectively behaves as
\beal
\label{app:Voigt_cent_app}
\phi_{\rm V}(\xD)
&\approx\frac{1}{\sqrt{\pi}}\,\frac{e^{-\tilde{x}_{\rm D}^2}}{[1+\xD\frac{\Delta\nu_{\rm D}}{\nu_{21}}]}
\nonumber\\
&\approx\frac{1}{\sqrt{\pi}}\,e^{-x_{\rm D}^2}\left[1-\xD\frac{\Delta\nu_{\rm D}}{\nu_{21}}(1+2 x_{\rm D}^2)\right]
\end{align}
close to the line center. 
At $\xD\sim \pm1$ this implies a relative correction of $\Delta\phi/\phi\sim \mp3\frac{\Delta\nu_{\rm D}}{\nu_{21}}\approx \mp\pot{6}{-5}$ at $z\sim 1100$. However, at $\xD\sim 3$ this corrections is expected to reach the $0.1\%$ level.
In Fig.~\ref{fig:Dphi_Voigt} the frequency dependence of this correction is shown in more detail, confirming these statements.
Note that as expected the behavior in distant wings ($|\xD|\gg~1$) is not changed.

\begin{figure}
\centering 
\includegraphics[width=0.90\columnwidth]
{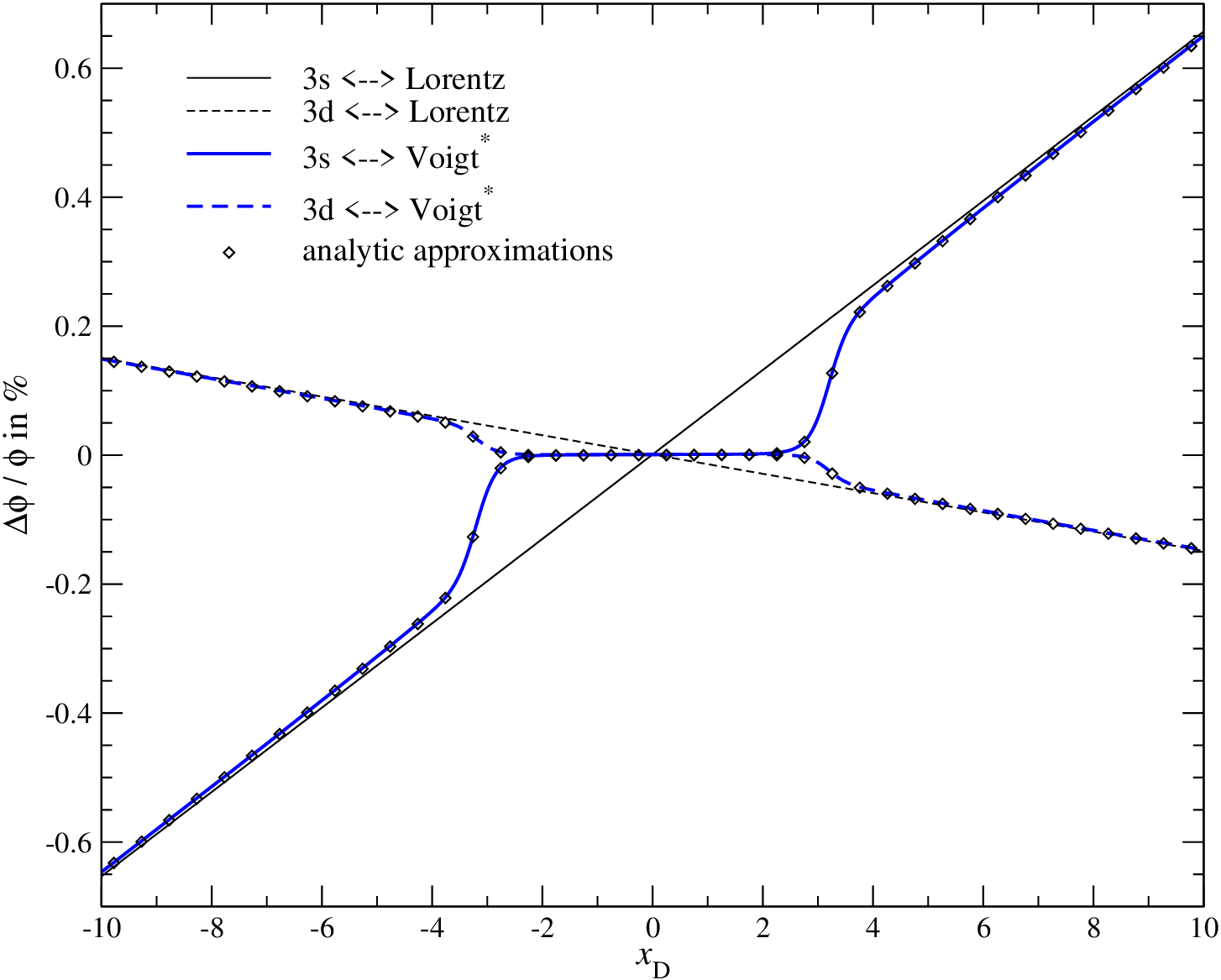}
\caption
{Relative difference of the 3s and 3d two-photon profiles with respect to the Lorentzian (thin line) and the Voigt profile (thick lines). In the first case the motion of the atoms was neglected, while for the comparison with the Voigt profile it was included. The boxes show the analytic approximation for the correction including the motion.}
\label{fig:Dphi_details}
\end{figure}
\subsection{Two-photon profiles for moving atoms}
As mentioned above, to include the motion of the atoms in the computation of the two-photon profiles one can in principle directly use the expression~\eqref{app:therm_av}. However, the computation of this integral is rather time-consuming, and in the very distant wings also is not necessary\footnote{There the motion of the atoms can be neglected since the two-photon profiles normally vary very slowly with frequency. Although this is not entirely true for the $n$s two-photon profiles close to the frequencies where $\phi^{2\gamma}$ vanishes, one expects a negligible additional correction due to this approximation.}. 
Therefore we use a different approach in which we utilize the fact that the relative difference, $\Delta\phi^{2\gamma}_{{\rm r}, i}(\nu)/\phi_{\Lambda}(\nu)$, of the restframe two photon profile, $\phi^{2\gamma}_{{\rm r}, i}(\nu)$, and the normal Lorentzian, $\phi_{\Lambda}(\nu)$, varies very slowly with frequency, as compared to $\phi_{\Lambda}(\nu)$.
Therefore, to lowest order for moving atoms the two-photon profile will be given by
$\phi^{2\gamma}_{{\rm m}, i}(\xD, a)\approx \phi_{\rm V}(\xD, a)\left [ 1+ \Delta\phi^{2\gamma}_{{\rm r}, i}(\nu)/\phi_{\Lambda}(\nu)\right]$.

However, close to the line center this approximation due to the steepness of the Lorentzian again becomes inaccurate at the percent level.
The lowest order correction can be computed approximating the rest frame two-photon profiles close to the Lyman $\alpha$ line center by
%
\beal
\label{app:Phi_expansion}
\phi^{2\gamma}_{{\rm r}, i}(\nu)
\approx \phi_{\Lambda}(\nu)[1+\alpha_i+\beta_i\,\xD].
\end{align}
Here  is the 
The coefficients $\alpha_i$ and $\beta_i$ for the 3s and 3d level are given in Table~\ref{tab:alphas}.
\begin{table}
\caption{Parameters for the 3s and 3d two photon profiles around the line center. These were determined in the range $-10\lesssim \xD \lesssim 10$. For explanation see Appendix~\ref{app:profiles}}
\label{tab:alphas}
\centering
{
\begin{tabular}{@{}ccc}
\hline
\hline
  & $\alpha_i$ & $\beta_i$ \\
\hline
3s & $\pot{3.73335}{-6}$ & $\pot{+6.5325}{-4}$
\\
3d & $\pot{7.45559}{-7}$ & $\pot{-1.5129}{-4}$ 
\\
\hline
\hline
\end{tabular}
}
\end{table}

%
Using the same method as described in Sect. 9.2 of \citet{Mihalas1978} it is then easy to show that for $|\xD|\leq 1000$
\beal
\label{app:Phi_expansion_m}
\phi^{2\gamma}_{{\rm m}, i}(\xD, a)
\!\approx\!
\phi_{\rm V}(\xD, a)\left [ 1+ \frac{\Delta\phi^{2\gamma}_{{\rm r}, i}(\nu)}{\phi_{\Lambda}(\nu)}
+\frac{\beta_i}{2}\,
\frac{\partial_{\tilde{x}_{\rm D}}H(\tilde{x}_{\rm D}, \tilde{a})}{\tilde{H}(\tilde{x}_{\rm D}, \tilde{a})}
\right].
\end{align}
%
Close to the line-center we will use this expression for the two-photon profiles. However, far away from the Lyman $\alpha$ resonance we will neglect the effect due to the motion of the atoms, and simply use the rest frame two-photon profiles. 

\changeI{However, we find that the correction in connection with the motion of the atom are not important for the cosmological recombination problem.}

\end{appendix}


\bibliographystyle{aa} 
\bibliography{Lit}

\begin{thebibliography}{38}
\expandafter\ifx\csname natexlab\endcsname\relax\def\natexlab#1{#1}\fi

\bibitem[{{Chluba} {et~al.}(2007){Chluba}, {Rubi{\~n}o-Mart{\'{\i}}n}, \&
  {Sunyaev}}]{Chluba2007}
{Chluba}, J., {Rubi{\~n}o-Mart{\'{\i}}n}, J.~A., \& {Sunyaev}, R.~A. 2007,
  \mnras, 374, 1310

\bibitem[{{Chluba} \& {Sunyaev}(2006{\natexlab{a}})}]{Chluba2006b}
{Chluba}, J. \& {Sunyaev}, R.~A. 2006{\natexlab{a}}, \aap, 458, L29

\bibitem[{{Chluba} \& {Sunyaev}(2006{\natexlab{b}})}]{Chluba2006}
{Chluba}, J. \& {Sunyaev}, R.~A. 2006{\natexlab{b}}, \aap, 446, 39

\bibitem[{{Chluba} \& {Sunyaev}(2007)}]{Chluba2007b}
{Chluba}, J. \& {Sunyaev}, R.~A. 2007, \aap, 475, 109

\bibitem[{{Chluba} \& {Sunyaev}(2008{\natexlab{a}})}]{Chluba2008c}
{Chluba}, J. \& {Sunyaev}, R.~A. 2008{\natexlab{a}}, ArXiv e-prints

\bibitem[{{Chluba} \& {Sunyaev}(2008{\natexlab{b}})}]{Chluba2008b}
{Chluba}, J. \& {Sunyaev}, R.~A. 2008{\natexlab{b}}, ArXiv e-prints

\bibitem[{{Chluba} \& {Sunyaev}(2008{\natexlab{c}})}]{Chluba2008a}
{Chluba}, J. \& {Sunyaev}, R.~A. 2008{\natexlab{c}}, \aap, 480, 629

\bibitem[{{Chluba} \& {Sunyaev}(2009)}]{Chlubaprep}
{Chluba}, J. \& {Sunyaev}, R.~A. 2009, in Preparation

\bibitem[{{Cresser} {et~al.}(1986){Cresser}, {Tang}, {Salamo}, \&
  {Chan}}]{Cresser1986}
{Cresser}, J.~D., {Tang}, A.~Z., {Salamo}, G.~J., \& {Chan}, F.~T. 1986, \pra,
  33, 1677

\bibitem[{{de Bernardis} {et~al.}(2009){de Bernardis}, {Bean}, {Galli},
  {Melchiorri}, {Silk}, \& {Verde}}]{deBernardis2009}
{de Bernardis}, F., {Bean}, R., {Galli}, S., {et~al.} 2009, \prd, 79, 043503

\bibitem[{{Dubrovich} \& {Grachev}(2005)}]{Dubrovich2005}
{Dubrovich}, V.~K. \& {Grachev}, S.~I. 2005, Astronomy Letters, 31, 359

\bibitem[{{Eisenstein}(2005)}]{EisensteinRev2005}
{Eisenstein}, D.~J. 2005, New Astronomy Review, 49, 360

\bibitem[{{Eisenstein} {et~al.}(2005){Eisenstein}, {Zehavi}, {Hogg},
  {Scoccimarro}, {Blanton}, {Nichol}, {Scranton}, {Seo}, {Tegmark}, {Zheng},
  {Anderson}, {Annis}, {Bahcall}, {Brinkmann}, {Burles}, {Castander},
  {Connolly}, {Csabai}, {Doi}, {Fukugita}, {Frieman}, {Glazebrook}, {Gunn},
  {Hendry}, {Hennessy}, {Ivezi{\'c}}, {Kent}, {Knapp}, {Lin}, {Loh}, {Lupton},
  {Margon}, {McKay}, {Meiksin}, {Munn}, {Pope}, {Richmond}, {Schlegel},
  {Schneider}, {Shimasaku}, {Stoughton}, {Strauss}, {SubbaRao}, {Szalay},
  {Szapudi}, {Tucker}, {Yanny}, \& {York}}]{Eisenstein2005}
{Eisenstein}, D.~J., {Zehavi}, I., {Hogg}, D.~W., {et~al.} 2005, \apj, 633, 560

\bibitem[{{Fendt} {et~al.}(2008){Fendt}, {Chluba}, {Rubino-Martin}, \&
  {Wandelt}}]{Fendt2008}
{Fendt}, W.~A., {Chluba}, J., {Rubino-Martin}, J.~A., \& {Wandelt}, B.~D. 2008,
  ArXiv e-prints, 807

\bibitem[{{G{\"o}ppert-Mayer}(1931)}]{Mayer1931}
{G{\"o}ppert-Mayer}. 1931, Annalen der Physik, 9, 273

\bibitem[{{Grachev} \& {Dubrovich}(2008)}]{Grachev2008}
{Grachev}, S.~I. \& {Dubrovich}, V.~K. 2008, Astronomy Letters, 34, 439

\bibitem[{{Hirata}(2008)}]{Hirata2008}
{Hirata}, C.~M. 2008, ArXiv e-prints

\bibitem[{{Hu} {et~al.}(1995){Hu}, {Scott}, {Sugiyama}, \& {White}}]{Hu1995}
{Hu}, W., {Scott}, D., {Sugiyama}, N., \& {White}, M. 1995, \prd, 52, 5498

\bibitem[{{H{\"u}tsi}(2006)}]{Gert2006}
{H{\"u}tsi}, G. 2006, \aap, 449, 891

\bibitem[{{Jones} \& {Wyse}(1985)}]{Jones1985}
{Jones}, B.~J.~T. \& {Wyse}, R.~F.~G. 1985, \aap, 149, 144

\bibitem[{{Karshenboim} \& {Ivanov}(2008)}]{Karshenboim2008}
{Karshenboim}, S.~G. \& {Ivanov}, V.~G. 2008, Astronomy Letters, 34, 289

\bibitem[{{Kholupenko} \& {Ivanchik}(2006)}]{Kholu2006}
{Kholupenko}, E.~E. \& {Ivanchik}, A.~V. 2006, Astronomy Letters, 32, 795

\bibitem[{{Mihalas}(1978)}]{Mihalas1978}
{Mihalas}, D. 1978, {Stellar atmospheres /2nd edition/} (San Francisco,
  W.~H.~Freeman and Co., 1978.~650 p.)

\bibitem[{{Peebles}(1968)}]{Peebles68}
{Peebles}, P.~J.~E. 1968, \apj, 153, 1

\bibitem[{{Peebles} {et~al.}(2000){Peebles}, {Seager}, \& {Hu}}]{Peebles2000}
{Peebles}, P.~J.~E., {Seager}, S., \& {Hu}, W. 2000, \apjl, 539, L1

\bibitem[{{Rubi{\~n}o-Mart{\'{\i}}n} {et~al.}(2006){Rubi{\~n}o-Mart{\'{\i}}n},
  {Chluba}, \& {Sunyaev}}]{Jose2006}
{Rubi{\~n}o-Mart{\'{\i}}n}, J.~A., {Chluba}, J., \& {Sunyaev}, R.~A. 2006,
  \mnras, 371, 1939

\bibitem[{{Rubi{\~n}o-Mart{\'{\i}}n} {et~al.}(2008){Rubi{\~n}o-Mart{\'{\i}}n},
  {Chluba}, \& {Sunyaev}}]{Jose2008}
{Rubi{\~n}o-Mart{\'{\i}}n}, J.~A., {Chluba}, J., \& {Sunyaev}, R.~A. 2008,
  \aap, 485, 377

\bibitem[{{Seager} {et~al.}(1999){Seager}, {Sasselov}, \&
  {Scott}}]{SeagerRecfast1999}
{Seager}, S., {Sasselov}, D.~D., \& {Scott}, D. 1999, \apjl, 523, L1

\bibitem[{{Seager} {et~al.}(2000){Seager}, {Sasselov}, \& {Scott}}]{Seager2000}
{Seager}, S., {Sasselov}, D.~D., \& {Scott}, D. 2000, \apjs, 128, 407

\bibitem[{{Seljak} {et~al.}(2003){Seljak}, {Sugiyama}, {White}, \&
  {Zaldarriaga}}]{Seljak2003}
{Seljak}, U., {Sugiyama}, N., {White}, M., \& {Zaldarriaga}, M. 2003, \prd, 68,
  083507

\bibitem[{{Silk}(1968)}]{Silk1968}
{Silk}, J. 1968, \apj, 151, 459

\bibitem[{{Sunyaev} \& {Chluba}(2007)}]{Sunyaev2007}
{Sunyaev}, R.~A. \& {Chluba}, J. 2007, Nuovo Cimento B Serie, 122, 919

\bibitem[{{Sunyaev} \& {Chluba}(2008)}]{Sunyaev2008}
{Sunyaev}, R.~A. \& {Chluba}, J. 2008, in Astronomical Society of the Pacific
  Conference Series, Vol. 395, Frontiers of Astrophysics: A Celebration of
  NRAO's 50th Anniversary, ed. A.~H. {Bridle}, J.~J. {Condon}, \& G.~C. {Hunt},
  35--+

\bibitem[{{Sunyaev} \& {Zeldovich}(1970)}]{Sunyaev1970}
{Sunyaev}, R.~A. \& {Zeldovich}, Y.~B. 1970, Astrophysics and Space Science, 7,
  3

\bibitem[{{Switzer} \& {Hirata}(2008)}]{Switzer2007I}
{Switzer}, E.~R. \& {Hirata}, C.~M. 2008, \prd, 77, 083006

\bibitem[{{Wong} {et~al.}(2008){Wong}, {Moss}, \& {Scott}}]{Wong2008}
{Wong}, W.~Y., {Moss}, A., \& {Scott}, D. 2008, \mnras, 386, 1023

\bibitem[{{Wong} \& {Scott}(2007)}]{Wong2007}
{Wong}, W.~Y. \& {Scott}, D. 2007, \mnras, 375, 1441

\bibitem[{{Zeldovich} {et~al.}(1968){Zeldovich}, {Kurt}, \&
  {Syunyaev}}]{Zeldovich68}
{Zeldovich}, Y.~B., {Kurt}, V.~G., \& {Syunyaev}, R.~A. 1968, Zhurnal
  Eksperimental noi i Teoreticheskoi Fiziki, 55, 278

\end{thebibliography}

\end{document}